\documentclass[%
reprint,prd,
superscriptaddress,
nofootinbib,
amsmath,amssymb,
]{revtex4-2}

\usepackage{graphicx}
\usepackage{dcolumn}
\usepackage{bm}
\usepackage{xcolor}
\usepackage{mathrsfs}
\usepackage{lipsum}
\usepackage{hyperref}
\hypersetup{
	pdftitle={},
	colorlinks=true,     
	linkcolor=blue,      
	citecolor=teal,      
	filecolor=black,      
	urlcolor=teal        
}
\usepackage{cleveref}
\usepackage{amsmath,amssymb}
\usepackage{amsfonts}

\newcommand{\pp}{\mathbf{p}}
\newcommand{\qq}{\mathbf{q}}

\begin{document}

\title{Nonthermal fixed point for massless scalar field theories}
	
	\author{Viktoria Noel}
    \affiliation{Institute for Theoretical Physics, Heidelberg University, Philosophenweg 16, 69120 Heidelberg, Germany}
    \affiliation{\mbox{Institute for Theoretical Physics, University of Tübingen, Auf der Morgenstelle 14, 72076 T\"ubingen, Germany}}

	\author{Chuang Huang}
    \email{huang@thphys.uni-heidelberg.de}
	\affiliation{Institute for Theoretical Physics, Heidelberg University, Philosophenweg 16, 69120 Heidelberg, Germany}
	\author{Aleksandr N.\ Mikheev}
	\affiliation{Institute for Theoretical Physics, Heidelberg University, Philosophenweg 16, 69120 Heidelberg, Germany}
    \affiliation{\mbox{Department of Physics, University of Konstanz, Universit{\"a}tsstra{\ss}e  10, 78464 Konstanz, Germany}}
    \affiliation{Institute of Physics, Johannes Gutenberg University Mainz, Staudingerweg 7, 55128 Mainz, Germany}
	\author{Jürgen Berges}
	\affiliation{Institute for Theoretical Physics, Heidelberg University, Philosophenweg 16, 69120 Heidelberg, Germany}

	\begin{abstract}
    We compute universal scaling exponents and scaling functions for massless scalar field theories far from equilibrium. Using a large-$N$ expansion to next-to-leading order, we investigate the nonperturbative infrared behaviour associated with the transport of a) energy or b) effective particle number towards low frequencies $\omega$ and momenta $\bar p$. For massless dispersion $\omega = \bar{p}$, we determine the scaling form of the distribution function whose universal power-law tail $f_S(\bar p) \sim 1/\bar{p}^\kappa$ is found to be a)~$\kappa = d+1$ and b)~$\kappa = d$ in $d=3$ spatial dimensions. Our results establish that relativistic and nonrelativistic theories do not belong to the same universality class in this case, which opens up new applications to far-from-equilibrium phenomena in high-energy and (quantum) many-body physics with linear dispersions.	
\end{abstract}

\maketitle

\section{Introduction}\label{sec:intro}

Thermal fixed points and their associated universality classes, which capture how macroscopic behaviour near criticality becomes insensitive to microscopic details, serve as well-established cornerstones in our understanding of equilibrium physics \cite{wilson1971renormalization, wilson1971renormalization2, Hohenberg:1977ym}. 
Significant progress has also been made in identifying universal features in nonequilibrium systems, for instance, wave turbulence \cite{Frisch1995a, Zakharov1992a}, coarsening dynamics \cite{Bray:1994zz}, ageing \cite{Calabrese_2005, gambassi2015}, and dynamical phase transitions \cite{Heyl:2017blm}.
While each of these exhibits characteristic scaling behaviour largely independent of microscopic details, a systematic classification of nonequilibrium scaling dynamics remains an open challenge.

In isolated quantum many-body systems, nonthermal fixed points \cite{Berges:2008wm, Berges:2008sr, Schmied:2018upn, Schmied:2018mte, Nowak:2013juc, Berges:2015kfa, Glidden:2020qmu, Berges:2014bba,Mikheev:2023juq} have been established as emergent attractor solutions that feature self-similar scaling and arise from a broad range of initial conditions in far-from-equilibrium systems.
They are associated with the transport of conserved quantities, leading to a redistribution of excitations across momentum scales.
Notably, this universal behaviour is not dependent on fine-tuning external parameters but rather develops dynamically as the system evolves far from equilibrium.
A variety of theoretical approaches have been employed to study nonthermal fixed points and the associated scaling behaviour, such as effective kinetic theories \cite{Berges:2010ez, Scheppach:2009wu, Chantesana:2018qsb, PineiroOrioli:2015cpb, Mikheev:2018adp, Walz:2017ffj, Rosenhaus:2024iqw, Rosenhaus:2025mgj, Rosenhaus2024a.PhysRevE.109.064127, Rosenhaus2024a.PhysRevLett.133.244002, Hu:2025bqi}, real-time classical statistical simulations \cite{Berges:2008wm, Gasenzer:2011by, Berges:2013eia, Nowak:2011sk, PineiroOrioli:2015cpb, Heinen:2022rew, Noel:2023oyz, Schole:2012kt, Karl:2013kua, Ewerz:2014tua, Berges:2016nru, Karl:2016wko, Siovitz:2024aqi, Noel:2025mtb, Rasch:2025hth, Rasch:2025kna, Mikheev:2024pur, Siovitz:2023ius, Walz:2017ffj}, as well as methods based on two-particle irreducible (2PI) quantum effective actions \cite{Berges:2015kfa, Shen:2019jhl, Chantesana:2018qsb, Scheppach:2009wu, Walz:2017ffj, Heinen:2022ham, Berges:2010ez, Berges:2016nru, Berges:2010ez, Gasenzer:2010rq, PineiroOrioli:2015cpb}. 

Many studies focus on the time evolution of correlation functions by preparing a highly overoccupied initial state and observing the system as it dynamically evolves towards a nonthermal fixed point.
In this case, relativistic and nonrelativistic scalar field theories are often argued to fall into the same universality class \cite{PineiroOrioli:2015cpb}. 
In relativistic scalar field theories, even for a vanishing mass parameter in the underlying Hamiltonian or action there is a nonzero effective mass gap arising due to quantum-statistical corrections. As a consequence, the late-time dynamics is effectively nonrelativistic \cite{Deng:2018xsk, namjoo2018relativistic}.
This naturally raises the question of whether a genuinely relativistic scaling regime can exist, where the mass parameter is tuned to be massless. This is very similar in spirit to equilibrium critical phenomena, where the mass parameter or temperature is adjusted to its critical value.  

Addressing this is particularly challenging in strongly correlated regimes, where perturbative expansions break down. 
The 2PI effective action provides a powerful tool in this case.
When combined with a $1/N$ expansion in terms of a large number of field components $N$ beyond leading order~\cite{Berges:2001fi}, this method offers a nonperturbative approach that does not rely on a small coupling parameter, enabling a controlled and systematic study of nonperturbative scaling phenomena. 

In this work, we apply the 2PI effective action framework to self-interacting scalar fields with $\mathrm{O}(N)$ symmetry in three spatial dimensions, focusing on nonthermal scaling solutions with massless dispersion. This has been suggested in~Ref.~\cite{PineiroOrioli:2015cpb} based on general scaling assumptions; however, explicit solutions and the scaling form of the distribution functions have not been established so far.
Here we provide these solutions based on a large-$N$ expansion to next-to-leading order (NLO). 
We extract a universal nonthermal scaling function in the infrared, which is markedly different from a Bose--Einstein distribution. The nonthermal scaling distribution exhibits a universal power-law tail $\sim 1/\bar{p}^\kappa$ for momenta $\bar p$ in the infrared regime with scaling exponent $\kappa \ge d$ for $d=3$ spatial dimensions.
This reflects the genuinely nonequilibrium nature of the solution, for which we establish a generalised fluctuation–dissipation-type relation as an asymptotic property. 

The paper is organised as follows. 
Sec.~\ref{sec:theory} introduces the theory and the nonperturbative large-$N$ approximation to NLO. 
Sec.~\ref{sec:universal} presents the fixed-point equations in the scaling regime, based on an effective kinetic description. We analyse the on-shell scaling behaviour of the self-energies involved and give the results for scaling functions and exponents in Sec.~\ref{sec:explicit}.
A conclusion is presented in Sec.~\ref{sec:conclusion}, followed by appendices on further calculational details. 
    
\section{Scalar field theory}
\label{sec:theory}
We consider a relativistic scalar field theory for real fields $\phi_a(x)$ with components $a=1,\ldots,N$, depending on the variable $x=\left(x^0, \mathbf{x}\right)$ for time $x^0$ and space $\mathbf{x}$ in $d$ spatial dimensions. The $\mathrm{O}(N)$-symmetric classical action reads
\begin{equation}
\begin{aligned}
S[\phi] & =\frac{1}{2} \int_{x y} \phi_a(x) i G_{0,a b}^{-1}(x, y) \phi_b(y) \\
& -\frac{\lambda}{4!N} \int_{x} \phi_a(x) \phi_a(x) \phi_b(x) \phi_b(x),
\label{eq:classaction}
\end{aligned}
\end{equation}
where $\int_x = \int d^{d+1}x$. Here the classical inverse propagator is
\begin{equation}
i G_{0,a b}^{-1}(x, y)=-\left(\square_x+m^2\right) \delta_{a b} \delta^{(d+1)}(x-y)
\label{eq:G0}
\end{equation}
with $\square_x=\partial_{x^0}^2-\partial_{\mathbf{x}}^2$ and summation over repeated indices is implied.

We are interested in nonequilibrium phenomena, which can be characterised in terms of universal scaling behaviour of correlation functions for Heisenberg field operators $\hat{\phi}_a(x)$ of the corresponding quantum theory. For nonequilibrium two-point functions, there are two linearly independent correlation functions, which can be associated with the expectation values of the commutator and the anti-commutator of field operators~\cite{Berges:2015kfa}:
\begin{subequations}
\begin{align}
    \rho_{ab}(x,y) &= i\,\langle [\hat{\phi}_a(x),\hat{\phi}_b(y)] \rangle\,, \label{eq:rho}\\
    F_{ab}(x,y) &= \frac{1}{2}\langle \{ \hat{\phi}_a(x),\hat{\phi}_b(y)\}\rangle - \langle\hat{\phi}_a(x)\rangle\langle\hat{\phi}_b(x)\rangle\,, \label{eq:F}
\end{align}
\end{subequations}
where $[\hat{\phi}_a(x),\hat{\phi}_b(y)] = \hat{\phi}_a(x) \hat{\phi}_b(y) - \hat{\phi}_b(y) \hat{\phi}_a(x)$ and $\{\hat{\phi}_a(x),\hat{\phi}_b(y)\}= \hat{\phi}_a(x) \hat{\phi}_b(y) + \hat{\phi}_b(y) \hat{\phi}_a(x)$. The two functions \eqref{eq:rho} and \eqref{eq:F} are commonly referred to as the spectral function and the statistical function, respectively. Alternatively to the spectral function \eqref{eq:rho}, one can introduce the retarded two-point function $G^R_{ab}(x,y)$ and the advanced function $G^A_{ab}(x,y)$ with~\cite{Berges:2015kfa}
\begin{equation}
    \rho_{ab}(x,y) = G^R_{ab}(x,y) - G^A_{ab}(x,y) \, .
\end{equation}
For the real scalar field theory we have $G^A_{ab}(x,y) = G^R_{ba}(y,x)$. 

While the statistical function \eqref{eq:F} and the spectral function \eqref{eq:rho} are connected in thermal equilibrium by the fluctuation--dissipation relation, this is in general not the case out of equilibrium~\cite{Berges:2008wm}. Out of equilibrium, $F$ and $G^{R,A}$ or $\rho$ form a basis set for all possible two-point correlation functions of the fields. 

We consider in this work the symmetric regime, where the expectation value of the field operator vanishes, $\langle \hat{\phi}_a(x)\rangle =0$. By virtue of the $O(N)$ symmetry, we write $G^R_{ab}(x,y)=G^R(x,y)\delta_{ab}$, $F_{ab}(x,y)=F(x,y)\delta_{ab}$, and also denote $G_{0,a b}^{-1}(x, y)=G_{0}^{-1}(x, y)\delta_{ab}$. 

Retarded (advanced) and statistical correlation functions can also be defined in classical field theory for statistical averages of fields, where the commutator is replaced by the Poisson bracket and the anti-commutator becomes two times the product of the classical fields~\cite{Berges:2015kfa}. It turns out that the quantum and classical field theories belong to the same universality class for the nonequilibrium scaling phenomena we will consider. This relation is also summarised in App.~\ref{app:keldysh},
and forms the basis for the description employed throughout the remainder of this work. 

\subsection{Self-energies at NLO large-$N$}
\label{sec:2pi}

The retarded self-energy $\Sigma^{R}$ encodes the difference between the classical inverse propagator (\ref{eq:G0}) and the inverse of the full retarded correlation function $(G^R)^{-1}$ involving all quantum-statistical corrections. More precisely, 
\begin{equation}
		{(G^{R})}^{-1}{(x, y)}  =-i G_{0}^{-1} (x, y)+\Sigma^{R}{(x, y)}.
\label{eq:realgammar}
\end{equation}
Similar to the above discussion of the retarded and statistical correlation functions, there is also a statistical self-energy function, which is given by~\cite{Berges:2001fi} 
\begin{equation}
\Sigma^F(x,y) = \int_{zw} {(G^{R})}^{-1}{(x, z)}\, F(z,w)\, {(G^{A})}^{-1}{(w, y)}\, .
	\label{eq:realgammaf}
\end{equation}

For known self-energies $\Sigma^R$ and $\Sigma^F$, (\ref{eq:realgammar}) and (\ref{eq:realgammaf}) would represent an exact set of equations for the determination of $G^{R,A}$ and $F$. However, in practice their solution requires approximations. In the following, we consider a self-consistent large-$N$ expansion to next-to-leading order (NLO) to determine the self-energies. In this nonperturbative description, the self-energies are given in terms of the two-point correlation functions themselves, i.e.~$\Sigma^F=\Sigma^F(F,G^{R,A})$ etc., such that (\ref{eq:realgammar}) and (\ref{eq:realgammaf}) represent a closed set of equations for $G^{R,A}$ and $F$ at any given order in the expansion.

\begin{figure}
    \centering
    \includegraphics[width=0.99\linewidth]{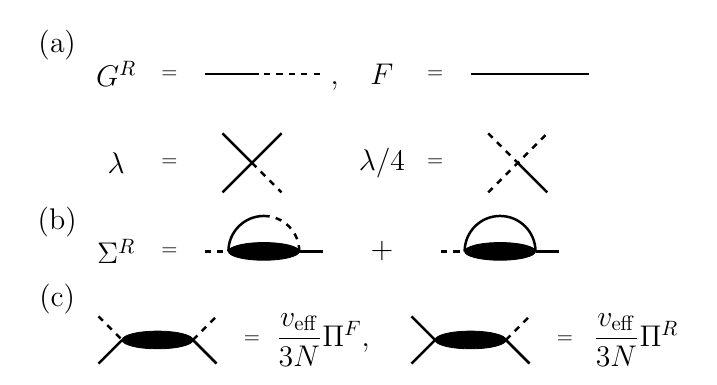}
    \caption{Diagrammatically, the retarded propagator $G^R$ and $F$ can be represented as shown in (a), along with the relevant interaction vertex. The diagrammatic expression for the self-energy is shown in (b) with (c) showing the resummations. The effective vertex $v_{\mathrm{eff}}$ sums the infinite chain of diagrams appearing at NLO in the $1/N$ expansion.}
    \label{fig:2pilargeN}
\end{figure}

The self-energies at NLO have been determined for the quantum field theory in Ref.~\cite{Berges:2015kfa}. We consider in the following the classical-statistical limit of the NLO self-energies, which neglects subleading quantum corrections for the scaling phenomena as explained in App.~\ref{app:keldysh}. We separate from the self-energy contributions at NLO a local part, which, for notational purposes, can be combined with the mass term in (\ref{eq:G0}) with the replacement
\begin{equation}
    m^2 \, \rightarrow \, M^2(x) = m^2 + \lambda \frac{N+2}{6N} F(x,x) \, .
\end{equation}
With this notation, the retarded self-energy in (\ref{eq:realgammar}) only contains the nonlocal contributions
\begin{equation}
\Sigma^R(x, y)  =-\frac{\lambda}{3 N}\left[F(x, y) \,I^R(x, y)+G^R(x, y)\, I^F(x, y)\right].
\label{eq:sigmaR}
\end{equation}
The statistical self-energy is given by
\begin{equation}
\Sigma^F(x, y)  =-\frac{\lambda}{3 N} F(x, y)\,I^F(x, y) \, .
\label{eq:statself}
\end{equation}
Here the retarded and statistical summation functions, $I^R(x,y)$ and $I^F(x,y)$, encoding the NLO ``chain" diagrams read 
\begin{subequations}
    \begin{align}
	I^R(x, y) & =  \Pi^R(x, y)-\int_{z} I^R(x, z)\, \Pi^R(z, y),
    \label{eq:IR}\\
	I^F(x, y) &=  \Pi^F(x, y) - \int_{z} I^R(x, z)\, \Pi^F(z, y) \nonumber\\
	& - \int_{z} I^F(x, z)\, \Pi^A(z, y),
    \label{eq:IF}
\end{align}
\end{subequations}
where the retarded and statistical ``one-loop" functions are
\begin{subequations}
    \begin{align}
    \Pi^R(x, y) &=  \frac{\lambda}{3} F(x, y)\,G^R(x, y) \, , 
    \label{eq:pirho}\\
	\Pi^F(x, y) & =  \frac{\lambda}{6}F(x, y)\,F(x, y) \, .
    \label{eq:piff}
    \end{align}
\end{subequations}
A diagrammatic representation of the contributions is given in Fig.~\ref{fig:2pilargeN}.
All of the above integrals extend from $-\infty$ to $\infty$ in time and space for the infinite volume limit considered. Compared to the initial-value problems discussed in Ref.~\cite{Berges:2015kfa}, this implies that the initial time is sent to the remote past when considering scaling solutions.

\section{Nonequilibrium scaling}
\label{sec:universal}
We consider spatially homogeneous systems, such that the correlation functions only depend on relative spatial coordinates, i.e.\ $G^R(x,y) = G^R(x^0,y^0;\mathbf{x}-\mathbf{y})$ and $F(x,y) = F(x^0,y^0;\mathbf{x}-\mathbf{y})$. We also assume spatial isotropy, such that these functions only depend on the absolute values $|\mathbf{x}-\mathbf{y}|$. Since time-translation invariance does not hold out of equilibrium in general, we take into account the dependence of the correlation functions on central time $t$ and relative time $\Delta t$ given by
\begin{equation}
    t = \frac{x^0+y^0}{2}\,, \quad \Delta t = x^0-y^0 \, .
\end{equation}

In order to discuss possible scaling solutions of equations (\ref{eq:realgammar}) and (\ref{eq:realgammaf}), it is convenient to Fourier transform the correlation functions with respect to relative coordinates,
\begin{equation}
    F(t,\omega,\mathbf{p}) = \int_{\Delta t,\mathbf x} e^{i (\omega \Delta t - \mathbf{p}\mathbf{x})} F\left(t+\frac{\Delta t}{2}, t- \frac{\Delta t}{2}; \mathbf{x}\right), 
\end{equation}
and similarly for the retarded correlator. While $F(t,\omega,\mathbf{p})$ is real, the retarded correlator $G^R(t,\omega,\mathbf{p})$ is in general a complex function for the real scalar field theory (\ref{eq:classaction}). Since for the complex conjugate of the retarded correlator $(G^{R}(t,\omega,\mathbf{p}))^*=G^{A}(t,\omega,\mathbf{p})$, it is convenient to introduce the real-valued 
\begin{equation}
    \tilde{\rho}(t,\omega,\mathbf{p}) = -i \left[G^R(t,\omega,\mathbf{p}) - G^A(t,\omega,\mathbf{p}) \right],
    \label{eq:tilderho}
\end{equation}
which represents the spectral function (\ref{eq:rho}) after Fourier transforming relative coordinates, and the additional factor $-i$ is introduced to have a real quantity.

\subsection{Gradient expansion}
\label{sec:gradexp}

Using the NLO large-$N$ approximation for the equations (\ref{eq:realgammar}) and (\ref{eq:realgammaf}), it would still be a formidable task to identify nonequilibrium scaling solutions for correlation functions. To proceed, we introduce the following additional approximations. 

We assume that variations with central time are sufficiently slow for the power-law behaviour of scaling solutions such that higher gradients in convolutions may be neglected. Specifically, we approximate equation (\ref{eq:IR}) with~\cite{Berges:2014bba, Chantesana:2018qsb}
\begin{eqnarray}
 &&   \int_{\Delta t_{xy},\mathbf{x}-\mathbf{y}} e^{i \left(\omega \Delta t_{xy} - \mathbf{p}(\mathbf{x}-\mathbf{y})\right)} \int_{z} I^R(x, z) \Pi^R(z, y) \nonumber\\
 &=&   e^{\frac{i}{2}( \frac{\partial}{\partial \omega}\frac{\partial}{\partial t^\prime_{xy}} - \frac{\partial}{\partial \omega^\prime}\frac{\partial}{\partial t_{xy}})} I^R(t_{xy},\omega,\mathbf{p}) \Pi^R(t^\prime_{xy},\omega^\prime,\mathbf{p})\big|_{\overset{t_{xy}=t^\prime_{xy},}{\omega = \omega^\prime}} \nonumber\\
 &=& I^R(t_{xy},\omega,\mathbf{p}) \Pi^R(t_{xy},\omega,\mathbf{p}) + \mathcal{O} \left( \frac{\partial}{\partial \omega}\frac{\partial}{\partial t_{xy}} \right) \, ,
\end{eqnarray}
and equivalently for (\ref{eq:IF}), using the lowest order in the derivative expansion. 
With this approximation, and $p$ denoting the momentum four-vector, (\ref{eq:IF}) can be written as~\cite{Berges:2010ez} 
\begin{equation}
I^F(t,\omega,\pp) = \Pi^F(t,\omega,\pp)\, v_{\mathrm{eff}}(t,\omega,\pp)\, ,
\label{eq:summations1}
\end{equation}
where the function $v_{\mathrm{eff }}(t,\omega,\pp)$, summing the geometric series of loop-diagrams appearing at NLO, is defined as 
\begin{equation}
	v_{\mathrm{eff }}(t,\omega,\pp) =\frac{1}{\left|1+ \Pi^R(t,\omega,\pp)\right|^2}.
    \label{eq:effv}
\end{equation}
Applying the lowest-order gradient expansion to (\ref{eq:pirho}) and (\ref{eq:piff}) yields
\begin{align}
    \Pi^R(t,\omega,\pp)&=\frac{\lambda}{3}\int_{\nu,\qq} F(t,\omega-\nu,\pp-\qq) G^R(t,\nu,\qq)\,,
    \label{eq:pifgrw}\\
	\Pi^F(t,\omega,\pp)&=\frac{\lambda}{6}\int_{\nu,\qq} F(t,\omega-\nu,\pp-\qq)F(t,\nu,\qq) \,,
    \label{eq:piffw}
\end{align}
where $\int_{\nu,\qq} \equiv \int d\nu d^dq/(2\pi)^{d+1}$.
Starting from (\ref{eq:IR}), it is again convenient to introduce the real combination 
\begin{equation}
I^{\tilde{\rho}}(t,\omega,\pp) = -i \left[I^R(t,\omega,\pp)-I^A(t,\omega,\pp)\right]
\end{equation}
for which
\begin{equation}
I^{\tilde{\rho}}(t,\omega,\pp) = \Pi^{\tilde{\rho}}(t,\omega,\pp)\, v_{\mathrm{eff}}(t,\omega,\pp)
\label{eq:Irho}
\end{equation}
with $\Pi^{\tilde{\rho}}(t,\omega,\pp)= -i [\Pi^R(t,\omega,\pp)-\Pi^A(t,\omega,\pp)]$ given by
\begin{equation}
\Pi^{\tilde{\rho}}(t,\omega,\pp)=\frac{\lambda}{3}\int_{\nu,\qq} F(t,\omega-\nu,\pp-\qq) \tilde{\rho}(t,\qq) \, .
\label{eq:Pirhotilde}
\end{equation} 

According to (\ref{eq:sigmaR}), the spectral self-energy defined as
\begin{equation}
\Sigma^{\tilde{\rho}}(t,\omega,\pp) = -i \left[\Sigma^R(t,\omega,\pp)-\Sigma^A(t,\omega,\pp)\right]
\label{eq:sigmatilderhodef}
\end{equation}
is at lowest order in the gradient expansion
\begin{align}
\label{eq:selfenergieswignerR}
\Sigma^{\tilde{\rho}}(t,\omega,\pp) &= 
-\frac{\lambda}{3N}\int_{\nu,\qq} \left[F(t,\omega-\nu,\pp-\qq)\, \Pi^{\tilde{\rho}}(t,\qq)\right. \nonumber\\
 &+
 \left.\tilde{\rho}(t,\omega-\nu,\pp-\qq)\, \Pi^F(t,\qq)\right] v_{\mathrm{eff}}(t,\qq).
\end{align}
Likewise, (\ref{eq:statself}) becomes
\begin{align}
\label{eq:selfenergieswigner1}
\Sigma^F(t,\omega,\pp)=& -\frac{\lambda}{3N}\int_{\nu,\qq} F(t,\omega-\nu,\pp-\qq)\, \Pi^F(t,\qq)  \nonumber\\ & \qquad \qquad \;\;\times\, v_{\text {eff}}(t,\qq) \, 
\end{align}
for the statistical part of the self-energy.

The lowest-order equations are obtained by 
neglecting ${\cal O}\left(\partial_{\omega} 
\partial_{t}\right)$
and higher contributions in the gradient expansion. To this
order the equations of motion for the statistical and spectral functions read~\cite{Berges:2008wm}:
\begin{subequations}
\begin{align}
2 \omega\, \partial_t F ( t, \omega, \pp) 
&=
\Sigma^{\tilde{\rho}} \left( t, \omega, \pp \right) 
F \left( t, \omega, \pp \right) \nonumber\\
&- 
\Sigma^F \left( t, \omega, \pp \right)
\tilde{\rho} \left( t, \omega, \pp \right),
\label{eq:LOgradF}\\[0.2cm]
2 \omega\, \partial_t \tilde{\rho} ( t, \omega, \pp) 
&=  0 \, .
\label{eq:LOgradrho}
\end{align}
\end{subequations}
At leading order in the gradient expansion, the spectral function has no dependence on central time $t$, i.e.~$\tilde{\rho}(t,\omega,\pp)=\tilde{\rho}(\omega,\pp)$.
This is in contrast to the statistical function $F(t,\omega,\pp)$, which at this order can depend on $t$ for general nonequilibrium solutions. The possible explicit dependence of the statistical correlation function on the central time is crucial to identify the nonequilibrium scaling solutions we are considering. By contrast, scaling phenomena in thermal equilibrium would be \mbox{$t$-independent} due to time-translation invariance.

\subsection{Effective kinetic description}
\label{sec:quasiparticle}

The self-energies~\eqref{eq:sigmaR} and~\eqref{eq:statself} are $\mathcal{O}(1/N)$ in the large-$N$ expansion. When computing these self-energies at this next-to-leading order in terms of the statistical and retarded propagators, we may neglect contributions to $F$ and $G^R$ or $\tilde{\rho}$ beyond leading order since these give only corrections beyond $\mathcal{O}(1/N)$ to the self-energies. 

Specifically, for the retarded propagator, we use the leading-order form
\begin{equation}
G^R(\omega, \mathbf{p})=\frac{1}{2\omega_{\mathbf{p}}}\left(\frac{1}{\omega+\omega_{\mathbf{p}}+i \epsilon}-\frac{1}{\omega-\omega_{\mathbf{p}}+i \epsilon}\right)\, ,
\label{eq:retprop}
\end{equation}
where we consider the massless dispersion relation
\begin{equation}
\omega_{\pp}=|\pp| \, 
    \label{eq:disp}
\end{equation}
for scaling solutions. Therefore, the spectral function~\eqref{eq:tilderho} reads
\begin{equation}
\tilde{\rho}(\omega, \mathbf{p}) = \frac{\pi}{\omega_{\mathbf{p}}}\big[\delta(\omega-\omega_{\mathbf{p}})-\delta(\omega+\omega_{\mathbf{p}})\big]\,.
\label{eq:rhoLO}
\end{equation}
Correspondingly, the leading-order form of the statistical function is
\begin{equation}
F(t,\omega, \mathbf{p}) = \frac{\pi}{\omega_{\mathbf{p}}} \, f(t,\mathbf{p})\big[\delta(\omega-\omega_{\mathbf{p}})+\delta(\omega+\omega_{\mathbf{p}})\big]\,,
\label{eq:FLO}
\end{equation}
where $f(t,\mathbf{p})$ denotes the time and momentum dependent ``on-shell'' occupation number distribution function. Here again the classical-statistical or high-occupancy assumption is used to neglect a ``quantum-half''. This kinetic description does, of course, not capture all aspects of the fully self-consistent field-theoretical treatment at NLO in the large-$N$ expansion. In particular, possible additional collective phenomena such as scaling instabilities~\cite{Preis:2022uqs} are not taken into account with the employed leading-order form~\eqref{eq:rhoLO} of the spectral function.

Using~\eqref{eq:retprop} and~\eqref{eq:FLO}, the retarded self-energy~\eqref{eq:pifgrw} is given by
\begin{align}
&\Pi^R(t,\omega, \mathbf{p})=  \lim _{\epsilon \rightarrow 0} \frac{\lambda}{12} \int_{\mathbf{q}} \frac{f(t, \mathbf{p}-\mathbf{q})}{\omega_{\mathbf{q}} \omega_{\mathbf{p}-\mathbf{q}}}\nonumber\\
& \times\left(\,\frac{1}{\omega_{\mathbf{q}}+\omega_{\mathbf{p}-\mathbf{q}}-\omega-i \epsilon}+\frac{1}{\omega_{\mathbf{q}}-\omega_{\mathbf{p}-\mathbf{q}}-\omega-i \epsilon}\right.\nonumber\\
& \left.+\frac{1}{\omega_{\mathbf{q}}-\omega_{\mathbf{p}-\mathbf{q}}+\omega+i \epsilon}+\frac{1}{\omega_{\mathbf{q}}+\omega_{\mathbf{p}-\mathbf{q}}+\omega+i \epsilon}\,\right),
\label{eq:pir}
\end{align}
while with~\eqref{eq:rhoLO} the statistical self-energy~\eqref{eq:piffw} reads  
\begin{align}
&\Pi^F(t,\omega,\mathbf{p}) \,=\, \frac{\lambda\pi}{12}\int_{\mathbf{q}}
\frac{f(t,\mathbf{p-q}) f(t,\mathbf{q})}{\omega_{\mathbf{q}}\omega_{\mathbf{p-q}}}
\nonumber\\
& \times\Big(\delta(\omega_{\mathbf{q}}+\omega_{\mathbf{p-q}}-\omega)
+ \delta(\omega_{\mathbf{q}}-\omega_{\mathbf{p-q}}-\omega) \nonumber\\
& + \delta(\omega_{\mathbf{q}}-\omega_{\mathbf{p-q}}+\omega)
+ \delta(\omega_{\mathbf{q}}+\omega_{\mathbf{p-q}}+\omega)
\Big).
\label{eq:pif-onsh}
\end{align}
These enter the remaining quantities like~\eqref{eq:selfenergieswignerR} and~\eqref{eq:selfenergieswigner1} with~\eqref{eq:effv}. 

This yields
\begin{align}
\Sigma^{\tilde{\rho}}(t,\omega,\pp) =& -\frac{\lambda}{6N}\int_{\qq}\frac{1}{\omega_{\pp-\qq}}\Big\{v_{\text{eff}}(t,\omega-\omega_{\pp-\qq},\qq) \nonumber\\
&\times\Big[f(t,\pp-\qq)\,\Pi^{\tilde{\rho}}(t,\omega-\omega_{\pp-\qq},\qq) \nonumber\\
&\qquad+\Pi^F(t,\omega-\omega_{\pp-\qq},\qq)\Big] \nonumber\\
&+v_{\text{eff}}(t,\omega+\omega_{\pp-\qq},\qq) \nonumber\\
&\times\Big[f(t,\pp-\qq)\,\Pi^{\tilde{\rho}}(t,\omega+\omega_{\pp-\qq},\qq) \nonumber\\
&\qquad-\Pi^F(t,\omega+\omega_{\pp-\qq},\qq)\Big]\Big\},
\label{eq:sigmatilderhoint}
\end{align}
where $\Pi^{\tilde{\rho}}=-i(\Pi^R-\Pi^A)=2\,\mathrm{Im}[\Pi^R]$ for real frequencies and momenta. Likewise, the on-shell statistical self-energy is given by
\begin{align}
&\Sigma^F(t,\omega,\pp) = -\frac{\lambda}{6N}\int_{\qq}\frac{f(t,\pp-\qq)}{\omega_{\pp-\qq}} \nonumber\\
& \times \Big[\Pi^F(t,\omega-\omega_{\pp-\qq},\qq)v_{\text{eff}}(t,\omega-\omega_{\pp-\qq},\qq) \nonumber\\
& + \Pi^F(t,\omega+\omega_{\pp-\qq},\qq)v_{\text{eff}}(t,\omega+\omega_{\pp-\qq},\qq)\Big].
\end{align}

With~\eqref{eq:FLO} the evolution equation~\eqref{eq:LOgradF} corresponds to 
\begin{equation}
	\frac{\partial f(t,\pp)}{\partial t}  \, = \, C[f](t,\pp) 
	\, 
	\label{eq:neffchange}
\end{equation}
with the collision integral
\begin{align}
	C[f](t,\pp) 
	= \int_0^\infty \frac{{\mathrm d} \omega}{2\pi} &\left[\Sigma^{\tilde{\rho}} ( t, \omega, \pp) F ( t, \omega, \pp)\right.\nonumber\\
	- &\left.\Sigma^F ( t, \omega, \pp ) \tilde{\rho}( \omega, \pp ) \right]. \qquad
	\label{eq:collisionint}
\end{align}
Using the on-shell forms~\eqref{eq:rhoLO} and~\eqref{eq:FLO} for $\tilde\rho$ and $F$ in the collision integral~\eqref{eq:collisionint}, the $\omega$-integral collapses onto $\omega=\omega_{\pp}$: the $\delta(\omega+\omega_{\pp})$ pieces of both~\eqref{eq:rhoLO} and~\eqref{eq:FLO} lie outside the integration range $\omega\in[0,\infty)$ and drop out, leaving
\begin{equation}
C[f](t,\pp) = \frac{1}{2\omega_{\pp}}\Big[f(t,\pp)\,\Sigma^{\tilde\rho}(t,\omega_{\pp},\pp) - \Sigma^F(t,\omega_{\pp},\pp)\Big].
\label{eq:Conshellselfenergies}
\end{equation}
This makes the gain-minus-loss structure of $C[f]$ explicit: the first term is built from $\Sigma^{\tilde\rho}$ together with the incoming occupation number $f(t,\pp)$, the second from $\Sigma^F$ alone.

The expression can be further rewritten to make the underlying two-to-two scattering processes with a time- and momentum-dependent effective vertex explicit. Following Ref.~\cite{Berges:2010ez}, for the considered high occupation numbers, this gives 
\begin{align}
&C[f](t,\mathbf{p})  =   \int_\mathbf{lqr} \frac{\lambda^2_{\rm eff}(t,{\mathbf p},{\mathbf l},{\mathbf q},{\mathbf r})}{6 N}  \nonumber\\
&\times 
(2\pi)^{d+1}\delta^{(d)}(\mathbf{p+l-q-r})\frac{\delta(\omega_\mathbf{p} + \omega_\mathbf{l} - \omega_\mathbf{q} - \omega_\mathbf{r} )}{2 \omega_\mathbf{p}\,2 \omega_\mathbf{l}\,2 \omega_\mathbf{q}\,2 \omega_\mathbf{r}} \nonumber\\
& \times 
\left\{ \left[ f(t,\mathbf{p}) + f(t,\mathbf{l}) \right] f(t,\mathbf{q})  f(t,\mathbf{r})\right.
\nonumber\\	
&- 
\left.f(t,\mathbf{p}) f(t,\mathbf{l}) \left[ f(t,\mathbf{q}) + f(t,\mathbf{r}) \right]\right\}.
\label{eq_collision_integral_rel_perp_NLO}
\end{align}
Here we have defined the time- and momentum-dependent effective coupling function  
\begin{eqnarray}
	\lambda^2_{\rm eff}(t,\mathbf{p},{\mathbf l},{\mathbf q},{\mathbf r}) & \equiv& \frac{\lambda^2}{3}
	\big[ v_{\text {eff}}(t,\omega_{\mathbf p} + \omega_{\mathbf l},{\mathbf p}+{\mathbf l})   \nonumber\\
	& &+\, v_{\text {eff}}(t,\omega_{\mathbf p} - \omega_{\mathbf q},{\mathbf p}-{\mathbf q})  
	\nonumber\\
	& &+\,   v_{\text {eff}}(t,\omega_{\mathbf p} - \omega_{\mathbf r},{\mathbf p}-{\mathbf r})\big]  \,, \quad
	\label{eq:leffrel}
\end{eqnarray}
which incorporates the vertex corrections for the different scattering channels. While in the underlying relativistic scalar field theory particle number is not conserved, the conserving nature of the two-to-two scattering processes arises since number changing processes are suppressed on shell at NLO of the large-$N$ expansion.  

\subsection{Scaling analysis}
\label{sec:scaling}

In the following, we focus on scaling properties of the nonequilibrium evolution. 
Near nonthermal fixed points, the dynamics is governed by self-similar evolutions in which correlation functions at different times are related by universal scaling. 
This allows one to characterise the dynamics through time-independent scaling functions and scaling exponents.

Specifically, the occupation number distribution $f(t,\mathbf{p})$ in~\eqref{eq:FLO} exhibits self-similar behaviour, such that 
\begin{equation}
f(t, \mathbf{p})=t^\alpha f_S\left( t^\beta \mathbf{p}\right)
\label{eq:dist2}
\end{equation}
in terms of the universal scaling exponents $\alpha$ and $\beta$ with scaling function $f_S$.
This scaling form represents a strong restriction for how the dynamics can depend on time and momentum. Since $t^{-\alpha} f(t, \pp)$ depends via the scaling function only on the combination $t^\beta \mathbf{p}$, rather than on time and momentum separately, the system’s characteristic time evolution is already captured by its momentum dependence for fixed time and vice versa. 

A positive exponent $\beta$ describes the scaling towards lower characteristic momenta as time proceeds:
For given time-independent scaling function $f_S(\mathbf{q})$ with $\mathbf{q} = t^\beta \mathbf{p}$, the momenta
$\mathbf{p}$ have to decrease as $t^\beta$ increases for larger times as $\beta > 0$. Correspondingly, a negative value for $\beta$ implies increasing characteristic momenta. Furthermore, a positive/negative  $\alpha$ leads to an increasing/decreasing distribution function with time according to~\eqref{eq:dist2}. Here we are interested in scaling solutions with $\alpha >0$ and $\beta >0$, corresponding to high occupation numbers growing $\sim t^\alpha$ for typical infrared momenta decreasing $\sim t^{-\beta}$. 

Further insights can be obtained by either imposing energy conservation or particle number conservation if applicable.
For constant total particle number 
\begin{equation}
 n \,=\, \int \frac{d^dp}{(2\pi)^d}\, f(t,\pp) \,=\, t^{\alpha - \beta d} \int \frac{d^dq}{(2\pi)^d}\, f_S(\qq) 
\end{equation}
one obtains the relation for 
\begin{equation} 
\!\!\!\!\!\!\mbox{\it particle conservation:} \quad  \alpha = \beta d\,.
\label{eq:pcrel}
\end{equation}
Similarly, with
\begin{eqnarray}
\label{eq:energyconserv}
\epsilon & \!=\! & \int \frac{d^dp}{(2\pi)^d}\, \omega_\pp\, f(t,\pp) \nonumber\\
&\!=\! & t^{\alpha - \beta (d+1)} \int \frac{d^dq}{(2\pi)^d}\, \omega_\qq\, f_S(\qq) 
\end{eqnarray}
one obtains from
\begin{equation} 
\!\!\!\!\!\!\mbox{\it energy conservation:} \quad   \alpha \,=\, \beta (d+1)\, ,
\label{eq:ecrel}
\end{equation}
where the linear dispersion~\eqref{eq:disp} is taken. As a consequence, one has $\alpha/\beta \geq d$, which we will use in the following.

In terms of~\eqref{eq:dist2} with~\eqref{eq:disp}, the one-loop function~\eqref{eq:pir} scales as 
\begin{equation}
\label{eq:PiR_S}
    \Pi^{R}(t,\omega, \mathbf{p})=t^{\alpha +\beta(3-d)}\,\Pi^{R}_S\big(t^{\beta} \omega, t^{\beta} \mathbf{p}\big)
\end{equation}
and equivalently for $\Pi^{\tilde{\rho}}(t,\omega, \mathbf{p})$. For~\eqref{eq:pif-onsh} one has
\begin{equation}
\label{eq:PiF_S}
\Pi^F(t, \omega, \mathbf{p})=t^{2\alpha+\beta(3-d)}\, \Pi^F_S\big( t^{\beta}\omega, t^{\beta} \mathbf{p}\big).
\end{equation}
For the considered case of positive exponents, the retarded self-energy grows for characteristic momenta according to~\eqref{eq:PiR_S}. Thus from~\eqref{eq:effv} we have for sufficiently late times that
$v_{\mathrm{eff }}(t,\omega,\pp) \simeq \left|\Pi^R(t,\omega,\pp)\right|^{-2}$, such that its scaling behaviour becomes 
\begin{equation}
	v_{\mathrm{eff }}(t,\omega,\pp) = t^{-2\alpha -2\beta(3-d)}\,v_{\mathrm{eff},S}\big(t^{\beta} \omega, t^{\beta} \mathbf{p}\big).
    \label{eq:effvscaling}
\end{equation}
This can also be used to obtain the scaling properties of the NLO self-energies,
\begin{equation}
    \Sigma^{\tilde{\rho}}(t,\omega, \mathbf{p})=t^{-2\beta}\,\Sigma^{\tilde{\rho}}_S(t^{\beta} \omega, t^\beta \mathbf{p}).
	\label{eq:scalingsigmatilderho}
\end{equation}
and
\begin{equation}
    \Sigma^{F}(t,\omega, \mathbf{p})=t^{\alpha-2\beta}\,\Sigma^{F}_S(t^{\beta} \omega, t^\beta \mathbf{p}).
	\label{eq:scalingsigmaF}
\end{equation}
Moreover,
\begin{equation} 
	C[f](t,{\pp}) \,=\, t^{\alpha-\beta} \, C[f_S](1,t^\beta \pp).
\end{equation}
Substituting this scaling into the evolution equation~\eqref{eq:neffchange} leads to the time-independent fixed-point equation for the distribution function,
\begin{equation}
\left[ \alpha+\beta\,\qq \cdot\mathbf{\nabla}_{\qq} \right] f_S(\qq) \,=\, C[f_S](1,\qq) \, ,
\label{eq:fixedpointf}
\end{equation}
and the scaling relation $\alpha - 1 = \alpha-\beta$ by comparison of left and right hand sides. The latter gives $\beta =  1$ irrespective of the value of $\alpha$. Using in addition~\eqref{eq:pcrel}, one finds for the transport of particles $\alpha = d$. Correspondingly~\eqref{eq:ecrel} gives $\alpha = d+1$ for energy transport.

\subsection{Scaling form of self-energies}
\label{sec:scalingformselfenergies}

Inserting the scaling forms~\eqref{eq:dist2}, \eqref{eq:PiR_S}, \eqref{eq:PiF_S} and~\eqref{eq:effvscaling}, and rescaling the loop momentum as $\qq\to t^{-\beta}\qq$, reproduces the overall scaling $t^{-2\beta}$ of~\eqref{eq:scalingsigmatilderho} and identifies the scaling function as
\begin{align}
\Sigma^{\tilde{\rho}}_S(\omega,\pp) =& -\frac{\lambda}{6N}\int_{\qq}\frac{1}{\omega_{\pp-\qq}}\Big\{v_{\text{eff},S}(\omega-\omega_{\pp-\qq},\qq) \nonumber\\
&\times\Big[f_S(\pp-\qq)\,\Pi^{\tilde{\rho}}_S(\omega-\omega_{\pp-\qq},\qq) \nonumber\\
&\qquad+\Pi^F_S(\omega-\omega_{\pp-\qq},\qq)\Big] \nonumber\\
&+v_{\text{eff},S}(\omega+\omega_{\pp-\qq},\qq) \nonumber\\
&\times\Big[f_S(\pp-\qq)\,\Pi^{\tilde{\rho}}_S(\omega+\omega_{\pp-\qq},\qq) \nonumber\\
&\qquad-\Pi^F_S(\omega+\omega_{\pp-\qq},\qq)\Big]\Big\}.
\label{eq:sigmatilderhoS}
\end{align}
Likewise, for the scaling form of the statistical self-energy we have
\begin{align}
&\Sigma_S^F(\omega,\pp) = -\frac{\lambda}{6N}\int_{\qq}\frac{f_S(\pp-\qq)}{\omega_{\pp-\qq}} \nonumber\\
& \times \Big[\Pi_S^F(\omega-\omega_{\pp-\qq},\qq)\, v_{\text{eff},S}(\omega-\omega_{\pp-\qq},\qq) \nonumber\\
& + \Pi_S^F(\omega+\omega_{\pp-\qq},\qq)\, v_{\text{eff},S}(\omega+\omega_{\pp-\qq},\qq)\Big].
\label{eq:sigmaFS}
\end{align}
By spatial isotropy these are functions of the magnitude of the external momentum, i.e.~$\Sigma^{\tilde{\rho}}_S(\omega,\pp)= \Sigma^{\tilde{\rho}}_S(\omega,\bar{p})$ and $\Sigma^F_S(\omega,\pp)= \Sigma^F_S(\omega,\bar{p})$.

The effective vertex appearing in the integrand of the self-energies takes the form
\begin{equation}
v_{\mathrm{eff},S}(\omega,\bar{p}) = \frac{1}{\big(\mathrm{Re}[\Pi^R_S(\omega,\bar{p})]\big)^2 + \big(\mathrm{Im}[\Pi^R_S(\omega,\bar{p})]\big)^2}
\label{eq:scalingformveff}
\end{equation}
in the infrared scaling regime. The functions $\Pi^R_S$ and $\Pi^F_S$ are determined by~\eqref{eq:pir} and~\eqref{eq:pif-onsh} in scaling form.

Carrying out the angular integral for the one-loop retarded function is
\begin{align}
\Pi^R_S(\omega,\bar{p}) =& \frac{\lambda}{48\pi^2 \bar{p}} \int_0^{\infty}\mathrm{d}\bar{q}\, f_S(\bar{q}) \left[ \ln \left|\frac{(\bar{p}+2\bar{q})^2-\omega^2}{(\bar{p}-2\bar{q})^2-\omega^2}\right|\right. \nonumber\\
&+ \left.i\pi\,\Gamma_\Theta(\omega,\bar{p},\bar{q}) \vphantom{\ln \left|\frac{(\bar{p}+2\bar{q})^2-\omega^2}{(\bar{p}-2\bar{q})^2-\omega^2}\right|}\right],
\label{eq:pirS_general}
\end{align}
where
\begin{align}
\Gamma_\Theta(\omega,\bar{p},\bar{q}) &\equiv \Theta(E_+-\Delta)\Theta(P-E_+) \nonumber\\
&+\, \Theta(E_--\Delta)\Theta(P-E_-) \nonumber\\
&-\, \Theta(-E_+-\Delta)\Theta(P+E_+) \nonumber\\
&-\, \Theta(-E_--\Delta)\Theta(P+E_-),
\label{eq:GammaThetaExplicit}
\end{align}
with $E_+ \equiv \omega+\bar{q}$, $E_- \equiv \omega-\bar{q}$, $P \equiv \bar{p}+\bar{q}$, and $\Delta \equiv |\bar{p}-\bar{q}|$. The detailed steps of the angular and radial integrations are given in App.~\ref{app:effective_vertex} and App.~\ref{app:PiF}.

Since $\Pi^{\tilde{\rho}}_S=2\,\mathrm{Im}[\Pi^R_S]$, this directly gives
\begin{equation}
\Pi^{\tilde{\rho}}_S(\omega,\bar{p}) = \frac{\lambda}{24\pi \bar{p}}\int_0^\infty \mathrm{d}\bar{q}\, f_S(\bar{q})\,\Gamma_\Theta(\omega,\bar{p},\bar{q}).
\label{eq:pitilderhoS_general}
\end{equation}
Correspondingly, the real part follows directly from the same logarithmic kernel already displayed in~\eqref{eq:pirS_general}:
\begin{equation}
\mathrm{Re}[\Pi^R_S(\omega,\bar{p})] = \frac{\lambda}{48\pi^2 \bar{p}}\int_0^\infty \mathrm{d}\bar{q}\, f_S(\bar{q})\, \ln\left|\frac{(\bar{p}+2\bar{q})^2-\omega^2}{(\bar{p}-2\bar{q})^2-\omega^2}\right|.
\label{eq:pirealS_general}
\end{equation}

The statistical function is obtained analogously, but the $\delta$-functions in~\eqref{eq:pif-onsh} carry an extra factor of $f_S$ from the second statistical propagator rather than $\pm i\pi$, and its imaginary part vanishes identically as expected for a manifestly real function. 
This yields, with $E_+,E_-,P,\Delta$ as defined above,
\begin{align}
\Pi^F_S(\omega,\bar{p}) &= \frac{\lambda}{48\pi \bar{p}} \int_0^{\infty}\mathrm{d}\bar{q}\, f_S(\bar{q}) \nonumber\\
&\times \Big[ f_S(E_+)\,\Theta(E_+ - \Delta)\,\Theta(P - E_+) \nonumber\\
&+ f_S(E_-)\,\Theta(E_- - \Delta)\,\Theta(P - E_-) \nonumber\\
&+ f_S(-E_+)\,\Theta(-E_+ - \Delta)\,\Theta(P + E_+) \nonumber\\
&+ f_S(-E_-)\,\Theta(-E_- - \Delta)\,\Theta(P + E_-) \Big].
\label{eq:piFS_general}
\end{align}

\subsection{Scaling form of the distribution function}

According to~\eqref{eq:fixedpointf} and the above analysis leading to $\beta =1$, the scaling function is a solution of the fixed-point equation
\begin{equation}
\alpha\,f_S(\bar{p})+ \bar{p}\,\frac{\partial f_S(\bar{p})}{\partial \bar{p}} = C[f_S](1,\bar{p})
\label{eq:scalingfs}
\end{equation}
with $\alpha$ given by~\eqref{eq:pcrel} for particle transport and~\eqref{eq:ecrel} for energy transport. Here the scaling form of the collision integral~\eqref{eq:Conshellselfenergies} reads 
\begin{equation}
C[f_S](1,\bar p) = \frac{1}{2\bar p}\Big[f_S(\bar p)\,\Sigma^{\tilde\rho}_S(\bar p,\bar p) - \Sigma^F_S(\bar p,\bar p)\Big],
\label{eq:collisionintonshell}
\end{equation}
using the scaling functions~\eqref{eq:sigmatilderhoS},~\eqref{eq:sigmaFS} evaluated on shell.

We consider a scaling ansatz of the form
\begin{equation}
f_S(\bar{p})=\displaystyle \frac{A}{B+\bar{p}^{\kappa}},
    \label{eq:dist}
\end{equation}
where $\kappa$ is a universal scaling exponent, while $A$ and $B$ are nonuniversal constants. For notational convenience, all quantities are considered to be made dimensionless by appropriate powers of some arbitrary scale. A corresponding scaling form has also been investigated for nonrelativistic theories~\cite{Walz:2017ffj, PineiroOrioli:2015cpb, Chantesana:2018qsb}.

The ansatz~\eqref{eq:dist} allows us to compute the universal exponent $\kappa$ from the large $\bar{p} \gg B^{1/\kappa}$ behaviour.
Since the distribution function $f_S$ enters~\eqref{eq:scalingfs} in equal amounts in the numerator and denominator on both sides of the equation, the parameter $A$ drops out of computations. The parameter $B$ implies a rescaled infrared momentum 
\begin{equation}
k_{\mathrm{IR}} \sim B^{1/\kappa},
\label{eq:kf}
\end{equation}
marking the crossover between an infrared plateau $f_S(0)=A/B$ and the power-law behaviour $f_S(\bar{p})\to A\, \bar{p}^{-\kappa}$ for $\bar{p} \gg k_{\mathrm{IR}}$. Following the scaling analysis of Sec.~\ref{sec:scaling}, the corresponding physical transport momentum scales as $t^{-\beta} k_{\mathrm{IR}}$ to zero at late times for the considered $\beta >0$.

\section{On-shell scaling behaviour}
\label{sec:explicit}

\subsection{One-loop functions}

\begin{figure*}
    \centering
    \includegraphics[width=0.99\linewidth]{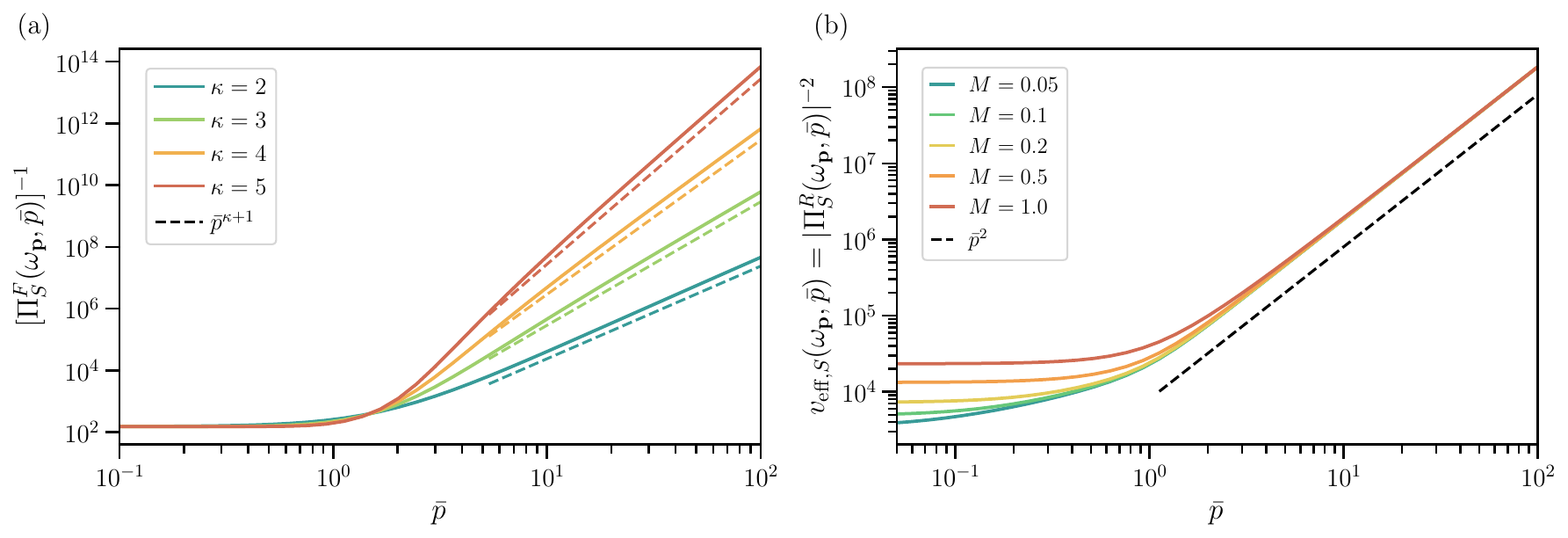}
    \caption{The inverse of the one-loop function (a), $[\Pi^F_S(\omega_\pp,\bar{p})]^{-1}$, is shown for $2\leq\kappa \leq5$. It exhibits a nearly constant form at low momenta, followed by a powerlaw at higher momenta. We determined this obeys $\sim \bar p^{1+\kappa}$. The effective vertex (b) for different mass values and $\kappa=4$, using $\lambda=1$. For low momenta $\bar p$, it is a mass-dependent constant which turns into a $\bar p^2$ powerlaw. This quadratic behaviour is independent of $\kappa$ for the range of $2 \leq\kappa \leq5$.}
    \label{fig:veffk}
\end{figure*}

Though we are interested in the massless case, it is instructive to evaluate first the expression for on-shell frequencies $\omega=\sqrt{\bar{p}^2+M^2}$ with mass parameter $M$. The Heaviside constraint in $\Gamma_\Theta$ \eqref{eq:GammaThetaExplicit} then reduces to a single window, $q_c<\bar{q}<\bar{p}+q_c$, with $q_c\equiv\big(\sqrt{\bar{p}^2+M^2}-\bar{p}\big)/2 = M^2\big/\big[2\big(\sqrt{\bar{p}^2+M^2}+\bar{p}\big)\big]$, which reduces to $q_c\simeq M^2/4\bar{p}$ only for $M\ll\bar{p}$, so that exactly
\begin{equation}
\Pi^{\tilde{\rho}}_S\big(\sqrt{\bar{p}^2+M^2},\bar{p}\big) = \frac{\lambda}{24\pi \bar{p}}\int_{q_c}^{\bar{p}+q_c}\mathrm{d}\bar{q}\, f_S(\bar{q}).
\label{eq:pitilderhoS_onshell}
\end{equation}
Unlike the imaginary part, the logarithmic kernel in~\eqref{eq:pirealS_general} does not collapse onto a finite window when going on shell; substituting $\omega=\sqrt{\bar{p}^2+M^2}$ merely fixes its argument, leaving the full radial integral over $\bar{q}\in(0,\infty)$.

Since we are ultimately interested in the massless case, we consider now the $M\to0$ explicitly at fixed $\bar{p}>0$. The window in~\eqref{eq:pitilderhoS_onshell} then collapses to $[0,\bar{p}]$, giving
\begin{equation}
\mathrm{Im}[\Pi^R_S(\bar{p},\bar{p})] = \tfrac12\Pi^{\tilde{\rho}}_S(\bar{p},\bar{p}) = \frac{\lambda}{48\pi \bar{p}}\int_0^{\bar{p}}\mathrm{d}\bar{q}\, f_S(\bar{q}),
\label{eq:imPiRS_massless}
\end{equation}
which in practice is manifestly finite for any $\bar{p}>0$. The real part, from~\eqref{eq:pirealS_general} at $\omega=\bar{p}$, reduces to
\begin{equation}
\mathrm{Re}[\Pi^R_S(\bar{p},\bar{p})] = \frac{\lambda}{48\pi^2 \bar{p}}\int_0^\infty \mathrm{d}\bar{q}\, f_S(\bar{q})\,\ln\left|\frac{\bar{q}+\bar{p}}{\bar{q}-\bar{p}}\right|.
\label{eq:rePiRS_massless}
\end{equation}
This is also finite for any fixed $\bar{p}>0$: the logarithmic kernel has only an integrable singularity at $\bar{q}=\bar{p}$, and vanishes smoothly as $\bar{q}\to0$. Both~\eqref{eq:imPiRS_massless} and~\eqref{eq:rePiRS_massless} are thus regular at $M=0$ for any fixed, nonzero external momentum.

The two behave very differently, however, as the external momentum itself is taken to zero. Since $f_S(\bar{q})$ in~\eqref{eq:imPiRS_massless} in practice does not vanish as $\bar{q}\to0$, $\lim_{\bar{p}\to0}\mathrm{Im}[\Pi^R_S(\bar{p},\bar{p})]\to \lambda\, f_S(0)/(48\pi)$ stays finite. By contrast, the kernel in~\eqref{eq:rePiRS_massless} Taylor-expands for $\bar{q}\gg\bar{p}$ as $\ln|(\bar{q}+\bar{p})/(\bar{q}-\bar{p})|\simeq 2\bar{p}/\bar{q}$, so that
\begin{equation}
\lim_{\bar{p}\to0}\mathrm{Re}[\Pi^R_S(\bar{p},\bar{p})] \simeq \frac{\lambda}{24\pi^2}\int\mathrm{d}\bar{q}\,\frac{f_S(\bar{q})}{\bar{q}},
\label{eq:rePiRS_smallp}
\end{equation}
which log-diverges at its own lower limit $\bar{q}\to0$, since $f_S(\bar{q})\to f_S(0)$ there. This is an infrared sensitivity of the loop integral to soft internal momenta $\bar{q}\to0$, which is absent for $\bar{p} > 0$. 
In addition, a nonzero mass $M > 0$ provides a regulator  
such that a finite $\bar{p}\to0$ limit of $\mathrm{Re}[\Pi^R_S]$ is observed.

Evaluating~\eqref{eq:piFS_general} for on-shell frequencies $\omega=\sqrt{\bar{p}^2+M^2}$, only the $\Theta(E_--\Delta)\Theta(P-E_-)$ term survives, over the same window $q_c<\bar{q}<\bar{p}+q_c$ established above for $\Pi^{\tilde{\rho}}_S$; the other three terms are absent due to the Heaviside constraints once $\omega\geq\bar{p}$. Keeping $E_-=\sqrt{\bar{p}^2+M^2}-\bar{q}$, 
this gives
\begin{equation}
\Pi^F_S\big(\sqrt{\bar{p}^2+M^2},\bar{p}\big) = \frac{\lambda}{48\pi \bar{p}}\int_{q_c}^{\bar{p}+q_c} \mathrm{d}\bar{q}\, f_S(\bar{q})\, f_S(\omega-\bar{q}).
\label{eq:piFS_onshell}
\end{equation}

Just as for $\Pi^{\tilde{\rho}}_S$, it is instructive to take $M\to0$ directly at fixed $\bar{p}>0$: the window in~\eqref{eq:piFS_onshell} collapses to $[0,\bar{p}]$ and $\omega\to\bar{p}$, giving
\begin{equation}
\Pi^F_S(\bar{p},\bar{p}) = \frac{\lambda}{48\pi \bar{p}}\int_0^{\bar{p}}\mathrm{d}\bar{q}\, f_S(\bar{q})\, f_S(\bar{p}-\bar{q}),
\label{eq:piFS_massless}
\end{equation}
again manifestly finite for any $\bar{p}>0$, since the integrand is bounded throughout the finite range of integration. Moreover, unlike $\mathrm{Re}[\Pi^R_S]$, $\Pi^F_S$ stays finite even as $\bar{p}\to0$
throughout the shrinking integration window,
$\lim_{\bar{p}\to0} \Pi^F_S(\bar{p},\bar{p}) \to \lambda\, f_S^2(0)/(48\pi)$.
In this respect $\Pi^F_S$ behaves like $\mathrm{Im}[\Pi^R_S]$ rather than $\mathrm{Re}[\Pi^R_S]$: of the three functions entering~\eqref{eq:sigmatilderhoS}, \eqref{eq:sigmaFS} and~\eqref{eq:effv}, only $\mathrm{Re}[\Pi^R_S]$ requires a regulator to investigate the properties of the full self-energies as $\bar{p}\to 0$.

For $\bar{p} \gg k_\mathrm{IR}$, the integral is dominated by $\bar{q} \sim k_\mathrm{IR}$ where $f_S(\bar{p}-\bar{q}) \approx f_S(\bar{p}) \sim \bar{p}^{-\kappa}$, yielding $\Pi^F_S(\bar{p},\bar{p}) \sim \bar{p}^{-(1+\kappa)}$, as derived in App.~\ref{app:PiF}. In turn, for $\bar{p} \ll k_\mathrm{IR}$ both distribution functions stay close to the plateau value $f_S(0)$ throughout the shrinking integration window. 
Altogether, combining the low-momentum limit with the high-momentum asymptotics, we have
\begin{equation}
	\Pi^F_S(\bar{p},\bar{p}) \sim
	\begin{cases}
    \displaystyle{\frac{\lambda}{48\pi}}f_S^2(0) & \text{for } \bar{p} \ll k_{\mathrm{IR}}, \\[8pt]
 \displaystyle{\frac{k_\mathrm{IR}\, f_S(k_\mathrm{IR})}{\bar{p}^{1+\kappa}}} & \text{for } \bar{p} \gg k_{\mathrm{IR}}.
\end{cases}
\label{eq:feffnum}
\end{equation}
The full numerical result is shown for the inverse, $(\Pi^F_S)^{-1}$, in Fig.~\ref{fig:veffk}(a) for different $\kappa$ exponents. 
The solution exhibits the flat region at low momenta, which depends on $f_S(0)$ only, followed by the power-law regime at higher momenta.  

\subsection{Effective vertex}

\begin{figure*}
    \centering
    \includegraphics[width=1.0\linewidth]{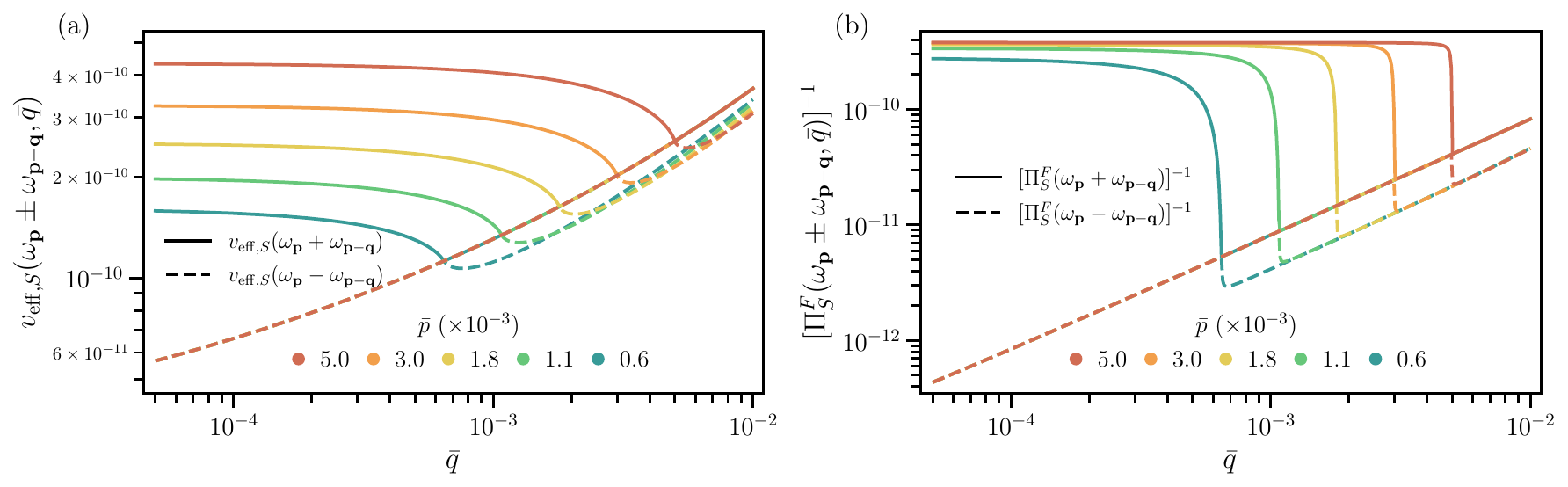}
    \caption{The quantities entering the self-energy integrand~\eqref{eq:sigmatilderhoS}, evaluated as functions of the loop-momentum $\bar q$ for several values of the external momentum $\bar p$ and $\kappa=4$, on shell ($\omega_\pp=\bar p$). (a): The effective vertex $v_{\mathrm{eff},S}(\omega_\pp \pm \omega_{\pp-\qq}, \bar q)$ for the $\omega_\pp + \omega_{\pp-\qq}$ (solid) and $\omega_\pp - \omega_{\pp-\qq}$ (dashed) branches. (b): The inverse of the statistical one-loop function $[\Pi^F_S(\omega_\pp \pm \omega_{\pp-\qq}, \bar q)]^{-1}$, with the same convention. Both quantities develop a steep edge near $\bar q \sim \bar p$, which is responsible for the peaked structure of the self-energy integrand shown in Fig.~\ref{fig:sigmaneqint}(a). }
    \label{fig:intermediate}
\end{figure*}

In the following, we discuss the scaling form of the effective vertex $v_{\text{eff},S}$ for the relativistic theory.
This has been extensively studied in nonrelativistic theories \cite{PineiroOrioli:2015cpb, Chantesana:2018qsb, Berges:2010ez, Walz:2017ffj}, where it has been shown to follow a universal scaling form in the infrared \cite{Chantesana:2018qsb, Walz:2017ffj}.

In the infrared scaling regime, the effective vertex takes the form~\eqref{eq:scalingformveff},
with $\mathrm{Re}[\Pi^R_S]$ and $\mathrm{Im}[\Pi^R_S]$ given on shell, at $M=0$, by~\eqref{eq:rePiRS_massless} and~\eqref{eq:imPiRS_massless}. For $\bar{p}\gtrsim k_{\mathrm{IR}}$, both parts are already finite at $M=0$: $\mathrm{Re}[\Pi^R_S(\bar{p},\bar{p})]\sim k_{\mathrm{IR}}^2 f_S(k_{\mathrm{IR}})/\bar{p}^2$, dominated by loop momenta $\bar{q}\sim k_{\mathrm{IR}}$ as in~\eqref{eq:kf}, while $\mathrm{Im}[\Pi^R_S(\bar{p},\bar{p})]\sim k_{\mathrm{IR}}f_S(k_{\mathrm{IR}})/\bar{p}$. Since the imaginary part decays more slowly, it dominates, giving
\begin{equation}
v_{\mathrm{eff},S}(\bar{p},\bar{p}) \sim \bar{p}^2 \qquad (\bar{p}\gtrsim k_{\mathrm{IR}}).
\label{eq:veffhighp}
\end{equation}

For $\bar{p}\lesssim k_{\mathrm{IR}}$, both parts likewise remain finite at $M=0$ for any fixed $\bar{p}>0$ (cf.\ the discussion following Eqs.~\eqref{eq:imPiRS_massless} and~\eqref{eq:rePiRS_massless}) but the two behave very differently as $\bar{p}\to0$ itself. $\mathrm{Im}[\Pi^R_S]\to\lambda f_S(0)/(48\pi)$ stays finite, while $\mathrm{Re}[\Pi^R_S]$ develops the mild logarithmic infrared sensitivity of Eq.~\eqref{eq:rePiRS_smallp}, growing without bound as $\bar{p}\to0$. The real part therefore comes to dominate $v_{\mathrm{eff},S}$ in this joint limit, giving
\begin{equation}
\lim_{\bar{p}\to 0} v_{\mathrm{eff},S}(\bar{p},\bar{p}) \sim \lim_{\bar{p}\to 0} \big[f_S(0)\,\ln(k_{\mathrm{IR}}/\bar{p})\big]^{-2} \to 0.
\label{eq:veffsmallp}
\end{equation}
Therefore, the effective vertex vanishes only in this limit, and only logarithmically slowly, not for any fixed $\bar{p}>0$, where it remains finite throughout, including growing as $\bar{p}^2$ once $\bar{p}\gtrsim k_{\mathrm{IR}}$ by~\eqref{eq:veffhighp}.

Numerically, it is convenient to regulate the $\bar{p}\to0$ endpoint with a small mass, evaluating $v_{\mathrm{eff},S}(\sqrt{\bar{p}^2+M^2},\bar{p})$ at $\bar{p}=0$ as a function of $M$ instead, as detailed in App.~\ref{app:effective_vertex}.
This is how Fig.~\ref{fig:veffk}(b) is obtained. 
Depending on whether $M$ sits below or above $k_{\mathrm{IR}}$, this reproduces the joint limit~\eqref{eq:veffsmallp} (for $M\ll k_{\mathrm{IR}}$, with $M$ in place of $\bar{p}$), or a different, quartic suppression $v_{\mathrm{eff},S}\sim M^4/[k_{\mathrm{IR}}^2f_S(k_{\mathrm{IR}})]^2$ once $M\gtrsim k_{\mathrm{IR}}$.
There, the real part's $M^{-2}$ scaling persists as in App.~\ref{app:effective_vertex}, while the imaginary part is no longer set by the infrared plateau $f_S(0)$ but by $f_S$ evaluated deep in its ultraviolet tail, which, assuming\footnote{This corresponds to $\kappa > 2$, cf.~App.~\ref{app:effective_vertex} for a discussion of $\kappa \le 2$.} $f_S$ falls off faster than $\bar{q}^{-2}$ at large $\bar{q}$, decays strictly faster than the real part's $M^{-2}$ and is correspondingly subdominant. Altogether,
\begin{eqnarray}
	&& \!\!\!\! v_{\mathrm{eff},S}(\sqrt{\bar{p}^2+M^2},\bar{p}) \nonumber\\
	&&\sim  \begin{cases}
\left[f_S(0)\,\ln\!\left(k_{\mathrm{IR}}/M\right)\right]^{-2} & \bar{p} \ll k_{\mathrm{IR}},\ M \ll k_{\mathrm{IR}}, \\
M^{4}/[k_{\mathrm{IR}}^2 f_S(k_{\mathrm{IR}})]^2 & \bar{p} \ll k_{\mathrm{IR}},\ M \gtrsim k_{\mathrm{IR}}, \\
	\bar{p}^{2} & \bar{p} \gg k_{\mathrm{IR}},
\end{cases} \qquad
\label{eq:veffnum}
\end{eqnarray}
as shown in Fig.~\ref{fig:veffk}(b).
Since the mass values plotted there span both sides of $k_{\mathrm{IR}}$, the resulting plateau is expected to interpolate between the two low-momentum regimes above rather than following either one exactly throughout. The quadratic high-momentum scaling~\eqref{eq:veffhighp} is confirmed numerically to hold independently of the detailed shape of $f_S$'s ultraviolet tail, confirming the universality of the effective vertex in the scaling regime.

\subsection{Self-energies}
\label{sec:scalingselfenergies}

\begin{figure*}
    \centering
    \includegraphics[width=0.95\linewidth]{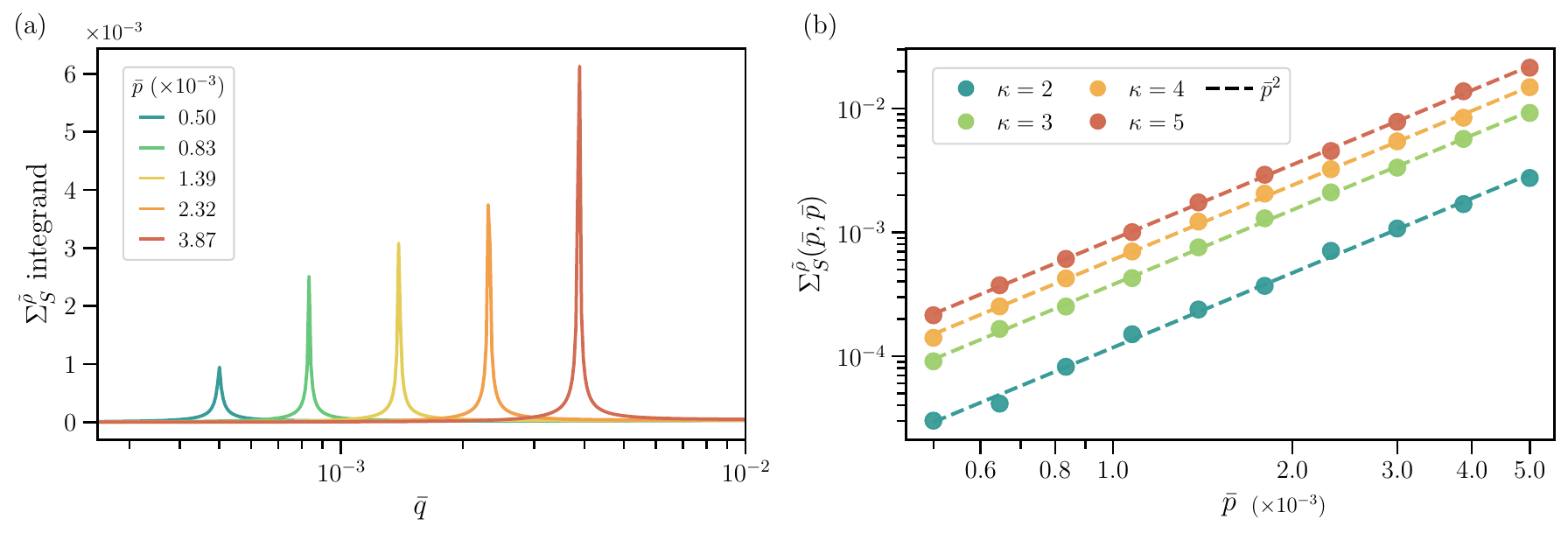}
    \caption{(a): The integrand of the retarded self-energy for $\kappa=4$, showing the sharp peaks developed at $\bar{p} \sim \bar{q}$, and how the dominant integration region shifts gradually towards the infrared for decreasing external momentum $\bar{p}$. (b): The universal scaling of the spectral self-energy $\sim \bar{p}^2$ for $\bar{p}\gtrsim k_{\mathrm{IR}}$. The data also verify that there is no $\kappa$-dependence in the scaling for the considered range of $2 \leq\kappa \leq5$.
    }
    \label{fig:sigmaneqint}
\end{figure*}

The scaling forms of the self-energies $\Sigma^{\tilde{\rho}}_S$ and $\Sigma^F_S$, given by~\eqref{eq:scalingsigmatilderho} and~\eqref{eq:scalingsigmaF}, are described in terms of convolution integrals involving the one-loop functions $\Pi^{\tilde{\rho}}_S$, $\Pi^F_S$ and the effective vertex. In this section, we show how the scaling of the building blocks leads to the scaling behaviour of the on-shell spectral self-energy in more detail as an example.

In accordance with~\eqref{eq:LOgradrho} at the considered leading order in the gradient expansion, the spectral self-energy is expected to have no dependence on central time $t$, such that $\Sigma^{\tilde{\rho}}(t,\omega,\pp) = \Sigma^{\tilde{\rho}}(\omega,\pp)$. This $t$-independence implies restrictions on the scaling behaviour of $\Sigma^{\tilde{\rho}}_S$.
Since the spectral self-energy is related to its scaling form by~\eqref{eq:scalingsigmatilderho}, 
the overall scaling factor $t^{-2\beta}$ must be compensated by a corresponding dependence of $\Sigma^{\tilde{\rho}}_S(t^{\beta} \omega,t^{\beta}\pp)$ on its arguments. The dependence on rescaled on-shell momenta follows as 
\begin{equation}
    \Sigma^{\tilde{\rho}}_S(\bar p,\bar p) \sim \bar{p}^2 \qquad (\bar p\gtrsim k_{\mathrm{IR}}) \, .
    \label{eq:sigmatilderhohighp}
\end{equation}
This relation has to hold for the self-similar spectral self-energy with the scaling form of the distribution~\eqref{eq:dist} in the regime where $\bar p$ is sufficiently large compared to $k_{\mathrm{IR}}$.

The quadratic dependence on external momenta for $\bar p\gtrsim k_{\mathrm{IR}}$ can be verified numerically for~\eqref{eq:sigmatilderhoS}. The inner convolutions involve the $\Pi^{\tilde{\rho}}_S$ and $\Pi^F_S$ loop functions and the effective vertex $v_{\mathrm{eff},S}$, which itself depends on $\Pi^R_S (\Pi^{\tilde{\rho}}_S)$.
While the nested convolutions in the self-energy can be evaluated directly, it is computationally more efficient to first evaluate $\Pi^R_S$ and $\Pi^F_S$ on a grid and then use these as input functions for evaluating the full self-energy.
As previously discussed in the context of the effective vertex, each $\Pi^R_S$ and $\Pi^F_S$ comes with a factor of $\lambda$ which for $\Pi^R_S\gg1$ drops out in the end.
Inserting the numerical expressions we obtained for these into the self-energy, we can investigate its momentum dependence.

We plot the integrand of $\Sigma^{\tilde{\rho}}_S$ in Fig.~\ref{fig:sigmaneqint}(a), for external momenta $\bar p\gtrsim k_{\mathrm{IR}}$.
For the range of momenta shown, we observe a distinct contribution to the self-energy peaked at $\bar{p} \sim \bar{q}$, which shifts towards lower loop momenta as the external momentum $\bar p$ is decreased towards $k_{\mathrm{IR}}$.
The origin of this peak can be traced back to the behaviour of $v_{\textrm{eff},S} (\omega_\pp\pm\omega_{\pp-\qq},\bar{q})$ (which depends on $\Pi^{R}_S(\omega_\pp\pm\omega_{\pp-\qq},\bar{q})$) and $\Pi^F_S(\omega_\pp\pm\omega_{\pp-\qq},\bar{q})$, which we plot in Fig.~\ref{fig:intermediate}(a), and (b) for the inverse of $\Pi^F_S$.
As seen there, both the $\omega_\pp + \omega_{\pp-\qq}$ and
$\omega_\pp - \omega_{\pp-\qq}$ branches of $v_{\mathrm{eff},S}$ and $\Pi^F_S$ develop a steep edge near $\bar{q} \sim \bar{p}$, where the functions considerably change over a narrow range of integration momenta.
It is this localised enhancement that produces the sharp peaks observed in the self-energy  , with the position of the edge tracking the external momentum $\bar{p}$ and thus shifting towards the infrared as $\bar{p}$ decreases.
Physically, this edge corresponds to the opening of the phase space for the intermediate scattering processes encoded in $\Pi^F_S$ and $\Pi^{R}_S$: as $\bar{q}$ increases past $\sim \bar{p}$, the frequency arguments $\omega_{\pp} \pm \omega_{\pp-\qq}$ enter the region where the on-shell energy conservation conditions in~\eqref{eq:pif-onsh} and~\eqref{eq:pir} can be satisfied. 
This can be seen explicitly for the collinear configuration $\pp\parallel\qq$, where $\omega_{\pp-\qq}=|\bar p-\bar q|$.
The two branches then become exactly on shell, i.e. their frequency argument equals their own momentum argument $\bar q$, precisely at $\bar q=\bar p$, with the ``$+$'' branch on shell for $\bar q>\bar p$ and the ``$-$'' branch on shell for $\bar q<\bar p$, so that the threshold sits at the same point $\bar q=\bar p$ for both branches, approached from opposite sides.
The self-energy at a given external momentum $\bar{p}$ is therefore dominated by scattering contributions at the same momentum scale, with each value of $\bar{p}$ probing processes at its own characteristic scale $\bar{q} \sim \bar{p}$.
Explicitly evaluating this integral for $\bar p\gtrsim k_{\mathrm{IR}}$, we find from the numerical solution the expected quadratic scaling behaviour~\eqref{eq:sigmatilderhohighp} in the considered range $2 \leq \kappa \leq 5$, as illustrated in Fig.~\ref{fig:sigmaneqint}(b).

\subsection{Universal scaling exponent $\kappa$}

\begin{figure*}
    \centering
    \includegraphics[width=0.95\linewidth]{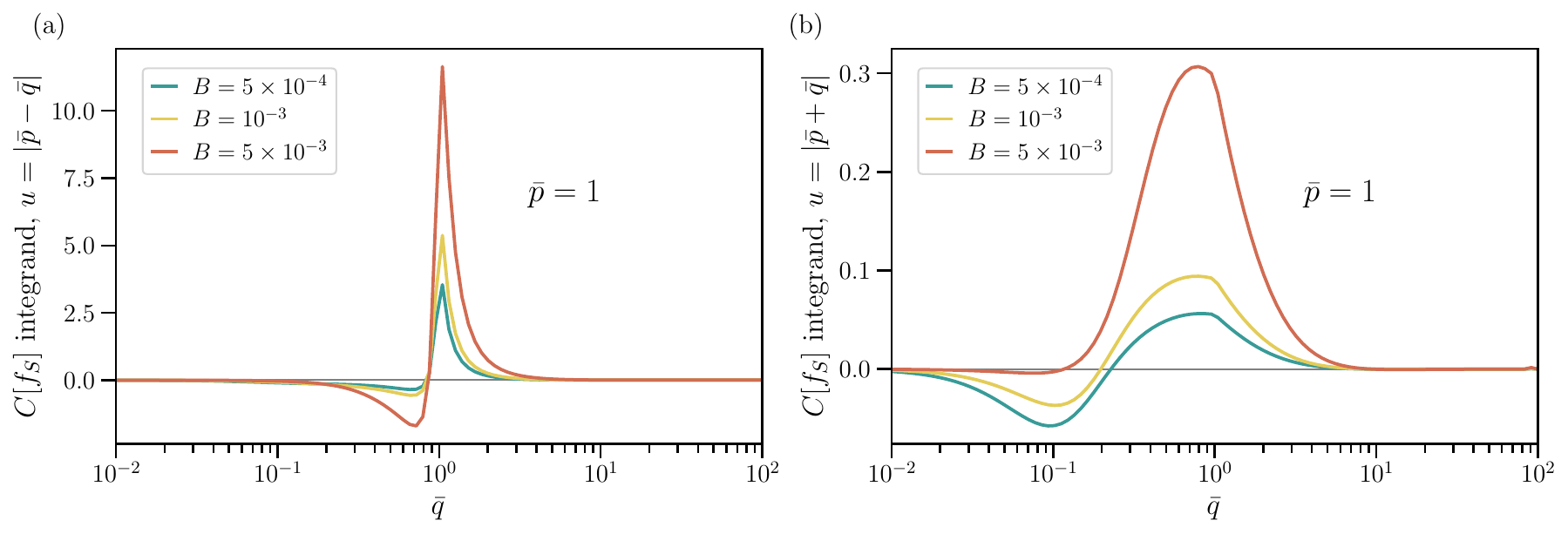}
    \caption{The integrand of the collision integral~\eqref{eq:collisionintdouble} evaluated
    as a function of the loop-momentum $\bar{q}$ at the two endpoints of the $u$-integration, (a) $u=|\bar{p}-\bar{q}|$ and (b) $u=|\bar{p}+\bar{q}|$, for $\kappa=3$ at fixed $\bar{p}=1$ and three values of $B$.}
    \label{fig:C_integrand}
\end{figure*}

For the scaling solutions~\eqref{eq:dist}, the fixed-point equation~\eqref{eq:scalingfs} can be written as
\begin{equation}
\left(\alpha - \kappa \right) f_S(\bar{p}) = C[f_S](1,\bar{p}) \quad \mathrm{for} \quad \bar{p}/k_{\mathrm{IR}} \to \infty \, ,
\label{eq:scalingfshighp}
\end{equation}
where $f_S(\bar{p}) \to A \bar{p}^{-\kappa}$ with~\eqref{eq:dist}. 
The scaling form for the collision integral~\eqref{eq:collisionintonshell} involves $\Sigma^{\tilde\rho}_S(\bar p,\bar p)$ and $\Sigma^F_S(\bar p,\bar p)$, which are themselves loop integrals over $\qq$ given by~\eqref{eq:sigmatilderhoS},~\eqref{eq:sigmaFS}. As in App.~\ref{app:effective_vertex}, changing variables from the angle between $\pp$ and $\qq$ to $u\equiv|\pp-\qq|$ at fixed $\bar q=|\qq|$ turns the explicit prefactor $1/\omega_{\pp-\qq}=1/u$ into a manifestly regular double integral over $(\bar q,u)$, and substituting into~\eqref{eq:collisionintonshell} gives
\begin{align}
&C[f_S](1,\bar p) = -\frac{\lambda}{48\pi^2N\bar p^2}\int_0^\infty \mathrm{d}\bar q\,\bar q \nonumber\\
&\times \int_{|\bar p-\bar q|}^{\bar p+\bar q}\mathrm{d}u\,\Big\{v_{\mathrm{eff},S}(\bar p-u,\bar q)\Big[f_S(\bar p)f_S(u)\,\Pi^{\tilde\rho}_S(\bar p-u,\bar q) \nonumber\\
&+\big(f_S(\bar p)-f_S(u)\big)\,\Pi^F_S(\bar p-u,\bar q)\Big] +v_{\mathrm{eff},S}(\bar p+u,\bar q) \nonumber\\
&\times \Big[f_S(\bar p)f_S(u)\,\Pi^{\tilde\rho}_S(\bar p+u,\bar q) 
-\big(f_S(\bar p)+f_S(u)\big)\nonumber\\
&\times \Pi^F_S(\bar p+u,\bar q)\Big]\Big\}.
\label{eq:collisionintdouble}
\end{align}
The grouping of terms in~\eqref{eq:collisionintdouble} reflects the Bose statistics of the underlying processes. 
The one-loop function $\Pi^{\tilde\rho}_S$ given by~\eqref{eq:pitilderhoS_general} is built from a single power of $f_S$, so $f_S(\bar p)f_S(u)\,\Pi^{\tilde\rho}_S$ effectively carries three powers of the occupation number, with the stimulated, Bose-enhanced combination matching the quartic gain term of the two-to-two collision integral~\eqref{eq_collision_integral_rel_perp_NLO}. By contrast, $\Pi^F_S$ given by~\eqref{eq:piFS_general} is already built from a product of two occupation numbers, i.e.\ it is itself a symmetrised two-particle object. The prefactor $f_S(\bar p)\mp f_S(u)$ multiplying it is then the residual imbalance between the external and momentum-transfer occupations, not a further stimulation factor.
This residual is a difference for the energy-subtracting (``$-u$'') branch (the detailed-balance-type combination controlling the net direction of a soft exchange process) and a sum for the energy-adding (``$+u$'') branch. The integrand of the collision integral~\eqref{eq:collisionintdouble} is shown in Fig.~\ref{fig:C_integrand} for $\bar{p} \gg k_{\mathrm{IR}}$ using different values of $B$ by specifying $\kappa = 3$. One observes that the integrand decreases as $B$ is lowered at fixed $\bar{p}$, which will be analysed in more detail in the following. 

Equation~\eqref{eq:scalingfshighp} holds for sufficiently large $\bar p\gg k_{\mathrm{IR}}$. This can be achieved by either $\bar p$ large at fixed $k_{\mathrm{IR}}$, or for small $k_{\mathrm{IR}}$ at fixed $\bar p$. These two limiting cases are not equivalent, since $C[f_S]$ is in general not a function of the ratio $\bar p/ k_{\mathrm{IR}}$ alone. In the following, we analyse the dependence of $C[f_S]$ on $\bar p$ and $k_{\mathrm{IR}}$ or equivalently $B^{1/\kappa}$ according to~\eqref{eq:kf}.

The one-loop building blocks of~\eqref{eq:collisionintdouble}, $\Pi^{\tilde\rho}_S = 2\mathrm{Im}[\Pi^R_S]$ and $\Pi^F_S$, 
are invariant under a uniform rescaling of the frequency and momenta entering them, 
and so is $v_{\mathrm{eff},S}=1/(\mathrm{Re}[\Pi^R_S]^2+\mathrm{Im}[\Pi^R_S]^2)$. Combined with the elementary property $f_S(c\bar q;A,B,\kappa)=f_S(\bar q;A/c^\kappa,B/c^\kappa,\kappa)$ of~\eqref{eq:dist} for any real $c>0$, substituting $\bar q\to c\bar q$ in the (single) loop integrals defining each building block shows that all three scale identically under a joint rescaling of external frequency, external momentum, and the nonuniversal parameters $A$ and $B$. Thus, with $X\in\{\Pi^{\tilde\rho}_S,\Pi^F_S,v_{\mathrm{eff},S}\}$ we have
\begin{equation}
X(c\omega,c\bar p;A,B,\kappa) = X(\omega,\bar p;A/c^\kappa,B/c^\kappa,\kappa).
\label{eq:buildingblockscov}
\end{equation}
Substituting $\bar q\to c\bar q,\,u\to cu$ in~\eqref{eq:collisionintdouble} at external momentum $c\bar p$, using~\eqref{eq:buildingblockscov} for each building block and taking into account the two-fold loop measures  
and the prefactor $1/\bar p^2\to1/(c\bar p)^2$ of~\eqref{eq:collisionintdouble} itself, yields 
the identity
\begin{equation}
C[f_S](1,c\bar p;A,B,\kappa) = c\,C[f_S](1,\bar p;A/c^\kappa,B/c^\kappa,\kappa).
\label{eq:exactcov}
\end{equation}
Since $v_{\mathrm{eff},S}\sim A^{-2}$, while every term of the bracket in~\eqref{eq:collisionintdouble} is cubic in $f_S\sim A$, $C[f_S]$ is exactly linear in $A$. Relabelling $\bar p\to\bar p/c$ in~\eqref{eq:exactcov} gives $C[f_S](1,\bar p;A,B,\kappa)=c\,C[f_S](1,\bar p/c;A/c^\kappa,B/c^\kappa,\kappa)$ valid for any $c$. Choosing $c=k_{\mathrm{IR}}\equiv B^{1/\kappa}$, so that $B/c^\kappa=1$, and pulling out the remaining $A/c^\kappa=A/B$ via linearity then gives
\begin{equation}
	C[f_S](1,\bar p;A,B,\kappa) = A\,B^{1/\kappa\,-1}\,\bar{C}(\bar p/k_{\mathrm{IR}};\kappa) \, ,
\label{eq:universalscaling}
\end{equation}
with $\bar{C}(x;\kappa)\equiv C[f_S](1,x;A{=}1,B{=}1,\kappa)$. Therefore, $C[f_S]$ depends on the nonuniversal parameter $B$ through the explicit prefactor $B^{1/\kappa\,-1}=k_{\mathrm{IR}}^{1-\kappa}$ together with the ratio $\bar p/k_{\mathrm{IR}}$ inside $\bar{C}$, rather than through $\bar p/k_{\mathrm{IR}}$ alone.

We now first consider the limit of large $\bar{p}$ at fixed $B$, and consider the other case of small $B$ at fixed $\bar{p}$ subsequently to determine $\kappa$. The leading-order structure of~\eqref{eq:collisionintdouble} for the limit of large $\bar p$ at fixed $k_{\mathrm{IR}}$ can be read off by matching each branch's frequency argument $\bar p\mp u$ against its own momentum argument $\bar q$.
Near the diagonal $\bar q\to\bar p$, $u\to0$, both $v_{\mathrm{eff},S}$ and the spectral functions $\Pi^{\tilde\rho}_S,\Pi^F_S$ remain smooth there, with the only nonanalytic, logarithmic behaviour at threshold appearing in $\mathrm{Re}[\Pi^R_S]$, which is subdominant to $\mathrm{Im}[\Pi^R_S]$ throughout this region for $\bar p\gg k_{\mathrm{IR}}$. 
Therefore, all three one-loop functions can be evaluated at their on-shell diagonal values $(\bar p,\bar p)$. 
Collecting the two branches of~\eqref{eq:collisionintdouble} at this order, the curly-bracketed term in the integrand reduces to $2f_S(u)\,v_{\mathrm{eff},S}(\bar p,\bar p)\big[f_S(\bar p)\,\Pi^{\tilde\rho}_S(\bar p,\bar p)-\Pi^F_S(\bar p,\bar p)\big]$, so the leading near-diagonal contribution to $C[f_S]$ is proportional to the combination $f_S(\bar p)\,\Pi^{\tilde\rho}_S(\bar p,\bar p)-\Pi^F_S(\bar p,\bar p)$, with the finite, $\bar p^2$-growing prefactor $v_{\mathrm{eff},S}(\bar p,\bar p)$ according to~\eqref{eq:veffnum}. 

By~\eqref{eq:imPiRS_massless} and~\eqref{eq:piFS_massless}, $\mathrm{Im}[\Pi^R_S(\bar p,\bar p)]\sim \int_0^{\bar p}\mathrm d\bar q\,f_S(\bar q)$ and $\Pi^F_S(\bar p,\bar p)\sim \int_0^{\bar p}\mathrm d\bar q\,f_S(\bar q)f_S(\bar p-\bar q)$ share the same prefactor $\lambda/(48\pi\bar p)$. 
For $\bar p\gg k_{\mathrm{IR}}$, the latter integral is dominated by its two endpoints, $\bar q\sim0$ and $\bar q\sim\bar p$, where one factor sits near the infrared plateau, while the other is approximately the constant $f_S(\bar p)$.
Each endpoint reproduces $f_S(\bar p)$ times the integral defining $\mathrm{Im}[\Pi^R_S(\bar p,\bar p)]$, and the two contribute equally, giving the asymptotic identity
\begin{equation}
	\Pi^F_S(\bar p,\bar p) = f_S(\bar p)\,\Pi^{\tilde\rho}_S(\bar p,\bar p) \quad \mathrm{for}\quad \bar p/k_{\mathrm{IR}}\to\infty ,
\label{eq:asymptoticFDT}
\end{equation}
using $\Pi^{\tilde\rho}_S=2\,\mathrm{Im}[\Pi^R_S]$.
We confirm~\eqref{eq:asymptoticFDT} numerically to hold with increasing accuracy as $\bar p/k_{\mathrm{IR}}\to\infty$, the correction vanishing one power of $\bar p$ faster than either side individually. 

Equation~\eqref{eq:asymptoticFDT} is a fluctuation--dissipation-type relation 
emerging as an asymptotic property of the resummed one-loop self-energies. 
The derivation follows purely from the endpoint structure of the $\Pi^F_S$ integral and the resulting cancellation of the leading near-diagonal contribution to $C[f_S]$ holds, in particular, independently of the nonuniversal constant $B$.
As a consequence of this cancellation, 
the scaling degree of $C[f_S](1,\bar p)$ is set by subleading terms. For generic $\kappa$, $C[f_S](1,\bar p)/f_S(\bar p)$ should saturate to a finite value as $\bar p\to\infty$ at fixed $B$, i.e.\ $\bar{C}(x;\kappa)\sim L(\kappa)\,x^{-\kappa}$ as $x\to\infty$, with $L(\kappa)$ generically nonzero. This follows if both sides of~\eqref{eq:scalingfshighp} scale as $\sim \bar{p}^{-\kappa}$ in general. 

Together with these arguments, we now determine the universal scaling exponent $\kappa$ from~\eqref{eq:scalingfshighp} using the scaling property~\eqref{eq:universalscaling} of the collision-integral functional. 
Equation~\eqref{eq:scalingfshighp} holds for any $\bar p\gg k_{\mathrm{IR}}$, which we realise in the following by considering small $k_{\mathrm{IR}}$ or $B^{1/\kappa}$ at fixed $\bar p$. Since $B$ is a nonuniversal quantity, the universal scaling exponent $\kappa$ must not depend on the specific value of $B$. 
In particular, $B>0$ can be small with $B\to 0$ at fixed $\bar p$ and~\eqref{eq:scalingfshighp} still has to hold, realising the required limit $\bar p/ k_{\mathrm{IR}}\to \infty$. 
In this limit $f_S(\bar p)\to A\,\bar p^{-\kappa}$ remains finite and nonzero for any $\kappa$, while~\eqref{eq:universalscaling} forces $C[f_S](1,\bar p)\to0$.
Since $\bar{C}(x;\kappa)$ is a function of $x$ alone and~\eqref{eq:universalscaling} is exact for any $\bar p,B$, the asymptotics $\bar{C}(x;\kappa)\sim L(\kappa)\,x^{-\kappa}$ argued above for $x\to\infty$ apply equally here. For $L(\kappa)$ generically nonzero,~\eqref{eq:universalscaling} then gives $C[f_S](1,\bar p)\sim B^{1/\kappa\,-1}(\bar p/k_{\mathrm{IR}})^{-\kappa}\sim B^{1/\kappa}\to0$. Equation~\eqref{eq:scalingfshighp} reduces in this limit to 
\begin{equation}
\left(\alpha - \kappa \right) A\,\bar{p}^{-\kappa} = 0 \quad \mathrm{for} \quad B \to 0\,.
\label{eq:scalingfshighpsmallB}
\end{equation}
For the considered case of fixed $\bar{p}$ this represents a genuine constraint on $\kappa$, 
in contrast to the alternative limit $\bar p\to\infty$ at fixed $B$, where $f_S(\bar p) \rightarrow A \bar{p}^{-\kappa}$ vanishes as well and the equation would be fulfilled for any $\kappa$. We thus find 
\begin{equation}
\kappa = \alpha \, .
\label{eq:kappaeqalpha}
\end{equation}

The argument leading to~\eqref{eq:kappaeqalpha} uses only the generic structure of~\eqref{eq:scalingfshighp}, not the specific value of $\alpha$, and so applies equally to either conservation law fixing $\alpha$ (Sec.~\ref{sec:scaling}). This yields two distinct self-similar exponents,
\begin{align}
\kappa &= d  \qquad\,\, \text{(particle-number conservation)}, \nonumber\\
\kappa &= d+1  \ \ \text{(energy conservation)},
\label{eq:kappavalues}
\end{align}
for the considered case of $d=3$, using~\eqref{eq:pcrel} and~\eqref{eq:ecrel} respectively. Since $\beta=1$ was found to hold irrespective of $\alpha$ (Sec.~\ref{sec:scaling}), both solutions describe self-similar evolution towards lower characteristic momenta with time. They differ in which conserved quantity is being redistributed by that process, not in its direction. 

That $C[f_S]$ vanishes at $\kappa=\alpha$ does not mean the collision integral switches off; it is specifically a statement about the tail. The vanishing $C[f_S]\sim B^{1/\kappa}\to0$ is the defining signature of a genuine transport solution with constant flux rather than a stationary equilibrium one. It is a statement about the leading order of the on-shell collision integral relative to $f_S(\bar p)$ itself: precisely the order that $\kappa=\alpha$ forces to cancel. A local collision rate that is suppressed relative to naive dimensional counting throughout the inertial range is what a constant-flux cascade requires: if $C[f_S]$ were $\mathcal{O}(f_S)$ at every scale, the distribution would be expected to relax locally rather than transport a conserved quantity across scales. So the actual particle or energy flux is carried by the subleading structure of $C[f_S]$. Taking $B\to0$ pushes the boundary $\bar p\lesssim k_{\mathrm{IR}}$ of the nonvanishing, transport-carrying region down towards $\bar p \to 0$ together with $k_{\mathrm{IR}}$ itself, such that the pure power law $f_S(\bar p)\to A\,\bar p^{-\kappa}$ prevails.

Our results can be compared with findings for nonrelativistic field theories. In particular, for large-$N$ theories with quadratic dispersion the value \mbox{$\kappa_{\mathrm{nonrel}} = d+1$} is reported for particle transport~\cite{Walz:2017ffj,Chantesana:2018qsb}
together with $\beta_{\mathrm{nonrel}} = 1/2$. By contrast, the relativistic theory with massless dispersion has $\kappa = d$ with $\beta = 1$. Thus, relativistic and nonrelativistic theories do not belong to the same universality class in this case. This is different for relativistic field theories in the presence of a nonzero mass gap $M > 0$, where infrared modes with sufficiently low momenta $\bar p \lesssim M$ scale as in corresponding nonrelativistic theories~\cite{PineiroOrioli:2015cpb}.

\section{Conclusion}
\label{sec:conclusion}

In this work, we established nonthermal scaling behaviour in massless scalar field theory based on NLO large-$N$ effective kinetic equations. The nonperturbative solutions describe transport of energy or effective particle number, which are distinguished by different values of the universal scaling exponent $\kappa=d(+1)$ for the distribution function, with the same central time exponent $\beta = 1$. 

While the scaling solutions of this work represent far-from-equilibrium behaviour, there are some remarkable similarities to equilibrium critical phenomena. The massless nonthermal theory has a corresponding infinite correlation length, which arises as a consequence of a careful adjustment that is similar to tuning an equilibrium system to its critical temperature. This is different for more conventional nonthermal fixed point solutions in nonrelativistic regimes, where scaling occurs without tuning even in the presence of a mass gap for a relativistic dispersion once the characteristic momenta drop below that gap. 

While the nonthermal scaling solutions are evolving in time, since they are in general not stationary or approximately stationary as e.g.~prethermalised states~\cite{Berges:2004ce}, their time-dependence can be scaled out of the fixed point equations themselves. 
After the proper rescalings, the nonthermal problem becomes rather similar to static equilibrium calculations. A major difference is the absence of a strict fluctuation--dissipation relation out of equilibrium. However, we found that there is a generalised fluctuation--dissipation-type relation with a nonthermal distribution emerging as an approximate or asymptotic property.  

While nonthermal fixed points in nonrelativistic regimes have been established experimentally in different systems~\cite{Prufer:2018hto, Erne:2018gmz, Glidden:2020qmu, Garcia-Orozco:2021hkx, Lannig:2023fzf, MorenoArmijos2024a.PhysRevLett.134.023401, sab2026universalbehaviorrelaxationdynamics, liang2025universalnonthermalfixedpoint}, it would be striking to discover the nonthermal critical behaviour of this work in experiments. This question is of direct relevance to early-universe cosmology, where relativistic scalar fields play a central role in processes such as heating after inflation \cite{Kofman:1994rk, Kofman:1997yn}. Axion-like models far from equilibrium provide another important example \cite{Sikivie:2009qn, Arias:2012az, Berges:2014xea}, yet the case of vanishing mass remains largely unexplored. Relativistic, massless dispersions may also be realised effectively using laboratory systems, such as those based on ultracold quantum gases providing very flexible testbeds~\cite{Bloch_2008}. This would give access to a wide range of far-from-equilibrium phenomena, such as new forms of strong wave-turbulence or coarsening-like behaviour in relativistic regimes.

\begin{acknowledgments}
We are grateful to M.~Tarpin for being involved in the initial stages of this project. We also thank T.~Gasenzer, J.~M.~Pawlowski, N.~Wink, J.~Wessely, and Y.~Y.~Tan for fruitful discussions and collaboration on related topics described here.  
This work is part of and funded by the Deutsche Forschungsgemeinschaft (DFG, German Research Foundation) under the Collaborative Research Centre, Project-ID No. 273811115, SFB 1225 ISOQUANT and Germany’s Excellence Strategy EXC 2181/1–390900948 (the Heidelberg STRUCTURES Excellence Cluster). 
VN acknowledges support by the ERC grant OPEN-2QS (Grant No. 101164443). ANM acknowledges financial support by the Swiss National Science Foundation (SNSF) through Sinergia Grant No. CRSII5 206008/1. 
\end{acknowledgments}

\appendix	
\onecolumngrid	
\section{Propagators on the Schwinger-Keldysh contour in a semiclassical approximation}
\label{app:keldysh}
The action \eqref{eq:classaction} can be rewritten as 
\begin{equation}
	\begin{aligned}
		S\left[\phi^{+}, \phi^{-}\right] & =\frac{1}{2} \int_{x y}\left(\phi_a^{+}(x), \phi_a^{-}(x)\right)\left(\begin{array}{cc}
			i D_{a b}^{-1}(x, y) & 0 \\
			0 & -i D_{a b}^{-1}(x, y)
		\end{array}\right)\left(\begin{array}{l}
			\phi_b^{+}(y) \\
			\phi_b^{-}(y)
		\end{array}\right) \\
		& -\frac{\lambda}{4 ! N}\left(\phi_a^{+}(x) \phi_a^{+}(x) \phi_b^{+}(x) \phi_b^{+}(x)-\phi_a^{-}(x) \phi_a^{-}(x) \phi_b^{-}(x) \phi_b^{-}(x)\right),
	\end{aligned}
\end{equation}
where the superscripts ‘$+$’ and ‘$-$’ indicate that the fields are taken on the forward ($\mathcal{C}^+$) and backward ($\mathcal{C}^{-}$) branch of the closed time path, respectively. The corresponding generating functional for the correlation functions is 
\begin{equation}
	\begin{aligned}
		Z[J, R] & =\int \mathcal{D} \phi^{+} \mathcal{D} \phi^{-} \exp \left\{i \left[S\left[\phi^{+}, \phi^{-}\right]+\int_x\left(\phi_a^{+}(x), \phi_a^{-}(x)\right)\left(\begin{array}{c}
			J_a^{+}(x) \\
			-J_a^{-}(x)
		\end{array}\right)\right.\right. \\
		& \left.\left.+\frac{1}{2} \int_{x y}\left(\phi_a^{+}(x), \phi_a^{-}(x)\right)\left(\begin{array}{cc}
			R_{a b}^{++}(x, y) & -R_{a b}^{+-}(x, y) \\
			-R_{a b}^{-+}(x, y) & R_{a b}^{--}(x, y)
		\end{array}\right)\left(\begin{array}{c}
			\phi_b^{+}(y) \\
			\phi_b^{-}(y)
		\end{array}\right)\right]\right\},
	\end{aligned}
\label{eq:generating}
\end{equation}
where $J^{\pm}$ are linear source terms, while $R^{++}$ etc. are sources bilinear in the fields. As a result of this notation, all spacetime integrations run from $-\infty$ to $+\infty$, therefore calculations are done in $d+1$ Minkowski spacetime. We obtain the connection to the spectral and statistical components by a linear transformation $A$ of the fields 
\begin{equation}
	\left(\begin{array}{c}
		\phi \\
		\tilde{\phi}
	\end{array}\right)=A\left(\begin{array}{c}
		\phi^{+} \\
		\phi^{-}
	\end{array}\right), \quad A=\left(\begin{array}{rr}
		\frac{1}{2} & \frac{1}{2} \\
		1 & -1
	\end{array}\right).
\end{equation}
By transforming the source terms according to
\begin{equation}
	\left(\begin{array}{c}
		J \\
		\tilde{J}
	\end{array}\right)=A\left(\begin{array}{c}
		J^{+} \\
		J^{-}
	\end{array}\right),\left(\begin{array}{cc}
		R^F & R^{\mathrm{R}} \\
		R^{\mathrm{A}} & R^{\tilde{F}}
	\end{array}\right)=A\left(\begin{array}{cc}
		R^{++} & R^{+-} \\
		R^{-+} & R^{--}
	\end{array}\right) A^T,
\end{equation}
and substituting these definitions into \eqref{eq:generating}, the resulting action functional can be written in terms of a free part 
\begin{equation}
	S_0[\phi, \tilde{\phi}]=\frac{1}{2} \int_{x y}\left(\phi_a(x), \tilde{\phi}_a(x)\right)\left(\begin{array}{cc}
		0 & i D_{a b}^{-1}(x, y) \\
		i D_{a b}^{-1}(x, y) & 0
	\end{array}\right)\left(\begin{array}{l}
		\phi_b(y) \\
		\tilde{\phi}_b(y)
	\end{array}\right),
\end{equation}
and an interaction part 
\begin{equation}
	S_{\mathrm{int}}[\phi, \tilde{\phi}]=-\frac{\lambda}{6 N} \int \tilde{\phi}_a(x) \phi_a(x) \phi_b(x) \phi_b(x)-\frac{\lambda}{24 N} \int \tilde{\phi}_a(x) \tilde{\phi}_a(x) \tilde{\phi}_b(x) \phi_b(x).
\end{equation}
The first term in the interaction part is equal to that of classical statistical field theory, while the second term is unique to quantum fields, i.e. a ``quantum vertex''. In this work, we neglect quantum corrections and so only keep the ``classical vertex''.  Defining $W=-i \ln Z$, the connected retarded/advanced propagators $G_{a b}^{\mathrm{R}, \mathrm{A}}(x, y)$, the statistical propagator $F_{a b}(x, y)$, and $\tilde{F}_{a b}(x, y)$ are given by 
\begin{equation}
	\begin{array}{ll}
		\dfrac{\delta^2 W}{\delta \tilde{J}_a(x) \delta J_b(y)}=G_{a b}^{\mathrm{R}}(x, y), & \dfrac{\delta^2 W}{\delta J_a(x) \delta \tilde{J}_b(y)}=G_{a b}^{\mathrm{A}}(x, y), \\[10pt]
		\dfrac{\delta^2 W}{\delta J_a(x) \delta J_b(y)}=i F_{a b}(x, y), & \dfrac{\delta^2 W}{\delta J_a(x) \delta J_b(y)}=i \tilde{F}_{a b}(x, y).
	\end{array}
\end{equation}
In the $\pm$ basis, the propagators are given by
\begin{equation}
	\left(\begin{array}{cc}
		F & -i G^{\mathrm{R}} \\
		-i G^{\mathrm{A}} & \tilde{F}
	\end{array}\right)=A\left(\begin{array}{cc}
		G^{++} & G^{+-} \\
		G^{-+} & G^{--}
	\end{array}\right) A^T,
\end{equation}
where the anomalous propagator $\tilde{F}$ vanishes. The inverse of the two-point function in this case is 
\begin{equation}
	\left(\begin{array}{cc}
		0 & i\left(G^{\mathrm{A}}\right)^{-1} \\
		i\left(G^{\mathrm{R}}\right)^{-1} & \left(G^{\mathrm{R}}\right)^{-1} \cdot F \cdot\left(G^{\mathrm{A}}\right)^{-1}
	\end{array}\right)=\left(\begin{array}{cc}
		0 & G_0^{-1} \\
		G_0^{-1} & 0
	\end{array}\right)-\left(\begin{array}{cc}
		0 & -i \Sigma^{\mathrm{A}} \\
		-i \Sigma^{\mathrm{R}} & \Sigma^F
	\end{array}\right),
\end{equation}
where the retarded, advanced and statistical self energies are obtained as 
\begin{equation}
	\left(\begin{array}{cc}
		0 & -i \Sigma^{\mathrm{A}} \\
		-i \Sigma^{\mathrm{R}} & \Sigma^F
	\end{array}\right)=\left(A^{-1}\right)^T\left(\begin{array}{cc}
		\Sigma^{++} & -\Sigma^{+-} \\
		-\Sigma^{-+} & \Sigma^{--}
	\end{array}\right) A^{-1}
\end{equation}
from the self energies in the $\pm$ basis.

\section{Effective vertex analytic computations}
\label{app:effective_vertex}
In the following, we perform the angular integrations analytically on $\Pi^R_S( \omega, \pp)$. For the nonrelativistic case, a similar computation has been performed in \cite{Walz:2017ffj}. First, let us change the integration variable $\pp-\qq \rightarrow \qq$ in \eqref{eq:pir} and introduce $y \equiv \mathrm{cos}(\theta)$, which is the polar angle between $\pp$ and $\qq$. In spherical polar coordinates, we arrive at 
\begin{align}
\Pi^{R}_S(\omega,\mathbf{p})
&= \frac{\lambda}{12}\,\frac{1}{4\pi^{2}}
   \int_{0}^{\infty} \mathrm{d}\bar{q}\, \bar{q}^{2} \int_{-1}^{1} \mathrm{d}y\,
   \frac{f_S(\bar{q})}{|\mathbf{p}-\mathbf{q}|\, \bar{q}}\nonumber
\\[4pt]
&\quad\times\left[
\frac{1}{|\mathbf{p}-\mathbf{q}|+\bar{q}-\omega-i\epsilon} +
\frac{1}{|\mathbf{p}-\mathbf{q}|-\bar{q}-\omega-i\epsilon} +
\frac{1}{|\mathbf{p}-\mathbf{q}|-\bar{q}+\omega+i\epsilon} +
\frac{1}{|\mathbf{p}-\mathbf{q}|+\bar{q}+\omega+i\epsilon}
\right],
\end{align}
with $|\qq|\equiv \bar{q}$, $|\pp|\equiv\bar{p}$, and $|\mathbf{p}-\mathbf{q}| = \sqrt{\bar{p}^2 + \bar{q}^2 - 2\bar{p}\bar{q}y}$.
Then, with the change of integration variables $u\equiv|\mathbf{p}-\mathbf{q}|$, one has $\mathrm{d}\bar{q}\mathrm{d}y = -(u/\bar{p}\bar{q})\mathrm{d}\bar{q}\mathrm{d}u$, and the new integration limits for $u$ are
\begin{equation}
    y = 1 \implies u_{\mathrm{min}} = \sqrt{\bar{p}^2+\bar{q}^2-2\bar{p}\bar{q}} = \sqrt{(\bar{p}-\bar{q})^2} = |\bar{p}-\bar{q}|,
\end{equation}
as well as
\begin{equation}
y = -1 \implies u_{\mathrm{max}} = \sqrt{\bar{p}^2+\bar{q}^2+2\bar{p}\bar{q}} = \sqrt{(\bar{p}+\bar{q})^2} = |\bar{p}+\bar{q}|,
\end{equation}
where we had a sign change in the integral which flipped the upper and lower boundaries.
Substituting all this back, the above equation becomes
\begin{equation}
\Pi^{R}_S(\omega,\bar{p})=\frac{\lambda}{48\pi^2 \bar{p}}\int_0^\infty \mathrm{d}\bar{q} f_S(\bar{q}) \int_{|\bar{p}-\bar{q}|}^{\bar{p}+\bar{q}} \mathrm{d}u\left[\frac{1}{u+\bar{q}-\omega-i\epsilon} + \frac{1}{u-\bar{q}-\omega-i\epsilon} + \frac{1}{u-\bar{q}+\omega+i\epsilon} + \frac{1}{u+\bar{q}+\omega+i\epsilon} \right],
\end{equation}
where the factor of $\bar{q}^2$ from the spherical integration measure was cancelled by the $\bar{q}^{-1}$ from $\omega_{\mathbf{q}}$ and the $\bar{q}^{-1}$ introduced by the substitution to $u$.
To evaluate the $u$ integral, we use \begin{equation}
    \lim_{\epsilon \to 0^+} \int_{|\bar{p}-\bar{q}|}^{\bar{p}+\bar{q}}\mathrm{d}u\frac{1}{u + C \pm i \epsilon} = \mathcal{PV} \int_{|\bar{p}-\bar{q}|}^{\bar{p}+\bar{q}}\mathrm{d}u\frac{1}{u+C} \mp i \pi  \int_{|\bar{p}-\bar{q}|}^{\bar{p}+\bar{q}}\mathrm{d}u \delta(u+C).
    \label{eq:plemelj}
\end{equation}
The real part of the integrals is given by
\begin{equation}
    \int_{|\bar{p}-\bar{q}|}^{\bar{p}+\bar{q}} \mathrm{d}u \frac{1}{u+C} = \ln|u+C| \Big|_{|\bar{p}-\bar{q}|}^{\bar{p}+\bar{q}} = \ln\left|\frac{\bar{p}+\bar{q}+C}{|\bar{p}-\bar{q}|+C}\right|,
\end{equation}
while the imaginary part is
\begin{equation}
    \pi \int_{|\bar{p}-\bar{q}|}^{\bar{p}+\bar{q}} \mathrm{d}u \delta(u+C) =
    \begin{cases}
    \pi, & -C \in [|\bar{p}-\bar{q}|, \bar{p}+\bar{q}], \\
    0, & \text{otherwise}.
    \end{cases}
\end{equation}
With this in mind, one can write down for the $\Pi^R_S$ function the expression
\begin{equation}
\Pi^R_S(\omega, \bar{p})=\frac{\lambda}{48\pi^2\bar{p}} \int_0^{\infty} \mathrm{d} \bar{q} f_S( \bar{q}) \Gamma_{\Pi}(\omega, \bar{p}, \bar{q}).
\label{eq:pirradial}
\end{equation}
After the appropriate substitution of integration limits, the kernel $\Gamma_{\Pi}$ has real part
\begin{equation}
\mathrm{Re} \Gamma_{\Pi}(\omega,\bar{p},\bar{q})=\ln \left| \frac{(\bar{p}+2\bar{q})^2 - \omega^2}{(\bar{p}-2\bar{q})^2 - \omega^2} \right|,
\end{equation}
and the imaginary part is given by $ \Theta(u_0 - |\bar{p}-\bar{q}|) \cdot \Theta(\bar{p}+\bar{q} - u_0)$, where $u_0$ is equal to $\omega - \bar{q}$, $\omega + \bar{q}$, $- \omega + \bar{q}$, and $-\omega - \bar{q}$ for the four terms respectively.
Altogether, we find
\begin{equation}
    \begin{aligned}
    \operatorname{Im} \Gamma_{\Pi}(\omega, \bar{p}, \bar{q})
    &=
    \pi\left[\Theta(\omega-\bar{q}-|\bar{p}-\bar{q}|) \cdot \Theta(\bar{p}+\bar{q}-(\omega-\bar{q})) 
    +  
    \Theta(\omega+\bar{q}-|\bar{p}-\bar{q}|) \cdot \Theta(\bar{p}+\bar{q}-(\omega+\bar{q}))\right. \\
    &- 
    \left.\Theta(\bar{q}-\omega-|\bar{p}-\bar{q}|) \cdot \Theta(\bar{p}+\bar{q}-(\bar{q}-\omega))
    -\Theta(-\omega-\bar{q}-|\bar{p}-\bar{q}|)\cdot \Theta(\bar{p}+\bar{q}-(-\omega-\bar{q}))\right].
    \end{aligned}
\end{equation}
We are interested in the on-shell limit behaviour of the effective vertex,
\begin{equation} v_{\mathrm{eff},S} \approx \left[(\mathrm{Re}\Pi^R_S)^2 + (\mathrm{Im}\Pi^R_S)^2\right]^{-1}.
\end{equation}
For this, we set $\omega = \sqrt{\bar{p}^2+M^2}$, and as we will see below, one has to be careful with the infrared limit of the real part of $\Pi^R_S$, where a finite mass, or small off-shell contribution is necessary to avoid a logarithmic divergence.

We now proceed to describe the behaviour of the effective vertex, based on analysing the limiting expressions of $\Pi^R_S$ in \eqref{eq:pirradial} for the cases $\bar{p} \lesssim k_{\mathrm{IR}}$ and $\bar{p} \gtrsim k_{\mathrm{IR}}$.
For the on-shell limit $\omega=\bar{p}$ of the massless theory, the imaginary part of the kernel $\Gamma_{\Pi}$ can be calculated by examining its behaviour as $\omega$ approaches $\bar{p}$ from above $\omega = \bar{p}+\epsilon$ and below $\omega = \bar{p}-\epsilon$, as follows.
\paragraph*{Approach $\omega=\bar{p}+\epsilon$, $\epsilon\to 0^+$.}
The root of term~(1) becomes $u_0^{(1)} = \bar{p}+\epsilon-\bar{q}$. For $\bar{q}<\bar{p}$ the lower-bound condition $\bar{p}-\bar{q}\le \bar{p}+\epsilon-\bar{q}$ reduces to $0\le\epsilon$ and is satisfied, while the upper bound $\bar{p}+\epsilon-\bar{q}\le \bar{p}+\bar{q}$ gives $\epsilon\le 2\bar{q}$ and is satisfied for any $\bar{q}>0$ as $\epsilon\to 0$.
For $\bar{q}>\bar{p}$ the lower bound $\bar{q}-\bar{p}\le \bar{p}+\epsilon-\bar{q}$ requires $\bar{q}\le \bar{p}+\epsilon/2$, which fails.
Hence term~(1) contributes $\Theta(\bar{p}-\bar{q})$. Term~(2) has $u_0^{(2)} = \bar{p}+\epsilon+\bar{q}$, whose upper-bound condition $\bar{p}+\epsilon+\bar{q}\le \bar{p}+\bar{q}$ requires $\epsilon\le 0$ and is violated.
Term~(3) has $u_0^{(3)} = \bar{q}-\bar{p}-\epsilon$; positivity together with the lower-bound condition forces $\bar{q}\ge \bar{p}+\epsilon$, but then the upper bound $\bar{q}-\bar{p}-\epsilon\le \bar{p}+\bar{q}$ requires $-\bar{p}-\epsilon\le \bar{p}$ which is satisfied, while the consistency condition $\bar{q}-\bar{p}-\epsilon\ge \bar{q}-\bar{p}$ requires $\epsilon\le 0$ and is again violated.
Combining the three terms with their signs gives $\mathrm{Im}\,\Gamma_{\Pi} = \pi\,\Theta(\bar{p}-\bar{q})$.

\paragraph*{Approach $\omega=\bar{p}-\epsilon$, $\epsilon\to 0^+$.}
Now term~(1) has $u_0^{(1)}=\bar{p}-\epsilon-\bar{q}$, whose lower bound
$\bar{p}-\bar{q}\le \bar{p}-\epsilon-\bar{q}$ requires $\epsilon\le 0$ and is violated; term~(1) therefore vanishes in this limit. Term~(2) has $u_0^{(2)}=\bar{p}-\epsilon+\bar{q}$
with upper bound $\bar{p}-\epsilon+\bar{q}\le \bar{p}+\bar{q}$ giving $-\epsilon\le 0$, satisfied for any $\bar{q}>0$, so term~(2) contributes a constant $1$. Term~(3) has $u_0^{(3)}=\bar{q}-\bar{p}+\epsilon$; positivity requires $\bar{q}\ge \bar{p}-\epsilon$, the lower bound $|\bar{p}-\bar{q}|\le \bar{q}-\bar{p}+\epsilon$ is satisfied for $\bar{q}\ge \bar{p}-\epsilon/2$, and the upper bound is automatic, so term~(3) contributes $\Theta(\bar{q}-\bar{p})$.
With the relative sign of term~(3) being negative, the sum is $0 + 1 - \Theta(\bar{q}-\bar{p}) - 0 = \Theta(\bar{p}-\bar{q})$, so that both limits converge on the same result,
\begin{equation}
    \mathrm{Im}\,\Gamma_{\Pi}(\omega=\bar{p},\bar{p},\bar{q}) = \pi\,\Theta(\bar{p}-\bar{q})\,.
\end{equation}

The massive on-shell limit shifts the on-shell frequency strictly above
$\bar{p}$, $\omega-\bar{p} = M^2/(\omega+\bar{p}) > 0$. Repeating the analysis with
$\omega>\bar{p}$ enforced strictly, term~(2) is absent due to $\omega+\bar{q}\le \bar{p}+\bar{q}\Leftrightarrow\omega\le \bar{p}$, and term~(3) by $\bar{q}-\omega\le \bar{q}-\bar{p}\Leftrightarrow \omega\ge \bar{p}$ combined with $\bar{q}-\omega\ge \bar{q}-\bar{p}$.
Only term~(1) survives, with support
\begin{equation}
    \frac{\omega-\bar{p}}{2}\le \bar{q} \le \frac{\omega+\bar{p}}{2}\,,
    \label{eq:pirbounds}
\end{equation}
where the lower bound originates from the upper $u$-limit and the upper
bound from the lower $u$-limit (after distinguishing $\bar{q}<\bar{p}$ and $\bar{q}>\bar{p}$). Both bounds are exact for any $M$: defining
\begin{equation}
    q_c\equiv\frac{\omega-\bar{p}}{2} = \frac{M^2}{2(\omega+\bar{p})}\,,
\end{equation}
the identity $\frac{\omega+\bar{p}}{2}=\bar{p}+q_c$ then holds exactly, with no expansion required anywhere in this step ($q_c$ reduces to $M^2/4\bar{p}$ only in the separate limit $M\ll\bar{p}$). So that
\begin{equation}
\mathrm{Im}\,\Gamma_{\Pi}\big(\omega=\sqrt{\bar{p}^2+M^2},\bar{p},\bar{q}\big)
    = \pi\,\Theta(\bar{q}-q_c)\,\Theta(\bar{p}+q_c-\bar{q})\,.
\end{equation}
As $M\to 0$ the lower edge $q_c\to 0$ and the upper edge $\bar{p}+q_c\to \bar{p}$,
recovering $\pi\,\Theta(\bar{p}-\bar{q})$.

Therefore, \eqref{eq:pirradial} simplifies to
\begin{equation}
    \Pi^R_S(\omega,\bar{p})=\frac{\lambda}{48\pi^2} \frac{1}{\bar{p}} \left[\int_0^{\infty}\mathrm{d}\bar{q}f_S(\bar{q})\ln \left|\frac{(\bar{p}+2 \bar{q})^2-\omega^2}{(\bar{p}-2 \bar{q})^2-\omega^2}\right|  +i\pi\int_{q_c}^{\bar{p}+q_c}\mathrm{d}\bar{q}f_S(\bar{q})\right].
\end{equation}
For the regime $\bar{p} \lesssim k_{\mathrm{IR}}$, we note that the distribution $f_S(\bar{q})$ is merely a constant, and therefore the radial integral becomes $\frac{1}{\bar{p}} \int_{0}^{\bar{p}} \mathrm{d}\bar{q}\, f_S(\bar{q})\simeq f_S(\bar{p})$.
Hence, in the limiting cases for the imaginary part of $\Pi^R_S$, we have
\begin{equation}
\begin{aligned}
& \operatorname{Im} \Pi^R_S\left(\bar{p},\bar{p}\right) \stackrel{\bar{p} \gtrsim k_{\mathrm{IR}}}{\simeq}  \frac{1}{\bar{p}} k_{\mathrm{IR}} f_S(k_{\mathrm{IR}}), \\
& \operatorname{Im} \Pi^R_S\left( \bar{p},\bar{p}\right) \stackrel{\bar{p} \lesssim k_{\mathrm{IR}}}{\simeq}  f_S(\bar{p}).
\end{aligned}
\end{equation}
For the real part of $\Pi^R_S$, in the case of $\omega = \bar{p}$, the kernel $\Gamma_{\Pi}$ simplifies to $\mathrm{ln}[(\bar{q}+\bar{p})/|\bar{q}-\bar{p}|]$.
For the regime $\bar{p} \lesssim k_{\mathrm{IR}}$, the integral $\int \mathrm{d}\bar{q}\, f_S(\bar{q})\,\Gamma_{\Pi}$ will be dominated by the integration variable $\bar{q}$ being near the scale $k_{\mathrm{IR}}$, therefore we need the case $\bar{q} \gg \bar{p}$.
The real part can be approximated by Taylor expanding the logarithmic term $\mathrm{ln}(\bar{p}+\bar{q})-\mathrm{ln}|\bar{p}-\bar{q}|\approx2\bar{p}/\bar{q}+\mathcal{O}((\bar{p}/\bar{q})^3)$.
The case $\bar{p} \gg \bar{q}$ then similarly leads to $\mathrm{ln}(\bar{q}+\bar{p})-\mathrm{ln}|\bar{q}-\bar{p}|\approx2\bar{q}/\bar{p}+\mathcal{O}((\bar{q}/\bar{p})^3)$.
For the case $\bar{p} \gtrsim k_{\mathrm{IR}}$, this predicts an overall $\sim \bar{p}^{-2}$ scaling for $\Pi^R_S$, however, for the $\bar{p} \lesssim k_{\mathrm{IR}}$, we are faced with a logarithmic divergence in the infrared as
\begin{equation}
    \operatorname{Re}\Pi^{R}_S \sim \frac{1}{\bar{p}} \left( \bar{p} \cdot \int \mathrm{d}\bar{q} \frac{1}{\bar{q}} f_S(\bar{q}) \right) \sim \int \mathrm{d}\bar{q} \frac{f_S(\bar{q})}{\bar{q}},
\end{equation}
since $f_S(\bar{q})$ is a constant in this momentum range.
Keeping a finite mass $M$, the real part of the kernel $\Gamma_{\Pi}$ is given by
\begin{equation}
    \operatorname{Re} \Gamma_{\Pi}(\omega=\sqrt{\bar{p}^2+M^2}, \bar{p}, \bar{q})=\ln \left|\frac{(2 \bar{q} + \bar{p})^2 - (\bar{p}^2 + M^2)}{(2 \bar{q} - \bar{p})^2 - (\bar{p}^2 + M^2)}\right|,
    \label{eq:pirln}
\end{equation}
which we can Taylor expand for $\bar{q}, \bar{p} \ll M/2$ to get
\begin{equation}
    \operatorname{Re} \Gamma_{\Pi}(\omega=\sqrt{\bar{p}^2+M^2}, \bar{p}, \bar{q}) \approx -\frac{8 \bar{p} \bar{q}}{M^2},
\end{equation}
while for $\bar{q} \gg M/2$ the massless behaviour $\operatorname{Re}\Gamma_{\Pi} \approx 2\bar{p}/\bar{q}$ is recovered.
The mass therefore acts as an infrared cutoff of the logarithmic divergence:
splitting the radial integral at $\bar{q} \sim M/2$,
\begin{equation}
\operatorname{Re}\Pi^R_S\big(\sqrt{\bar{p}^2+M^2},\bar{p}\to 0\big)
\approx \frac{\lambda}{48\pi^2}\left[-\frac{8}{M^2}\int_0^{M/2}\mathrm{d}\bar{q}\, \bar{q}\, f_S(\bar{q})
+ 2\int_{M/2}^{\infty}\mathrm{d}\bar{q}\,\frac{f_S(\bar{q})}{\bar{q}}\right].
\label{eq:repisplit}
\end{equation}
Which of the two terms dominates depends on where the crossover scale $M/2$ sits relative to $k_{\mathrm{IR}}$.
For $M \ll k_{\mathrm{IR}}$, the first (mass) term is confined to $\bar{q}\lesssim M/2 \ll k_{\mathrm{IR}}$, where $f_S(\bar{q})\approx f_S(0)$, and gives a finite, $M$-independent contribution, while the logarithmic divergence resurfaces in the second term as its lower limit $M/2\to0$ and is regulated there:
\begin{equation}
\operatorname{Re}\Pi^R_S\big(\sqrt{\bar{p}^2+M^2},\bar{p}\to 0\big)\simeq \frac{\lambda f_S(0)}{24\pi^2}\,\ln\!\left(\frac{c\,k_{\mathrm{IR}}}{M}\right) \qquad (M \ll k_{\mathrm{IR}}),
\label{eq:relogM}
\end{equation}
with $c$ an $\mathcal{O}(1)$ constant depending on the shape of $f_S$. Assuming $\kappa > 2$, for $M \gtrsim k_{\mathrm{IR}}$ the crossover scale $M/2$ already lies beyond the plateau of $f_S$, so the second (massless-tail) term in \eqref{eq:repisplit} is suppressed and the first term dominates instead:
\begin{equation}
\operatorname{Re}\Pi^R_S \approx -\frac{\lambda}{6\pi^2 M^2}\int_0^\infty\mathrm{d}\bar{q}\, \bar{q}\, f_S(\bar{q})
\sim -\frac{k_{\mathrm{IR}}^2 f_S(k_{\mathrm{IR}})}{M^2} \qquad (M \gtrsim k_{\mathrm{IR}}).
\label{eq:reM2}
\end{equation}

The imaginary part follows the same $\bar{p}\ll M$ hierarchy used above: expanding $\omega=\sqrt{\bar{p}^2+M^2}\simeq M+\bar{p}^2/2M$ for $\bar{p}\ll M$, the exact bounds \eqref{eq:pirbounds} on the surviving $\Theta$-function support become $(\omega-\bar{p})/2\simeq M/2-\bar{p}/2$ and $(\omega+\bar{p})/2\simeq M/2+\bar{p}/2$, i.e.~a window $\bar{q}\in[M/2-\bar{p}/2,\,M/2+\bar{p}/2]$ of width $\bar{p}$ centred at $\bar{q}\simeq M/2$, independently of how $M$ compares to $k_{\mathrm{IR}}$:
\begin{equation}
\operatorname{Im}\Pi^R_S\big(\sqrt{\bar{p}^2+M^2},\bar{p}\to 0\big) \simeq \frac{\lambda}{48\pi}\, f_S(M/2).
\label{eq:imMgeneral}
\end{equation}
For $M \ll k_{\mathrm{IR}}$ this reproduces the constant $\sim f_S(0)$ used in \eqref{eq:relogM}, since $f_S(M/2)\simeq f_S(0)$ there. For $M \gtrsim k_{\mathrm{IR}}$, however, \eqref{eq:imMgeneral} instead probes the ultraviolet tail of the scaling function, $f_S(M/2)\sim (M/2)^{-\kappa}$ [cf.\ \eqref{eq:dist}], so that $\operatorname{Im}\Pi^R_S\sim M^{-\kappa}$. For $\kappa>2$, this decays strictly faster than the real part's $M^{-2}$ in \eqref{eq:reM2}; the real part therefore dominates $|\Pi^R_S|^2$ in \emph{both} low-momentum regimes, albeit via different mechanisms -- growing without bound (logarithmically) as $M\to0$ for $M \ll k_{\mathrm{IR}}$, or simply outlasting a now strongly suppressed imaginary part for $M \gtrsim k_{\mathrm{IR}}$.

Collecting these results, for the real part we have
\begin{equation}
\begin{aligned}
& \operatorname{Re} \Pi^R_S\left(\bar{p},\bar{p}\right) \stackrel{\bar{p} \gtrsim k_{\mathrm{IR}}}{\simeq} \frac{k_{\mathrm{IR}}^2 f_S(k_{\mathrm{IR}})}{\bar{p}^2}, \\[4pt]
& \operatorname{Re} \Pi^R_S\left(\sqrt{\bar{p}^2+M^2},\bar{p}\right) \stackrel{\bar{p} \lesssim k_{\mathrm{IR}}}{\simeq}
\begin{cases}
f_S(0)\,\ln\!\left(k_{\mathrm{IR}}/M\right) & (M \ll k_{\mathrm{IR}}), \\
-k_{\mathrm{IR}}^2 f_S(k_{\mathrm{IR}})/M^2 & (M \gtrsim k_{\mathrm{IR}}).
\end{cases}
\end{aligned}
\label{eq:reallcases}
\end{equation}
Therefore, when considering the scaling of the effective vertex $v_{\mathrm{eff},S}=|\Pi^R_S|^{-2}$: at high momenta ($\bar{p}\gtrsim k_{\mathrm{IR}}$), the imaginary part decreases more slowly than the real part and dominates, giving $v_{\mathrm{eff},S}\sim \bar{p}^2$. At low momenta ($\bar{p}\lesssim k_{\mathrm{IR}}$), the real part instead dominates throughout, but with two distinct sub-cases set by the mass regulator: a logarithmically suppressed constant, $v_{\mathrm{eff},S}\sim[f_S(0)\ln(k_{\mathrm{IR}}/M)]^{-2}$, for $M \ll k_{\mathrm{IR}}$; and a quartic scaling, $v_{\mathrm{eff},S}\sim M^4/[k_{\mathrm{IR}}^2f_S(k_{\mathrm{IR}})]^2$, for $M \gtrsim k_{\mathrm{IR}}$. The effective vertex therefore vanishes only in the joint limit $\bar{p}\to0$, $M\to0$, and only logarithmically slowly; it remains finite for any fixed $\bar{p}>0$ at $M=0$, growing as $\bar{p}^2$ once $\bar{p}\gtrsim k_{\mathrm{IR}}$.

For $\kappa\le2$, both~\eqref{eq:reM2} and the resulting quartic scaling change character. The estimate~\eqref{eq:reM2} is obtained from the first term of~\eqref{eq:repisplit} by extending its finite upper limit $M/2$ to infinity -- legitimate only once the integrand $\bar{q}\,f_S(\bar{q})$ has already saturated by $\bar{q}\sim k_{\mathrm{IR}}\ll M/2$, i.e.~only for $\kappa>2$; the underlying integral $\int_0^{M/2}\mathrm{d}\bar{q}\,\bar{q}\,f_S(\bar{q})$ itself remains perfectly finite for any $\kappa$, being bounded by the finite limit $M/2$ rather than by infinity, so no divergence is at stake here. For $\kappa\le2$ this extension is simply no longer legitimate: the integral is instead dominated by its own upper limit, $\int_0^{M/2}\mathrm{d}\bar{q}\,\bar{q}\,f_S(\bar{q})\sim(M/2)^{2-\kappa}$, giving
\begin{equation}
\operatorname{Re}\Pi^R_S\big(\sqrt{\bar{p}^2+M^2},\bar{p}\to0\big)\sim M^{-\kappa}\qquad(\kappa\le2,\ M\gtrsim k_{\mathrm{IR}}),
\label{eq:reMkappa}
\end{equation}
in place of the $M^{-2}$ of~\eqref{eq:reM2}. Since~\eqref{eq:imMgeneral}'s $\operatorname{Im}\Pi^R_S\sim M^{-\kappa}$ never relied on $\kappa\gtrless2$ to begin with, the real and imaginary parts become comparable rather than one cleanly dominating, and the low-momentum, large-mass branch of the effective vertex turns explicitly $\kappa$-dependent,
\begin{equation}
v_{\mathrm{eff},S}(\bar{p}\ll M,\ M\gtrsim k_{\mathrm{IR}})\sim M^{2\kappa}\qquad(\kappa\le2),
\label{eq:veffkappadep}
\end{equation}
replacing the universal quartic branch of~\eqref{eq:veffnum}. The high-momentum branch~\eqref{eq:veffhighp} is more robust: although its own $\operatorname{Re}\Pi^R_S\sim k_{\mathrm{IR}}^2f_S(k_{\mathrm{IR}})/\bar{p}^2$ likewise assumes $\kappa>2$ by the identical mechanism (for $\kappa\le2$ it becomes $\operatorname{Re}\Pi^R_S\sim\bar{p}^{-\kappa}$ instead), the conclusion $v_{\mathrm{eff},S}\sim \bar{p}^2$ survives regardless, since $\operatorname{Im}\Pi^R_S\sim1/\bar{p}$ continues to dominate $\operatorname{Re}\Pi^R_S\sim\bar{p}^{-\min(\kappa,2)}$ for any $\kappa>1$.

\section{Computations for $\Pi^F_S(\omega,\bar{p})$}
\label{app:PiF}

To further push the analysis on \eqref{eq:pif-onsh}, we reuse some of the steps from App.~\ref{app:effective_vertex}. Representing the delta functions from \eqref{eq:pif-onsh} as
\begin{equation}
\delta(\nu \pm \omega_r \pm \omega_{q-r}) = \lim_{\varepsilon\to 0^+} \frac{1}{2\pi i}\left(\frac{1}{\omega_r \pm \omega_{q-r} - \nu - i\varepsilon} - \frac{1}{\omega_r \pm \omega_{q-r} - \nu + i\varepsilon}\right),
\label{eq:delta_identity}
\end{equation}
we can perform the same change of variables $u = |\pp - \qq|$ as in App.~\ref{app:effective_vertex}. The key difference with respect to $\Pi^R_S$ is the presence of an additional factor $f_S(u)$ under the integral, originating from the second distribution function in \eqref{eq:pif-onsh}. Using the identity \eqref{eq:delta_identity}, the real part vanishes identically, as expected for $\Pi^F$, and we are left with integrals of the form
\begin{equation}
\int_{|\bar{p}-\bar{q}|}^{\bar{p}+\bar{q}} \mathrm{d}u\, \frac{1}{1 + u^\kappa}\, \delta(u + C) = \frac{1}{1 + |C|^\kappa}\,,
\end{equation}
whenever $-C \in [|\bar{p}-\bar{q}|, \bar{p}+\bar{q}]$, and zero otherwise. Evaluating the four delta-function roots at $u_0 = \omega - \bar{q}$, $\omega + \bar{q}$, $\bar{q} - \omega$, and $-\omega - \bar{q}$, the kernel takes the form
\begin{align}
\Gamma_{\Pi^F}(\omega, \bar{p}, \bar{q}) = \pi\Big[&\, f_S(\omega - \bar{q})\,\Theta(\omega - \bar{q} - |\bar{p}-\bar{q}|)\,\Theta(\bar{p} + \bar{q} - \omega + \bar{q}) \nonumber\\
+&\, f_S(\omega + \bar{q})\,\Theta(\omega + \bar{q} - |\bar{p}-\bar{q}|)\,\Theta(\bar{p} + \bar{q} - \omega - \bar{q}) \nonumber\\
+&\, f_S(\bar{q} - \omega)\,\Theta(\bar{q} - \omega - |\bar{p}-\bar{q}|)\,\Theta(\bar{p} + \bar{q} - \bar{q} + \omega) \nonumber\\
+&\, f_S(-\omega - \bar{q})\,\Theta(-\omega - \bar{q} - |\bar{p}-\bar{q}|)\,\Theta(\bar{p} + \bar{q} + \omega + \bar{q})\Big],
\label{eq:GammaPiF}
\end{align}
which, using the shorthand $E_\pm = \omega \pm \bar{q}$, $P = \bar{p} + \bar{q}$, and
$\Delta = |\bar{p} - \bar{q}|$, can be written more compactly as in the main text.

The structure of the Heaviside constraints in \eqref{eq:GammaPiF} mirrors that encountered for $\mathrm{Im}\,\Gamma_{\Pi}$ in
App.~\ref{app:effective_vertex}.
The crucial difference is the sign pattern: whereas $\mathrm{Im}\,\Gamma_{\Pi}$ entered with relative signs $(+,+,-,-)$, all four terms in $\Gamma_{\Pi^F}$ enter with a positive sign.
Repeating the term-by-term analysis at $\omega=\bar{p}$ in the massless case from above and below yields, in contrast to the $\Pi^R$ result, a genuinely discontinuous answer: the boundary contributions of terms~2 and~3 no longer cancel against each other, and the two one-sided limits disagree.
The physical on-shell condition $\omega^2 - \bar{p}^2 = M^2 \ge 0$
selects the timelike side $\omega \to \bar{p}^+$, on which only term~1
contributes, giving
\begin{equation}
    \Pi^F_S(\bar{p},\bar{p})\Big|_{M=0}
    = \frac{\lambda}{48 \pi \bar{p}}\int_0^{\bar{p}} \mathrm{d}\bar{q}\, f_S(\bar{q})\,f_S(\bar{p}-\bar{q}).
    \label{eq:PiFmassless}
\end{equation}
For finite mass, in the limit $\omega = \sqrt{\bar{p}^2 + M^2}$ the Heaviside constraints simplify considerably, and only the first term in \eqref{eq:GammaPiF} contributes: its second condition, $\Theta(\bar{p}+2\bar{q}-\omega)$, gives the exact lower bound $\bar{q}\ge q_c\equiv(\omega-\bar{p})/2=M^2/[2(\omega+\bar{p})]$ for any $\bar{q}$; its first condition, $\Theta(\omega-\bar{q}-|\bar{p}-\bar{q}|)$, is automatically satisfied for $\bar{q}<\bar{p}$ (since $\omega\ge\bar{p}$ on shell), but for $\bar{q}>\bar{p}$ reduces to $\bar{q}\le(\omega+\bar{p})/2=\bar{p}+q_c$, giving the exact upper bound. Together, these give the exact integration range $q_c < \bar{q} < \bar{p}+q_c$.
Term~2 requires $P-E_+\ge0$, i.e.~$\bar{p}\ge\omega$, which is never satisfied on shell. Term~3 similarly reduces, on either side of $\bar{q}=\bar{p}$, to a condition equivalent to $\omega\le\bar{p}$, again never satisfied.
Term~4 vanishes identically for $\omega, \bar{q} > 0$. This yields the exact on-shell expression \eqref{eq:piFS_onshell}.
The limit $M\to 0$ of this expression is smooth: as $M\to0$, $q_c\to0$ and $\omega\to\bar{p}$, so the window $[q_c,\bar{p}+q_c]$ shrinks continuously to $[0,\bar{p}]$ and the integrand $f_S(\bar{q})f_S(\omega-\bar{q})$ relaxes continuously to $f_S(\bar{q})f_S(\bar{p}-\bar{q})$, both bounded throughout the integration range, with no logarithms or inverse powers of $M$ appearing anywhere in the process.
The massive on-shell expression therefore reduces smoothly to \eqref{eq:PiFmassless} as $M\to 0$, in contrast to $\mathrm{Re}\,\Pi^R_S$, where the mass cures a genuine logarithmic infrared divergence and the $M\to 0$ limit is nonuniform.

For the limiting behaviour, we note that in the regime $\bar{p} \gg k_\mathrm{IR}$, the integral in \eqref{eq:PiFmassless} is dominated by $\bar{q} \sim k_\mathrm{IR}$, where $f_S(\bar{q}) \approx f_S(0)$ is approximately constant, while $f_S(\bar{p} - \bar{q}) \approx f_S(\bar{p}) \sim \bar{p}^{-\kappa}$. This gives
\begin{equation}
\Pi^F_S(\bar{p}, \bar{p})\Big|_{\bar{p} \gg k_\mathrm{IR}} \sim \frac{k_\mathrm{IR}\, f_S(k_\mathrm{IR})}{\bar{p}^{1+\kappa}}\,.
\label{eq:piFS_highp}
\end{equation}
In the opposite limit $\bar{p} \ll k_\mathrm{IR}$, both $f_S(\bar{q})$ and $f_S(\bar{p} - \bar{q})$ stay close to the plateau value $f_S(0)$ throughout the integration range, so that
\begin{equation}
\Pi^F_S(\bar{p}, \bar{p})\Big|_{\bar{p} \ll k_\mathrm{IR}} \to \frac{\lambda}{48\pi}f_S^2(0)\,,
\end{equation}
which is independent of $\bar{p}$, consistent with the flat region observed numerically and with the exact limit~\eqref{eq:feffnum}.

\section{Self-energy computations}
\label{app:fpdetail}
At next-to-leading order in $1/N$, the nonlocal part of the spectral self-energy on the Keldysh contour is given by 
\begin{equation}
\Sigma^\rho(x, y)=-\frac{\lambda}{3 N}\left[F(x, y) I^\rho(x, y)+\rho(x, y) I^F(x, y)\right].
\end{equation}
Using a leading-order gradient expansion in time, the summation functions $I^F$ and $I^{\rho}$ can be majorly simplified in momentum space, as discussed in Refs.~\cite{Berges:2015kfa, Chantesana:2018qsb}.
For the spectral self-energy in Wigner space, this leads to the expression 
\begin{equation}
\Sigma^\rho(\omega, \mathbf{p})=-\frac{\lambda}{3N} \int_{\nu,\mathbf{q}} \left[F(t,\omega-\nu, \mathbf{p}-\mathbf{q}) \Pi^\rho(t,\nu, \mathbf{q}) +\rho(\omega-\nu, \mathbf{p}-\mathbf{q}) \Pi^F(t,\nu, \mathbf{q})\right] v_{\mathrm{eff}}(t,\nu, \mathbf{q}),
\label{eq:selfrho}
\end{equation}
where we did not put a time-dependence for the self energy, as the overall time dependence on the right hand side of the equation eventually drops out. 
The next step is to evaluate the frequency integrals, which is done by the quasiparticle approximation of Sec.~\ref{sec:quasiparticle}.
First, $\Pi^{\rho}$ can then be written as 
\begin{align}
&\Pi^\rho(t,\nu, \mathbf{q}) =\frac{\lambda}{3} \int_{\mu, \mathbf{r}} F(t,\nu-\mu, \mathbf{q}-\mathbf{r}) \rho(\mu, \mathbf{r})\nonumber\\
&=
\frac{\lambda}{3} \int_{\mu,\mathbf{r}}  f(t,\nu-\mu, \mathbf{q}-\mathbf{r})
\frac{2\pi}{2 \omega_{\mathbf{q}-\mathbf{r}}}\left[\delta\left(\nu-\mu-\omega_{\mathbf{q}-\mathbf{r}}\right)-\delta\left(\nu-\mu+\omega_{\mathbf{q}-\mathbf{r}}\right)\right] \frac{2 \pi i}{2 \omega_{\mathbf{r}}}
\left[\delta\left(\mu-\omega_{\mathbf{r}}\right)-\delta\left(\mu+\omega_{\mathbf{r}}\right)\right]\nonumber\\
&=
\frac{i\pi\lambda}{6}\int_{\mathbf{r}} \frac{f(t,\mathbf{q}-\mathbf{r})}{\omega_{\mathbf{q}-\mathbf{r}} \omega_{\mathbf{r}}}  \left[ 
\delta\left(\nu-\omega_{\mathbf{r}}+\omega_{\mathbf{q}-\mathbf{r}}\right)
+ \delta\left(\nu-\omega_{\mathbf{r}}-\omega_{\mathbf{q}-\mathbf{r}}\right)  - \delta\left(\nu+\omega_{\mathbf{r}}+\omega_{\mathbf{q}-\mathbf{r}}\right)
- \delta\left(\nu+\omega_{\mathbf{r}}-\omega_{\mathbf{q}-\mathbf{r}}\right) \right].
\label{eq:sigmarhoapp}
\end{align}
In a similar manner, $\Pi^F$ can also be computed as 
\begin{align}
&\Pi^F(t,\nu, \mathbf{q})
=
\frac{\lambda}{6}\int_{\mu, \mathbf{r}} F(t,\nu-\mu, \mathbf{q}-\mathbf{r}) F(t,\mu, \mathbf{r}) \nonumber\\
&=
\frac{\lambda}{6}\int_{\mu,\mathbf{r}} f(t,\nu-\mu, \mathbf{q}-\mathbf{r}) 
\frac{2 \pi}{2\omega_{\mathbf{q}-\mathbf{r}}}\left[ 
\delta\left(\nu-\mu-\omega_{\mathbf{q}-\mathbf{r}}\right) - \delta\left(\nu-\mu+\omega_{\mathbf{q}-\mathbf{r}}\right)\right] f(t,\mu, \mathbf{r})  \frac{2 \pi}{2\omega_{\mathbf{r}}}
\left[\delta\left(\mu-\omega_{\mathbf{r}}\right)-\delta\left(\mu+\omega_{\mathbf{r}}\right)\right]\nonumber\\
&=
\frac{\pi\lambda}{12}\int_{\mathbf{r}} \frac{f(t,\mathbf{q}-\mathbf{r}) f(t,\mathbf{r})}{\omega_{\mathbf{q}-\mathbf{r}} \omega_{\mathbf{r}}} \left[ 
\delta\left(\nu-\omega_{\mathbf{r}}+\omega_{\mathbf{q}-\mathbf{r}}\right) 
+ \delta\left(\nu-\omega_{\mathbf{r}}-\omega_{\mathbf{q}-\mathbf{r}}\right) + \delta\left(\nu+\omega_{\mathbf{r}}-\omega_{\mathbf{q}-\mathbf{r}}\right) 
+ \delta\left(\nu+\omega_{\mathbf{r}}+\omega_{\mathbf{q}-\mathbf{r}}\right) \right].
\label{eq:sigmafapp}
\end{align}
When explicitly writing out the self-energy \eqref{eq:selfrho}, for the first term, we get
\begin{align}
\Sigma^{\rho}_1&(\omega,\pp)
=
\int_{\nu,\mathbf{q}} F(t,\omega-\nu, \mathbf{p}-\mathbf{q})  \Pi^\rho(t,\nu, \mathbf{q}) v_{\mathrm{eff}}(t,\nu, \mathbf{q})\nonumber\\
&=
-\frac{\lambda}{3N}\int_{\nu,\mathbf{q}} \frac{2 \pi f(t,\omega-\nu, \mathbf{p}-\mathbf{q})}{2 \omega_{\mathbf{p}-\mathbf{q}}}\Pi^\rho(t,\nu, \mathbf{q}) v_{\mathrm{eff}}(t,\nu, \mathbf{q})\left[\delta\left(\omega-\nu-\omega_{\mathbf{p}-\mathbf{q}}\right)-\delta\left(\omega-\nu+\omega_{\mathbf{p}-\mathbf{q}}\right)\right]\nonumber\\
&=
-\frac{\lambda}{6N}\int_{\mathbf{q}} \frac{f(t,\mathbf{p}-\mathbf{q})}{\omega_{\mathbf{p}-\mathbf{q}}} \left[ v_{\mathrm{eff}}\left(t, \omega-\omega_{\mathbf{p}-\mathbf{q}}, \mathbf{q}\right) \Pi^\rho\left(t,\omega-\omega_{\mathbf{p}-\mathbf{q}}, \mathbf{q}\right) + v_{\mathrm{eff}}\left(t, \omega+\omega_{\mathbf{p}-\mathbf{q}}, \mathbf{q}\right) \Pi^\rho\left(t,\omega+\omega_{\mathbf{p}-\mathbf{q}}, \mathbf{q}\right) \right],
\end{align}
and for the second term, it is
\begin{align}
\Sigma^{\rho}_2(\omega,\pp)
&=
\int_{\nu,\mathbf{q}} \rho(w-\nu,\mathbf{p}-\mathbf{q}) \Pi^F(t,\nu, \mathbf{q}) v_{\mathrm{eff}}(t,\nu, \mathbf{q})\nonumber\\
&=
-\frac{\lambda}{3N}\int_{\nu,\qq} \frac{2\pi i}{2\omega_{\pp-\qq}} \Pi^F(t,\nu, \mathbf{q})v_{\mathrm{eff}}(t,\nu, \mathbf{q}) \left[\delta(\omega-\nu-\omega_{\pp-\qq}) - \delta(\omega-\nu+\omega_{\pp-\qq})\right] \nonumber\\
&= 
-\frac{i\lambda}{6N}\int_{\qq} \frac{1}{\omega_{\pp-\qq}} \left[ \Pi^F(t,\omega-\omega_{\pp-\qq},\qq) v_{\mathrm{eff}}(t,\omega-\omega_{\pp-\qq},\qq) - \Pi^F(t,\omega+\omega_{\pp-\qq},\qq) v_{\mathrm{eff}}(t,\omega+\omega_{\pp-\qq},\qq) \right].
\end{align}
Here, we have used $f(t,-\omega_{\mathbf{p}},\mathbf{p}) = -[f(t,\omega_{\mathbf{p}},\mathbf{p}) + 1] \approx -f(t,\omega_{\mathbf{p}},\mathbf{p}) \equiv f(t,\mathbf{p})$, together with the spatial isotropy assumption. Using 
\begin{equation}
f(t, \mathbf{p})=t^\alpha f_S\left(t^\beta|\mathbf{p}|\right)
\equiv t^{\alpha}\frac{A}{B+\left(t^\beta|\mathbf{p}|\right)^\kappa},
\label{eq:distlong}
\end{equation}
we see that the overall time dependence from the $t^{\alpha}$ terms drops out in the self-energy.
To consider the relationship between the retarded and spectral self-energies, we recall
\begin{equation}
\Sigma^R(\omega, \mathbf{p})=\lim _{\epsilon \rightarrow 0} \int \frac{\mathrm{~d} \omega^{\prime}}{2 \pi i} \frac{\Sigma^\rho(\omega', \mathbf{p})}{\omega'-\omega-i\epsilon} = \mathcal{PV}\int \frac{\mathrm{d}\omega'}{2\pi i}\frac{\Sigma^{\rho}(\omega',\pp)}{\omega'-\omega} + \frac{1}{2}\Sigma^{\rho}(\omega,\pp).
\end{equation}
Since $\Sigma^{\rho}(t,t',\mathbf{p})$ is an odd function of $t-t'$, its Fourier transform $\Sigma^{\rho}(\omega,\mathbf{p})$ is purely imaginary, so that the $\Sigma^{\rho}(\omega,\pp)/2$ term in the above decomposition is itself purely imaginary. Writing it as $\Sigma^{\rho}(\omega,\pp)/2 = i\,\mathrm{Im}[\Sigma^R(\omega,\pp)]$ and using the real combination $\Sigma^{\tilde{\rho}}(\omega,\pp)\equiv-i\big(\Sigma^R(\omega,\pp)-\Sigma^A(\omega,\pp)\big) = -i\Sigma^{\rho}(\omega,\pp)$ introduced in~\eqref{eq:sigmatilderhodef}, 
this becomes $\mathrm{Im}[\Sigma^R(\omega,\pp)]=\Sigma^{\tilde{\rho}}(\omega,\pp)/2$. Therefore, the real part of $\Sigma^R(\omega, \mathbf{p})$ is given by the principal value integral, while its imaginary part is given by the $\Sigma^{\rho}(\omega,\pp)/2$ term.
\begin{align}
\mathrm{Im}\left[\Sigma^R(\omega,\pp) \right] =  \frac{\lambda}{12N} & \int_{\qq} \left\{ \frac{ v_{\mathrm{eff}}\left(t, \omega-\omega_{\pp-\mathbf{q}}, \mathbf{q}\right)}{\omega_{\pp-\mathbf{q}}}  \left [if(t, \pp-\mathbf{q})\,\Pi^{\rho}\left(t, \omega-\omega_{\pp-\mathbf{q}},\mathbf{q}\right) - \Pi^F(t,\omega-\omega_{\pp-\qq}, \qq)\right] \right.\nonumber\\
&\ \ + 
\left.\frac{ v_{\mathrm{eff}}\left(t, \omega+\omega_{\pp-\mathbf{q}}, \mathbf{q}\right)}{\omega_{\pp-\mathbf{q}}}  \left [ i f(t, \pp-\mathbf{q})\,\Pi^{\rho} \left(t, \omega+\omega_{\pp-\mathbf{q}},\mathbf{q}\right) + \Pi^F(t,\omega+\omega_{\pp-\qq}, \qq)\right] \right\}.
\end{align}
For the real part of the retarded self-energy, the principal value integral gives 
\begin{align}
\mathrm{Re}&\!\left[\Sigma^R(\omega,\pp)\right]
= 
\int \frac{\mathrm{d}\omega'}{2\pi i}
\frac{\Sigma^{\rho}(\omega', \pp)}{\omega'-\omega}\nonumber\\
&= 
\frac{\lambda}{6N}
\int_{\omega',\qq}
\left\lbrace \frac{i f(t,\pp-\qq)}{(\omega'-\omega)\,\omega_{\pp-\qq}}
          \left[
          v_{\mathrm{eff}}\!\left(t, \omega'-\omega_{\pp-\qq},\qq\right)
          \Pi^{\rho}(\omega'-\omega_{\pp-\qq},\qq) 
        + v_{\mathrm{eff}}\!\left(t, \omega'+\omega_{\pp-\qq},\qq\right)
          \Pi^{\rho}(\omega'+\omega_{\pp-\qq},\qq)
          \right]\right.\nonumber\\
    &-
    \left.\frac{1}{\left(\omega'-\omega\right)\omega_{\pp-\qq}}
    \left[v_{\mathrm{eff}}\!\left(t, \omega'-\omega_{\pp-\qq},\qq\right)\Pi^F(\omega'-\omega_{\pp-\qq},\qq) - v_{\mathrm{eff}}\!\left(t, \omega'+\omega_{\pp-\qq},\qq\right)\Pi^F(\omega'+\omega_{\pp-\qq},\qq)\right]
\right\rbrace.
\end{align}

\twocolumngrid
\bibliography{main}

\begin{thebibliography}{66}%
\makeatletter
\providecommand \@ifxundefined [1]{%
 \@ifx{#1\undefined}
}%
\providecommand \@ifnum [1]{%
 \ifnum #1\expandafter \@firstoftwo
 \else \expandafter \@secondoftwo
 \fi
}%
\providecommand \@ifx [1]{%
 \ifx #1\expandafter \@firstoftwo
 \else \expandafter \@secondoftwo
 \fi
}%
\providecommand \natexlab [1]{#1}%
\providecommand \enquote  [1]{``#1''}%
\providecommand \bibnamefont  [1]{#1}%
\providecommand \bibfnamefont [1]{#1}%
\providecommand \citenamefont [1]{#1}%
\providecommand \href@noop [0]{\@secondoftwo}%
\providecommand \href [0]{\begingroup \@sanitize@url \@href}%
\providecommand \@href[1]{\@@startlink{#1}\@@href}%
\providecommand \@@href[1]{\endgroup#1\@@endlink}%
\providecommand \@sanitize@url [0]{\catcode `\\12\catcode `\$12\catcode
  `\&12\catcode `\#12\catcode `\^12\catcode `\_12\catcode `\%12\relax}%
\providecommand \@@startlink[1]{}%
\providecommand \@@endlink[0]{}%
\providecommand \url  [0]{\begingroup\@sanitize@url \@url }%
\providecommand \@url [1]{\endgroup\@href {#1}{\urlprefix }}%
\providecommand \urlprefix  [0]{URL }%
\providecommand \Eprint [0]{\href }%
\providecommand \doibase [0]{https://doi.org/}%
\providecommand \selectlanguage [0]{\@gobble}%
\providecommand \bibinfo  [0]{\@secondoftwo}%
\providecommand \bibfield  [0]{\@secondoftwo}%
\providecommand \translation [1]{[#1]}%
\providecommand \BibitemOpen [0]{}%
\providecommand \bibitemStop [0]{}%
\providecommand \bibitemNoStop [0]{.\EOS\space}%
\providecommand \EOS [0]{\spacefactor3000\relax}%
\providecommand \BibitemShut  [1]{\csname bibitem#1\endcsname}%
\let\auto@bib@innerbib\@empty
\bibitem [{\citenamefont
  {Wilson}(1971{\natexlab{a}})}]{wilson1971renormalization}%
  \BibitemOpen
  \bibfield  {author} {\bibinfo {author} {\bibfnamefont {K.~G.}\ \bibnamefont
  {Wilson}},\ }\bibfield  {title} {\bibinfo {title} {{Renormalization group and
  critical phenomena. I. Renormalization group and the Kadanoff scaling
  picture}},\ }\href {https://doi.org/10.1103/PhysRevB.4.3174} {\bibfield
  {journal} {\bibinfo  {journal} {Phys. Rev. B}\ }\textbf {\bibinfo {volume}
  {4}},\ \bibinfo {pages} {3174} (\bibinfo {year}
  {1971}{\natexlab{a}})}\BibitemShut {NoStop}%
\bibitem [{\citenamefont
  {Wilson}(1971{\natexlab{b}})}]{wilson1971renormalization2}%
  \BibitemOpen
  \bibfield  {author} {\bibinfo {author} {\bibfnamefont {K.~G.}\ \bibnamefont
  {Wilson}},\ }\bibfield  {title} {\bibinfo {title} {{Renormalization group and
  critical phenomena. II. Phase-space cell analysis of critical behavior}},\
  }\href {https://doi.org/10.1103/PhysRevB.4.3184} {\bibfield  {journal}
  {\bibinfo  {journal} {Phys. Rev. B}\ }\textbf {\bibinfo {volume} {4}},\
  \bibinfo {pages} {3184} (\bibinfo {year} {1971}{\natexlab{b}})}\BibitemShut
  {NoStop}%
\bibitem [{\citenamefont {Hohenberg}\ and\ \citenamefont
  {Halperin}(1977)}]{Hohenberg:1977ym}%
  \BibitemOpen
  \bibfield  {author} {\bibinfo {author} {\bibfnamefont {P.~C.}\ \bibnamefont
  {Hohenberg}}\ and\ \bibinfo {author} {\bibfnamefont {B.~I.}\ \bibnamefont
  {Halperin}},\ }\bibfield  {title} {\bibinfo {title} {{Theory of Dynamic
  Critical Phenomena}},\ }\href {https://doi.org/10.1103/RevModPhys.49.435}
  {\bibfield  {journal} {\bibinfo  {journal} {Rev. Mod. Phys.}\ }\textbf
  {\bibinfo {volume} {49}},\ \bibinfo {pages} {435} (\bibinfo {year}
  {1977})}\BibitemShut {NoStop}%
\bibitem [{\citenamefont {Frisch}(1995)}]{Frisch1995a}%
  \BibitemOpen
  \bibfield  {author} {\bibinfo {author} {\bibfnamefont {U.}~\bibnamefont
  {Frisch}},\ }\href {https://doi.org/10.1017/CBO9781139170666} {\emph
  {\bibinfo {title} {Turbulence: The Legacy of A. N. Kolmogorov}}}\ (\bibinfo
  {publisher} {CUP, Cambridge, UK},\ \bibinfo {year} {1995})\BibitemShut
  {NoStop}%
\bibitem [{\citenamefont {Zakharov}\ \emph {et~al.}(1992)\citenamefont
  {Zakharov}, \citenamefont {{L'vov}},\ and\ \citenamefont
  {Falkovich}}]{Zakharov1992a}%
  \BibitemOpen
  \bibfield  {author} {\bibinfo {author} {\bibfnamefont {V.~E.}\ \bibnamefont
  {Zakharov}}, \bibinfo {author} {\bibfnamefont {V.~S.}\ \bibnamefont
  {{L'vov}}},\ and\ \bibinfo {author} {\bibfnamefont {G.}~\bibnamefont
  {Falkovich}},\ }\href {https://doi.org/10.1007/978-3-642-50052-7} {\emph
  {\bibinfo {title} {Kolmogorov Spectra of Turbulence I: Wave Turbulence}}}\
  (\bibinfo  {publisher} {Springer, Berlin},\ \bibinfo {year}
  {1992})\BibitemShut {NoStop}%
\bibitem [{\citenamefont {Bray}(1994)}]{Bray:1994zz}%
  \BibitemOpen
  \bibfield  {author} {\bibinfo {author} {\bibfnamefont {A.~J.}\ \bibnamefont
  {Bray}},\ }\bibfield  {title} {\bibinfo {title} {{Theory of phase-ordering
  kinetics}},\ }\href {https://doi.org/10.1080/00018739400101505} {\bibfield
  {journal} {\bibinfo  {journal} {Adv. Phys.}\ }\textbf {\bibinfo {volume}
  {43}},\ \bibinfo {pages} {357} (\bibinfo {year} {1994})},\ \Eprint
  {https://arxiv.org/abs/cond-mat/9501089} {arXiv:cond-mat/9501089}
  \BibitemShut {NoStop}%
\bibitem [{\citenamefont {Calabrese}\ and\ \citenamefont
  {Gambassi}(2005)}]{Calabrese_2005}%
  \BibitemOpen
  \bibfield  {author} {\bibinfo {author} {\bibfnamefont {P.}~\bibnamefont
  {Calabrese}}\ and\ \bibinfo {author} {\bibfnamefont {A.}~\bibnamefont
  {Gambassi}},\ }\bibfield  {title} {\bibinfo {title} {Ageing properties of
  critical systems},\ }\href {https://doi.org/10.1088/0305-4470/38/18/r01}
  {\bibfield  {journal} {\bibinfo  {journal} {J. Phys. A}\ }\textbf {\bibinfo
  {volume} {38}},\ \bibinfo {pages} {R133–R193} (\bibinfo {year}
  {2005})}\BibitemShut {NoStop}%
\bibitem [{\citenamefont {Chiocchetta}\ \emph {et~al.}(2015)\citenamefont
  {Chiocchetta}, \citenamefont {Tavora}, \citenamefont {Gambassi},\ and\
  \citenamefont {Mitra}}]{gambassi2015}%
  \BibitemOpen
  \bibfield  {author} {\bibinfo {author} {\bibfnamefont {A.}~\bibnamefont
  {Chiocchetta}}, \bibinfo {author} {\bibfnamefont {M.}~\bibnamefont {Tavora}},
  \bibinfo {author} {\bibfnamefont {A.}~\bibnamefont {Gambassi}},\ and\
  \bibinfo {author} {\bibfnamefont {A.}~\bibnamefont {Mitra}},\ }\bibfield
  {title} {\bibinfo {title} {Short-time universal scaling in an isolated
  quantum system after a quench},\ }\href
  {https://doi.org/10.1103/PhysRevB.91.220302} {\bibfield  {journal} {\bibinfo
  {journal} {Phys. Rev. B}\ }\textbf {\bibinfo {volume} {91}},\ \bibinfo
  {pages} {220302} (\bibinfo {year} {2015})}\BibitemShut {NoStop}%
\bibitem [{\citenamefont {Heyl}(2018)}]{Heyl:2017blm}%
  \BibitemOpen
  \bibfield  {author} {\bibinfo {author} {\bibfnamefont {M.}~\bibnamefont
  {Heyl}},\ }\bibfield  {title} {\bibinfo {title} {{Dynamical quantum phase
  transitions: a review}},\ }\href {https://doi.org/10.1088/1361-6633/aaaf9a}
  {\bibfield  {journal} {\bibinfo  {journal} {Rept. Prog. Phys.}\ }\textbf
  {\bibinfo {volume} {81}},\ \bibinfo {pages} {054001} (\bibinfo {year}
  {2018})},\ \Eprint {https://arxiv.org/abs/1709.07461} {arXiv:1709.07461
  [cond-mat.stat-mech]} \BibitemShut {NoStop}%
\bibitem [{\citenamefont {Berges}\ \emph {et~al.}(2008)\citenamefont {Berges},
  \citenamefont {Rothkopf},\ and\ \citenamefont {Schmidt}}]{Berges:2008wm}%
  \BibitemOpen
  \bibfield  {author} {\bibinfo {author} {\bibfnamefont {J.}~\bibnamefont
  {Berges}}, \bibinfo {author} {\bibfnamefont {A.}~\bibnamefont {Rothkopf}},\
  and\ \bibinfo {author} {\bibfnamefont {J.}~\bibnamefont {Schmidt}},\
  }\bibfield  {title} {\bibinfo {title} {{Non-thermal fixed points: Effective
  weak-coupling for strongly correlated systems far from equilibrium}},\ }\href
  {https://doi.org/10.1103/PhysRevLett.101.041603} {\bibfield  {journal}
  {\bibinfo  {journal} {Phys. Rev. Lett.}\ }\textbf {\bibinfo {volume} {101}},\
  \bibinfo {pages} {041603} (\bibinfo {year} {2008})},\ \Eprint
  {https://arxiv.org/abs/0803.0131} {arXiv:0803.0131 [hep-ph]} \BibitemShut
  {NoStop}%
\bibitem [{\citenamefont {Berges}\ and\ \citenamefont
  {Hoffmeister}(2009)}]{Berges:2008sr}%
  \BibitemOpen
  \bibfield  {author} {\bibinfo {author} {\bibfnamefont {J.}~\bibnamefont
  {Berges}}\ and\ \bibinfo {author} {\bibfnamefont {G.}~\bibnamefont
  {Hoffmeister}},\ }\bibfield  {title} {\bibinfo {title} {{Nonthermal fixed
  points and the functional renormalization group}},\ }\href
  {https://doi.org/10.1016/j.nuclphysb.2008.12.017} {\bibfield  {journal}
  {\bibinfo  {journal} {Nucl. Phys. B}\ }\textbf {\bibinfo {volume} {813}},\
  \bibinfo {pages} {383} (\bibinfo {year} {2009})},\ \Eprint
  {https://arxiv.org/abs/0809.5208} {arXiv:0809.5208 [hep-th]} \BibitemShut
  {NoStop}%
\bibitem [{\citenamefont {Schmied}\ \emph
  {et~al.}(2019{\natexlab{a}})\citenamefont {Schmied}, \citenamefont
  {Mikheev},\ and\ \citenamefont {Gasenzer}}]{Schmied:2018upn}%
  \BibitemOpen
  \bibfield  {author} {\bibinfo {author} {\bibfnamefont {C.-M.}\ \bibnamefont
  {Schmied}}, \bibinfo {author} {\bibfnamefont {A.~N.}\ \bibnamefont
  {Mikheev}},\ and\ \bibinfo {author} {\bibfnamefont {T.}~\bibnamefont
  {Gasenzer}},\ }\bibfield  {title} {\bibinfo {title} {{Prescaling in a
  far-from-equilibrium Bose gas}},\ }\href
  {https://doi.org/10.1103/PhysRevLett.122.170404} {\bibfield  {journal}
  {\bibinfo  {journal} {Phys. Rev. Lett.}\ }\textbf {\bibinfo {volume} {122}},\
  \bibinfo {pages} {170404} (\bibinfo {year} {2019}{\natexlab{a}})},\ \Eprint
  {https://arxiv.org/abs/1807.07514} {arXiv:1807.07514} \BibitemShut {NoStop}%
\bibitem [{\citenamefont {Schmied}\ \emph
  {et~al.}(2019{\natexlab{b}})\citenamefont {Schmied}, \citenamefont
  {Mikheev},\ and\ \citenamefont {Gasenzer}}]{Schmied:2018mte}%
  \BibitemOpen
  \bibfield  {author} {\bibinfo {author} {\bibfnamefont {C.-M.}\ \bibnamefont
  {Schmied}}, \bibinfo {author} {\bibfnamefont {A.~N.}\ \bibnamefont
  {Mikheev}},\ and\ \bibinfo {author} {\bibfnamefont {T.}~\bibnamefont
  {Gasenzer}},\ }\bibfield  {title} {\bibinfo {title} {{Non-thermal fixed
  points: Universal dynamics far from equilibrium}},\ }\href
  {https://doi.org/10.1142/S0217751X19410069} {\bibfield  {journal} {\bibinfo
  {journal} {Int. J. Mod. Phys. A}\ }\textbf {\bibinfo {volume} {34}},\
  \bibinfo {pages} {1941006} (\bibinfo {year} {2019}{\natexlab{b}})},\ \Eprint
  {https://arxiv.org/abs/1810.08143} {arXiv:1810.08143 [cond-mat.quant-gas]}
  \BibitemShut {NoStop}%
\bibitem [{\citenamefont {Nowak}\ \emph {et~al.}(2016)\citenamefont {Nowak},
  \citenamefont {Erne}, \citenamefont {Karl}, \citenamefont {Schole},
  \citenamefont {Sexty},\ and\ \citenamefont {Gasenzer}}]{Nowak:2013juc}%
  \BibitemOpen
  \bibfield  {author} {\bibinfo {author} {\bibfnamefont {B.}~\bibnamefont
  {Nowak}}, \bibinfo {author} {\bibfnamefont {S.}~\bibnamefont {Erne}},
  \bibinfo {author} {\bibfnamefont {M.}~\bibnamefont {Karl}}, \bibinfo {author}
  {\bibfnamefont {J.}~\bibnamefont {Schole}}, \bibinfo {author} {\bibfnamefont
  {D.}~\bibnamefont {Sexty}},\ and\ \bibinfo {author} {\bibfnamefont
  {T.}~\bibnamefont {Gasenzer}},\ }\bibfield  {title} {\bibinfo {title}
  {Nonthermal fixed points: universality, topology, and turbulence in bose
  gases},\ }in\ \href
  {https://doi.org/10.1093/acprof:oso/9780198768166.003.0007} {\emph {\bibinfo
  {booktitle} {Strongly Interacting Quantum Systems out of Equilibrium: Lecture
  Notes of the Les Houches Summer School: Volume 99, August 2012}}},\ \bibinfo
  {editor} {edited by\ \bibinfo {editor} {\bibfnamefont {T.}~\bibnamefont
  {Giamarchi}}, \bibinfo {editor} {\bibfnamefont {A.~J.}\ \bibnamefont
  {Millis}}, \bibinfo {editor} {\bibfnamefont {O.}~\bibnamefont {Parcollet}},
  \bibinfo {editor} {\bibfnamefont {H.}~\bibnamefont {Saleur}}, \bibinfo
  {editor} {\bibfnamefont {L.~F.}\ \bibnamefont {Cugliandolo}}, \bibinfo
  {editor} {\bibfnamefont {T.}~\bibnamefont {Giamarchi}}, \bibinfo {editor}
  {\bibfnamefont {A.~J.}\ \bibnamefont {Millis}}, \bibinfo {editor}
  {\bibfnamefont {O.}~\bibnamefont {Parcollet}}, \bibinfo {editor}
  {\bibfnamefont {H.}~\bibnamefont {Saleur}},\ and\ \bibinfo {editor}
  {\bibfnamefont {L.~F.}\ \bibnamefont {Cugliandolo}}}\ (\bibinfo  {publisher}
  {Oxford University Press},\ \bibinfo {year} {2016})\ \Eprint
  {https://arxiv.org/abs/1302.1448} {arXiv:1302.1448 [cond-mat.quant-gas]}
  \BibitemShut {NoStop}%
\bibitem [{\citenamefont {Berges}(2016)}]{Berges:2015kfa}%
  \BibitemOpen
  \bibfield  {author} {\bibinfo {author} {\bibfnamefont {J.}~\bibnamefont
  {Berges}},\ }\bibfield  {title} {\bibinfo {title} {Nonequilibrium quantum
  fields: from cold atoms to cosmology},\ }in\ \href
  {https://doi.org/10.1093/acprof:oso/9780198768166.003.0002} {\emph {\bibinfo
  {booktitle} {Strongly Interacting Quantum Systems out of Equilibrium: Lecture
  Notes of the Les Houches Summer School: Volume 99, August 2012}}},\ \bibinfo
  {editor} {edited by\ \bibinfo {editor} {\bibfnamefont {T.}~\bibnamefont
  {Giamarchi}}, \bibinfo {editor} {\bibfnamefont {A.~J.}\ \bibnamefont
  {Millis}}, \bibinfo {editor} {\bibfnamefont {O.}~\bibnamefont {Parcollet}},
  \bibinfo {editor} {\bibfnamefont {H.}~\bibnamefont {Saleur}}, \bibinfo
  {editor} {\bibfnamefont {L.~F.}\ \bibnamefont {Cugliandolo}}, \bibinfo
  {editor} {\bibfnamefont {T.}~\bibnamefont {Giamarchi}}, \bibinfo {editor}
  {\bibfnamefont {A.~J.}\ \bibnamefont {Millis}}, \bibinfo {editor}
  {\bibfnamefont {O.}~\bibnamefont {Parcollet}}, \bibinfo {editor}
  {\bibfnamefont {H.}~\bibnamefont {Saleur}},\ and\ \bibinfo {editor}
  {\bibfnamefont {L.~F.}\ \bibnamefont {Cugliandolo}}}\ (\bibinfo  {publisher}
  {Oxford University Press},\ \bibinfo {year} {2016})\ \Eprint
  {https://arxiv.org/abs/1503.02907} {arXiv:1503.02907 [hep-ph]} \BibitemShut
  {NoStop}%
\bibitem [{\citenamefont {Glidden}\ \emph {et~al.}(2021)\citenamefont
  {Glidden}, \citenamefont {Eigen}, \citenamefont {Dogra}, \citenamefont
  {Hilker}, \citenamefont {Smith},\ and\ \citenamefont
  {Hadzibabic}}]{Glidden:2020qmu}%
  \BibitemOpen
  \bibfield  {author} {\bibinfo {author} {\bibfnamefont {J.~A.~P.}\
  \bibnamefont {Glidden}}, \bibinfo {author} {\bibfnamefont {C.}~\bibnamefont
  {Eigen}}, \bibinfo {author} {\bibfnamefont {L.~H.}\ \bibnamefont {Dogra}},
  \bibinfo {author} {\bibfnamefont {T.~A.}\ \bibnamefont {Hilker}}, \bibinfo
  {author} {\bibfnamefont {R.~P.}\ \bibnamefont {Smith}},\ and\ \bibinfo
  {author} {\bibfnamefont {Z.}~\bibnamefont {Hadzibabic}},\ }\bibfield  {title}
  {\bibinfo {title} {{Bidirectional dynamic scaling in an isolated Bose gas far
  from equilibrium}},\ }\href {https://doi.org/10.1038/s41567-020-01114-x}
  {\bibfield  {journal} {\bibinfo  {journal} {Nature Phys.}\ }\textbf {\bibinfo
  {volume} {17}},\ \bibinfo {pages} {457} (\bibinfo {year} {2021})},\ \Eprint
  {https://arxiv.org/abs/2006.01118} {arXiv:2006.01118 [cond-mat.quant-gas]}
  \BibitemShut {NoStop}%
\bibitem [{\citenamefont {Berges}\ \emph {et~al.}(2015)\citenamefont {Berges},
  \citenamefont {Boguslavski}, \citenamefont {Schlichting},\ and\ \citenamefont
  {Venugopalan}}]{Berges:2014bba}%
  \BibitemOpen
  \bibfield  {author} {\bibinfo {author} {\bibfnamefont {J.}~\bibnamefont
  {Berges}}, \bibinfo {author} {\bibfnamefont {K.}~\bibnamefont {Boguslavski}},
  \bibinfo {author} {\bibfnamefont {S.}~\bibnamefont {Schlichting}},\ and\
  \bibinfo {author} {\bibfnamefont {R.}~\bibnamefont {Venugopalan}},\
  }\bibfield  {title} {\bibinfo {title} {{Universality far from equilibrium:
  From superfluid Bose gases to heavy-ion collisions}},\ }\href
  {https://doi.org/10.1103/PhysRevLett.114.061601} {\bibfield  {journal}
  {\bibinfo  {journal} {Phys. Rev. Lett.}\ }\textbf {\bibinfo {volume} {114}},\
  \bibinfo {pages} {061601} (\bibinfo {year} {2015})},\ \Eprint
  {https://arxiv.org/abs/1408.1670} {arXiv:1408.1670 [hep-ph]} \BibitemShut
  {NoStop}%
\bibitem [{\citenamefont {Mikheev}\ \emph {et~al.}(2023)\citenamefont
  {Mikheev}, \citenamefont {Siovitz},\ and\ \citenamefont
  {Gasenzer}}]{Mikheev:2023juq}%
  \BibitemOpen
  \bibfield  {author} {\bibinfo {author} {\bibfnamefont {A.~N.}\ \bibnamefont
  {Mikheev}}, \bibinfo {author} {\bibfnamefont {I.}~\bibnamefont {Siovitz}},\
  and\ \bibinfo {author} {\bibfnamefont {T.}~\bibnamefont {Gasenzer}},\
  }\bibfield  {title} {\bibinfo {title} {{Universal dynamics and non-thermal
  fixed points in quantum fluids far from equilibrium}},\ }\href
  {https://doi.org/10.1140/epjs/s11734-023-00974-7} {\bibfield  {journal}
  {\bibinfo  {journal} {Eur. Phys. J. ST}\ }\textbf {\bibinfo {volume} {232}},\
  \bibinfo {pages} {3393} (\bibinfo {year} {2023})},\ \Eprint
  {https://arxiv.org/abs/2304.12464} {arXiv:2304.12464 [cond-mat.quant-gas]}
  \BibitemShut {NoStop}%
\bibitem [{\citenamefont {Berges}\ and\ \citenamefont
  {Sexty}(2011)}]{Berges:2010ez}%
  \BibitemOpen
  \bibfield  {author} {\bibinfo {author} {\bibfnamefont {J.}~\bibnamefont
  {Berges}}\ and\ \bibinfo {author} {\bibfnamefont {D.}~\bibnamefont {Sexty}},\
  }\bibfield  {title} {\bibinfo {title} {{Strong versus weak wave-turbulence in
  relativistic field theory}},\ }\href
  {https://doi.org/10.1103/PhysRevD.83.085004} {\bibfield  {journal} {\bibinfo
  {journal} {Phys. Rev. D}\ }\textbf {\bibinfo {volume} {83}},\ \bibinfo
  {pages} {085004} (\bibinfo {year} {2011})},\ \Eprint
  {https://arxiv.org/abs/1012.5944} {arXiv:1012.5944 [hep-ph]} \BibitemShut
  {NoStop}%
\bibitem [{\citenamefont {Scheppach}\ \emph {et~al.}(2010)\citenamefont
  {Scheppach}, \citenamefont {Berges},\ and\ \citenamefont
  {Gasenzer}}]{Scheppach:2009wu}%
  \BibitemOpen
  \bibfield  {author} {\bibinfo {author} {\bibfnamefont {C.}~\bibnamefont
  {Scheppach}}, \bibinfo {author} {\bibfnamefont {J.}~\bibnamefont {Berges}},\
  and\ \bibinfo {author} {\bibfnamefont {T.}~\bibnamefont {Gasenzer}},\
  }\bibfield  {title} {\bibinfo {title} {{Matter Wave Turbulence: Beyond
  Kinetic Scaling}},\ }\href {https://doi.org/10.1103/PhysRevA.81.033611}
  {\bibfield  {journal} {\bibinfo  {journal} {Phys. Rev. A}\ }\textbf {\bibinfo
  {volume} {81}},\ \bibinfo {pages} {033611} (\bibinfo {year} {2010})},\
  \Eprint {https://arxiv.org/abs/0912.4183} {arXiv:0912.4183
  [cond-mat.quant-gas]} \BibitemShut {NoStop}%
\bibitem [{\citenamefont {Chantesana}\ \emph {et~al.}(2019)\citenamefont
  {Chantesana}, \citenamefont {Pi\~neiro Orioli},\ and\ \citenamefont
  {Gasenzer}}]{Chantesana:2018qsb}%
  \BibitemOpen
  \bibfield  {author} {\bibinfo {author} {\bibfnamefont {I.}~\bibnamefont
  {Chantesana}}, \bibinfo {author} {\bibfnamefont {A.}~\bibnamefont {Pi\~neiro
  Orioli}},\ and\ \bibinfo {author} {\bibfnamefont {T.}~\bibnamefont
  {Gasenzer}},\ }\bibfield  {title} {\bibinfo {title} {{Kinetic theory of
  nonthermal fixed points in a Bose gas}},\ }\href
  {https://doi.org/10.1103/PhysRevA.99.043620} {\bibfield  {journal} {\bibinfo
  {journal} {Phys. Rev. A}\ }\textbf {\bibinfo {volume} {99}},\ \bibinfo
  {pages} {043620} (\bibinfo {year} {2019})},\ \Eprint
  {https://arxiv.org/abs/1801.09490} {arXiv:1801.09490 [cond-mat.quant-gas]}
  \BibitemShut {NoStop}%
\bibitem [{\citenamefont {Pi\~neiro Orioli}\ \emph {et~al.}(2015)\citenamefont
  {Pi\~neiro Orioli}, \citenamefont {Boguslavski},\ and\ \citenamefont
  {Berges}}]{PineiroOrioli:2015cpb}%
  \BibitemOpen
  \bibfield  {author} {\bibinfo {author} {\bibfnamefont {A.}~\bibnamefont
  {Pi\~neiro Orioli}}, \bibinfo {author} {\bibfnamefont {K.}~\bibnamefont
  {Boguslavski}},\ and\ \bibinfo {author} {\bibfnamefont {J.}~\bibnamefont
  {Berges}},\ }\bibfield  {title} {\bibinfo {title} {{Universal self-similar
  dynamics of relativistic and nonrelativistic field theories near nonthermal
  fixed points}},\ }\href {https://doi.org/10.1103/PhysRevD.92.025041}
  {\bibfield  {journal} {\bibinfo  {journal} {Phys. Rev. D}\ }\textbf {\bibinfo
  {volume} {92}},\ \bibinfo {pages} {025041} (\bibinfo {year} {2015})},\
  \Eprint {https://arxiv.org/abs/1503.02498} {arXiv:1503.02498 [hep-ph]}
  \BibitemShut {NoStop}%
\bibitem [{\citenamefont {Mikheev}\ \emph {et~al.}(2019)\citenamefont
  {Mikheev}, \citenamefont {Schmied},\ and\ \citenamefont
  {Gasenzer}}]{Mikheev:2018adp}%
  \BibitemOpen
  \bibfield  {author} {\bibinfo {author} {\bibfnamefont {A.~N.}\ \bibnamefont
  {Mikheev}}, \bibinfo {author} {\bibfnamefont {C.-M.}\ \bibnamefont
  {Schmied}},\ and\ \bibinfo {author} {\bibfnamefont {T.}~\bibnamefont
  {Gasenzer}},\ }\bibfield  {title} {\bibinfo {title} {{Low-energy effective
  theory of nonthermal fixed points in a multicomponent Bose gas}},\ }\href
  {https://doi.org/10.1103/PhysRevA.99.063622} {\bibfield  {journal} {\bibinfo
  {journal} {Phys. Rev. A}\ }\textbf {\bibinfo {volume} {99}},\ \bibinfo
  {pages} {063622} (\bibinfo {year} {2019})},\ \Eprint
  {https://arxiv.org/abs/1807.10228} {arXiv:1807.10228 [cond-mat.quant-gas]}
  \BibitemShut {NoStop}%
\bibitem [{\citenamefont {Walz}\ \emph {et~al.}(2018)\citenamefont {Walz},
  \citenamefont {Boguslavski},\ and\ \citenamefont {Berges}}]{Walz:2017ffj}%
  \BibitemOpen
  \bibfield  {author} {\bibinfo {author} {\bibfnamefont {R.}~\bibnamefont
  {Walz}}, \bibinfo {author} {\bibfnamefont {K.}~\bibnamefont {Boguslavski}},\
  and\ \bibinfo {author} {\bibfnamefont {J.}~\bibnamefont {Berges}},\
  }\bibfield  {title} {\bibinfo {title} {{Large-N kinetic theory for highly
  occupied systems}},\ }\href {https://doi.org/10.1103/PhysRevD.97.116011}
  {\bibfield  {journal} {\bibinfo  {journal} {Phys. Rev. D}\ }\textbf {\bibinfo
  {volume} {97}},\ \bibinfo {pages} {116011} (\bibinfo {year} {2018})},\
  \Eprint {https://arxiv.org/abs/1710.11146} {arXiv:1710.11146 [hep-ph]}
  \BibitemShut {NoStop}%
\bibitem [{\citenamefont {Rosenhaus}\ and\ \citenamefont
  {Schubring}(2026)}]{Rosenhaus:2024iqw}%
  \BibitemOpen
  \bibfield  {author} {\bibinfo {author} {\bibfnamefont {V.}~\bibnamefont
  {Rosenhaus}}\ and\ \bibinfo {author} {\bibfnamefont {D.}~\bibnamefont
  {Schubring}},\ }\bibfield  {title} {\bibinfo {title} {{Strong wave turbulence
  in strongly local large-N theories}},\ }\href
  {https://doi.org/10.1103/p7p2-6f27} {\bibfield  {journal} {\bibinfo
  {journal} {Phys. Rev. E}\ }\textbf {\bibinfo {volume} {114}},\ \bibinfo
  {pages} {035101} (\bibinfo {year} {2026})},\ \Eprint
  {https://arxiv.org/abs/2406.18475} {arXiv:2406.18475 [hep-th]} \BibitemShut
  {NoStop}%
\bibitem [{\citenamefont {Rosenhaus}\ and\ \citenamefont
  {Falkovich}()}]{Rosenhaus:2025mgj}%
  \BibitemOpen
  \bibfield  {author} {\bibinfo {author} {\bibfnamefont {V.}~\bibnamefont
  {Rosenhaus}}\ and\ \bibinfo {author} {\bibfnamefont {G.}~\bibnamefont
  {Falkovich}},\ }\href@noop {} {\bibinfo {title} {{Weak and strong turbulence
  in self-focusing and defocusing media}}},\ \Eprint
  {https://arxiv.org/abs/2501.12451} {arXiv:2501.12451 [physics.flu-dyn]}
  \BibitemShut {NoStop}%
\bibitem [{\citenamefont {Rosenhaus}\ and\ \citenamefont
  {Smolkin}(2024)}]{Rosenhaus2024a.PhysRevE.109.064127}%
  \BibitemOpen
  \bibfield  {author} {\bibinfo {author} {\bibfnamefont {V.}~\bibnamefont
  {Rosenhaus}}\ and\ \bibinfo {author} {\bibfnamefont {M.}~\bibnamefont
  {Smolkin}},\ }\bibfield  {title} {\bibinfo {title} {Wave turbulence and the
  kinetic equation beyond leading order},\ }\href
  {https://doi.org/10.1103/PhysRevE.109.064127} {\bibfield  {journal} {\bibinfo
   {journal} {Phys. Rev. E}\ }\textbf {\bibinfo {volume} {109}},\ \bibinfo
  {pages} {064127} (\bibinfo {year} {2024})}\BibitemShut {NoStop}%
\bibitem [{\citenamefont {Rosenhaus}\ and\ \citenamefont
  {Falkovich}(2024)}]{Rosenhaus2024a.PhysRevLett.133.244002}%
  \BibitemOpen
  \bibfield  {author} {\bibinfo {author} {\bibfnamefont {V.}~\bibnamefont
  {Rosenhaus}}\ and\ \bibinfo {author} {\bibfnamefont {G.}~\bibnamefont
  {Falkovich}},\ }\bibfield  {title} {\bibinfo {title} {Interaction
  renormalization and validity of kinetic equations for turbulent states},\
  }\href {https://doi.org/10.1103/PhysRevLett.133.244002} {\bibfield  {journal}
  {\bibinfo  {journal} {Phys. Rev. Lett.}\ }\textbf {\bibinfo {volume} {133}},\
  \bibinfo {pages} {244002} (\bibinfo {year} {2024})}\BibitemShut {NoStop}%
\bibitem [{\citenamefont {Hu}\ and\ \citenamefont
  {Rosenhaus}(2025)}]{Hu:2025bqi}%
  \BibitemOpen
  \bibfield  {author} {\bibinfo {author} {\bibfnamefont {X.-Y.}\ \bibnamefont
  {Hu}}\ and\ \bibinfo {author} {\bibfnamefont {V.}~\bibnamefont {Rosenhaus}},\
  }\bibfield  {title} {\bibinfo {title} {{Beyond the Boltzmann equation for
  weakly coupled quantum fields}},\ }\href
  {https://doi.org/10.1007/JHEP10(2025)070} {\bibfield  {journal} {\bibinfo
  {journal} {JHEP}\ }\textbf {\bibinfo {volume} {10}},\ \bibinfo {pages}
  {070}},\ \Eprint {https://arxiv.org/abs/2503.09932} {arXiv:2503.09932
  [hep-th]} \BibitemShut {NoStop}%
\bibitem [{\citenamefont {Gasenzer}\ \emph {et~al.}(2012)\citenamefont
  {Gasenzer}, \citenamefont {Nowak},\ and\ \citenamefont
  {Sexty}}]{Gasenzer:2011by}%
  \BibitemOpen
  \bibfield  {author} {\bibinfo {author} {\bibfnamefont {T.}~\bibnamefont
  {Gasenzer}}, \bibinfo {author} {\bibfnamefont {B.}~\bibnamefont {Nowak}},\
  and\ \bibinfo {author} {\bibfnamefont {D.}~\bibnamefont {Sexty}},\ }\bibfield
   {title} {\bibinfo {title} {{Charge separation in Reheating after
  Cosmological Inflation}},\ }\href
  {https://doi.org/10.1016/j.physletb.2012.03.031} {\bibfield  {journal}
  {\bibinfo  {journal} {Phys. Lett. B}\ }\textbf {\bibinfo {volume} {710}},\
  \bibinfo {pages} {500} (\bibinfo {year} {2012})},\ \Eprint
  {https://arxiv.org/abs/1108.0541} {arXiv:1108.0541 [hep-ph]} \BibitemShut
  {NoStop}%
\bibitem [{\citenamefont {Berges}\ \emph {et~al.}(2014)\citenamefont {Berges},
  \citenamefont {Boguslavski}, \citenamefont {Schlichting},\ and\ \citenamefont
  {Ve\-nu\-go\-pa\ lan}}]{Berges:2013eia}%
  \BibitemOpen
  \bibfield  {author} {\bibinfo {author} {\bibfnamefont {J.}~\bibnamefont
  {Berges}}, \bibinfo {author} {\bibfnamefont {K.}~\bibnamefont {Boguslavski}},
  \bibinfo {author} {\bibfnamefont {S.}~\bibnamefont {Schlichting}},\ and\
  \bibinfo {author} {\bibfnamefont {R.}~\bibnamefont {Ve\-nu\-go\-pa\ lan}},\
  }\bibfield  {title} {\bibinfo {title} {{Turbulent thermalization process in
  heavy-ion collisions at ultrarelativistic energies}},\ }\href
  {https://doi.org/10.1103/PhysRevD.89.074011} {\bibfield  {journal} {\bibinfo
  {journal} {Phys. Rev. D}\ }\textbf {\bibinfo {volume} {89}},\ \bibinfo
  {pages} {074011} (\bibinfo {year} {2014})},\ \Eprint
  {https://arxiv.org/abs/1303.5650} {arXiv:1303.5650 [hep-ph]} \BibitemShut
  {NoStop}%
\bibitem [{\citenamefont {Nowak}\ \emph {et~al.}(2012)\citenamefont {Nowak},
  \citenamefont {Schole}, \citenamefont {Sexty},\ and\ \citenamefont
  {Gasenzer}}]{Nowak:2011sk}%
  \BibitemOpen
  \bibfield  {author} {\bibinfo {author} {\bibfnamefont {B.}~\bibnamefont
  {Nowak}}, \bibinfo {author} {\bibfnamefont {J.}~\bibnamefont {Schole}},
  \bibinfo {author} {\bibfnamefont {D.}~\bibnamefont {Sexty}},\ and\ \bibinfo
  {author} {\bibfnamefont {T.}~\bibnamefont {Gasenzer}},\ }\bibfield  {title}
  {\bibinfo {title} {{Nonthermal fixed points, vortex statistics, and
  superfluid turbulence in an ultracold Bose gas}},\ }\href
  {https://doi.org/10.1103/PhysRevA.85.043627} {\bibfield  {journal} {\bibinfo
  {journal} {Phys. Rev. A}\ }\textbf {\bibinfo {volume} {85}},\ \bibinfo
  {pages} {043627} (\bibinfo {year} {2012})},\ \Eprint
  {https://arxiv.org/abs/1111.6127} {arXiv:1111.6127 [cond-mat.quant-gas]}
  \BibitemShut {NoStop}%
\bibitem [{\citenamefont {Heinen}\ \emph {et~al.}()\citenamefont {Heinen},
  \citenamefont {Mikheev}, \citenamefont {Schmied},\ and\ \citenamefont
  {Gasenzer}}]{Heinen:2022rew}%
  \BibitemOpen
  \bibfield  {author} {\bibinfo {author} {\bibfnamefont {P.}~\bibnamefont
  {Heinen}}, \bibinfo {author} {\bibfnamefont {A.~N.}\ \bibnamefont {Mikheev}},
  \bibinfo {author} {\bibfnamefont {C.-M.}\ \bibnamefont {Schmied}},\ and\
  \bibinfo {author} {\bibfnamefont {T.}~\bibnamefont {Gasenzer}},\ }\href@noop
  {} {\bibinfo {title} {{Non-thermal fixed points of universal sine-Gordon
  coarsening dynamics}}},\ \Eprint {https://arxiv.org/abs/2212.01162}
  {arXiv:2212.01162 [cond-mat.quant-gas]} \BibitemShut {NoStop}%
\bibitem [{\citenamefont {Noel}\ and\ \citenamefont
  {Spitz}(2024)}]{Noel:2023oyz}%
  \BibitemOpen
  \bibfield  {author} {\bibinfo {author} {\bibfnamefont {V.}~\bibnamefont
  {Noel}}\ and\ \bibinfo {author} {\bibfnamefont {D.}~\bibnamefont {Spitz}},\
  }\bibfield  {title} {\bibinfo {title} {{Detecting defect dynamics in
  relativistic field theories far from equilibrium using topological data
  analysis}},\ }\href {https://doi.org/10.1103/PhysRevD.109.056011} {\bibfield
  {journal} {\bibinfo  {journal} {Phys. Rev. D}\ }\textbf {\bibinfo {volume}
  {109}},\ \bibinfo {pages} {056011} (\bibinfo {year} {2024})},\ \Eprint
  {https://arxiv.org/abs/2312.04959} {arXiv:2312.04959 [hep-ph]} \BibitemShut
  {NoStop}%
\bibitem [{\citenamefont {Schole}\ \emph {et~al.}(2012)\citenamefont {Schole},
  \citenamefont {Nowak},\ and\ \citenamefont {Gasenzer}}]{Schole:2012kt}%
  \BibitemOpen
  \bibfield  {author} {\bibinfo {author} {\bibfnamefont {J.}~\bibnamefont
  {Schole}}, \bibinfo {author} {\bibfnamefont {B.}~\bibnamefont {Nowak}},\ and\
  \bibinfo {author} {\bibfnamefont {T.}~\bibnamefont {Gasenzer}},\ }\bibfield
  {title} {\bibinfo {title} {{Critical Dynamics of a Two-dimensional Superfluid
  near a Non-Thermal Fixed Point}},\ }\href
  {https://doi.org/10.1103/PhysRevA.86.013624} {\bibfield  {journal} {\bibinfo
  {journal} {Phys. Rev. A}\ }\textbf {\bibinfo {volume} {86}},\ \bibinfo
  {pages} {013624} (\bibinfo {year} {2012})},\ \Eprint
  {https://arxiv.org/abs/1204.2487} {arXiv:1204.2487 [cond-mat.quant-gas]}
  \BibitemShut {NoStop}%
\bibitem [{\citenamefont {Karl}\ \emph {et~al.}(2013)\citenamefont {Karl},
  \citenamefont {Nowak},\ and\ \citenamefont {Gasenzer}}]{Karl:2013kua}%
  \BibitemOpen
  \bibfield  {author} {\bibinfo {author} {\bibfnamefont {M.}~\bibnamefont
  {Karl}}, \bibinfo {author} {\bibfnamefont {B.}~\bibnamefont {Nowak}},\ and\
  \bibinfo {author} {\bibfnamefont {T.}~\bibnamefont {Gasenzer}},\ }\bibfield
  {title} {\bibinfo {title} {{Universal scaling at nonthermal fixed points of a
  two-component Bose gas}},\ }\href
  {https://doi.org/10.1103/PhysRevA.88.063615} {\bibfield  {journal} {\bibinfo
  {journal} {Phys. Rev. A}\ }\textbf {\bibinfo {volume} {88}},\ \bibinfo
  {pages} {063615} (\bibinfo {year} {2013})},\ \Eprint
  {https://arxiv.org/abs/1307.7368} {arXiv:1307.7368 [cond-mat.quant-gas]}
  \BibitemShut {NoStop}%
\bibitem [{\citenamefont {Ewerz}\ \emph {et~al.}(2015)\citenamefont {Ewerz},
  \citenamefont {Gasenzer}, \citenamefont {Karl},\ and\ \citenamefont
  {Samberg}}]{Ewerz:2014tua}%
  \BibitemOpen
  \bibfield  {author} {\bibinfo {author} {\bibfnamefont {C.}~\bibnamefont
  {Ewerz}}, \bibinfo {author} {\bibfnamefont {T.}~\bibnamefont {Gasenzer}},
  \bibinfo {author} {\bibfnamefont {M.}~\bibnamefont {Karl}},\ and\ \bibinfo
  {author} {\bibfnamefont {A.}~\bibnamefont {Samberg}},\ }\bibfield  {title}
  {\bibinfo {title} {{Non-Thermal Fixed Point in a Holographic Superfluid}},\
  }\href {https://doi.org/10.1007/JHEP05(2015)070} {\bibfield  {journal}
  {\bibinfo  {journal} {JHEP}\ }\textbf {\bibinfo {volume} {05}},\ \bibinfo
  {pages} {070}},\ \Eprint {https://arxiv.org/abs/1410.3472} {arXiv:1410.3472
  [hep-th]} \BibitemShut {NoStop}%
\bibitem [{\citenamefont {Berges}\ and\ \citenamefont
  {Wallisch}(2017)}]{Berges:2016nru}%
  \BibitemOpen
  \bibfield  {author} {\bibinfo {author} {\bibfnamefont {J.}~\bibnamefont
  {Berges}}\ and\ \bibinfo {author} {\bibfnamefont {B.}~\bibnamefont
  {Wallisch}},\ }\bibfield  {title} {\bibinfo {title} {{Nonthermal Fixed Points
  in Quantum Field Theory Beyond the Weak-Coupling Limit}},\ }\href
  {https://doi.org/10.1103/PhysRevD.95.036016} {\bibfield  {journal} {\bibinfo
  {journal} {Phys. Rev. D}\ }\textbf {\bibinfo {volume} {95}},\ \bibinfo
  {pages} {036016} (\bibinfo {year} {2017})},\ \Eprint
  {https://arxiv.org/abs/1607.02160} {arXiv:1607.02160 [hep-ph]} \BibitemShut
  {NoStop}%
\bibitem [{\citenamefont {Karl}\ and\ \citenamefont
  {Gasenzer}(2017)}]{Karl:2016wko}%
  \BibitemOpen
  \bibfield  {author} {\bibinfo {author} {\bibfnamefont {M.}~\bibnamefont
  {Karl}}\ and\ \bibinfo {author} {\bibfnamefont {T.}~\bibnamefont
  {Gasenzer}},\ }\bibfield  {title} {\bibinfo {title} {{Strongly anomalous
  non-thermal fixed point in a quenched two-dimensional Bose gas}},\ }\href
  {https://doi.org/10.1088/1367-2630/aa7eeb} {\bibfield  {journal} {\bibinfo
  {journal} {New J. Phys.}\ }\textbf {\bibinfo {volume} {19}},\ \bibinfo
  {pages} {093014} (\bibinfo {year} {2017})},\ \Eprint
  {https://arxiv.org/abs/1611.01163} {arXiv:1611.01163 [cond-mat.quant-gas]}
  \BibitemShut {NoStop}%
\bibitem [{\citenamefont {Siovitz}\ \emph {et~al.}(2025)\citenamefont
  {Siovitz}, \citenamefont {Gl{\"u}ck}, \citenamefont {Deller}, \citenamefont
  {Schmutz}, \citenamefont {Klein}, \citenamefont {Strobel}, \citenamefont
  {Oberthaler},\ and\ \citenamefont {Gasenzer}}]{Siovitz:2024aqi}%
  \BibitemOpen
  \bibfield  {author} {\bibinfo {author} {\bibfnamefont {I.}~\bibnamefont
  {Siovitz}}, \bibinfo {author} {\bibfnamefont {A.-M.~E.}\ \bibnamefont
  {Gl{\"u}ck}}, \bibinfo {author} {\bibfnamefont {Y.}~\bibnamefont {Deller}},
  \bibinfo {author} {\bibfnamefont {A.}~\bibnamefont {Schmutz}}, \bibinfo
  {author} {\bibfnamefont {F.}~\bibnamefont {Klein}}, \bibinfo {author}
  {\bibfnamefont {H.}~\bibnamefont {Strobel}}, \bibinfo {author} {\bibfnamefont
  {M.~K.}\ \bibnamefont {Oberthaler}},\ and\ \bibinfo {author} {\bibfnamefont
  {T.}~\bibnamefont {Gasenzer}},\ }\bibfield  {title} {\bibinfo {title}
  {{Double sine-Gordon class of universal coarsening dynamics in a spin-1 Bose
  gas}},\ }\href {https://doi.org/10.1103/df5w-3yfd} {\bibfield  {journal}
  {\bibinfo  {journal} {Phys. Rev. A}\ }\textbf {\bibinfo {volume} {112}},\
  \bibinfo {pages} {023304} (\bibinfo {year} {2025})},\ \Eprint
  {https://arxiv.org/abs/2412.13986} {arXiv:2412.13986 [cond-mat.quant-gas]}
  \BibitemShut {NoStop}%
\bibitem [{\citenamefont {Noel}\ \emph {et~al.}(2025)\citenamefont {Noel},
  \citenamefont {Gasenzer},\ and\ \citenamefont {Boguslavski}}]{Noel:2025mtb}%
  \BibitemOpen
  \bibfield  {author} {\bibinfo {author} {\bibfnamefont {V.}~\bibnamefont
  {Noel}}, \bibinfo {author} {\bibfnamefont {T.}~\bibnamefont {Gasenzer}},\
  and\ \bibinfo {author} {\bibfnamefont {K.}~\bibnamefont {Boguslavski}},\
  }\bibfield  {title} {\bibinfo {title} {{Kelvin waves in nonequilibrium
  universal dynamics of relativistic scalar field theories}},\ }\href
  {https://doi.org/10.1103/h4hn-r6kp} {\bibfield  {journal} {\bibinfo
  {journal} {Phys. Rev. Res.}\ }\textbf {\bibinfo {volume} {7}},\ \bibinfo
  {pages} {033220} (\bibinfo {year} {2025})},\ \Eprint
  {https://arxiv.org/abs/2503.01771} {arXiv:2503.01771 [cond-mat.quant-gas]}
  \BibitemShut {NoStop}%
\bibitem [{\citenamefont {Rasch}\ and\ \citenamefont
  {Gasenzer}(2026)}]{Rasch:2025hth}%
  \BibitemOpen
  \bibfield  {author} {\bibinfo {author} {\bibfnamefont {N.}~\bibnamefont
  {Rasch}}\ and\ \bibinfo {author} {\bibfnamefont {T.}~\bibnamefont
  {Gasenzer}},\ }\bibfield  {title} {\bibinfo {title} {{Decaying superfluid
  turbulence near an anomalous nonthermal fixed point}},\ }\href
  {https://doi.org/10.1103/p6wv-z621} {\bibfield  {journal} {\bibinfo
  {journal} {Phys. Rev. A}\ }\textbf {\bibinfo {volume} {113}},\ \bibinfo
  {pages} {L051302} (\bibinfo {year} {2026})},\ \Eprint
  {https://arxiv.org/abs/2509.21285} {arXiv:2509.21285 [cond-mat.quant-gas]}
  \BibitemShut {NoStop}%
\bibitem [{\citenamefont {Rasch}\ \emph {et~al.}(2025)\citenamefont {Rasch},
  \citenamefont {Chomaz},\ and\ \citenamefont {Gasenzer}}]{Rasch:2025kna}%
  \BibitemOpen
  \bibfield  {author} {\bibinfo {author} {\bibfnamefont {N.}~\bibnamefont
  {Rasch}}, \bibinfo {author} {\bibfnamefont {L.}~\bibnamefont {Chomaz}},\ and\
  \bibinfo {author} {\bibfnamefont {T.}~\bibnamefont {Gasenzer}},\ }\bibfield
  {title} {\bibinfo {title} {{Anomalous nonthermal fixed point in a
  quasi-two-dimensional dipolar Bose gas}},\ }\href
  {https://doi.org/10.1103/x2rj-ptgy} {\bibfield  {journal} {\bibinfo
  {journal} {Phys. Rev. A}\ }\textbf {\bibinfo {volume} {112}},\ \bibinfo
  {pages} {053310} (\bibinfo {year} {2025})},\ \Eprint
  {https://arxiv.org/abs/2506.01653} {arXiv:2506.01653 [cond-mat.quant-gas]}
  \BibitemShut {NoStop}%
\bibitem [{\citenamefont {Mikheev}\ \emph {et~al.}(2025)\citenamefont
  {Mikheev}, \citenamefont {Noel}, \citenamefont {Siovitz}, \citenamefont
  {Strobel}, \citenamefont {Oberthaler},\ and\ \citenamefont
  {Berges}}]{Mikheev:2024pur}%
  \BibitemOpen
  \bibfield  {author} {\bibinfo {author} {\bibfnamefont {A.~N.}\ \bibnamefont
  {Mikheev}}, \bibinfo {author} {\bibfnamefont {V.}~\bibnamefont {Noel}},
  \bibinfo {author} {\bibfnamefont {I.}~\bibnamefont {Siovitz}}, \bibinfo
  {author} {\bibfnamefont {H.}~\bibnamefont {Strobel}}, \bibinfo {author}
  {\bibfnamefont {M.~K.}\ \bibnamefont {Oberthaler}},\ and\ \bibinfo {author}
  {\bibfnamefont {J.}~\bibnamefont {Berges}},\ }\bibfield  {title} {\bibinfo
  {title} {{Extracting the symmetries of nonequilibrium quantum many-body
  systems}},\ }\href {https://doi.org/10.21468/SciPostPhys.18.2.044} {\bibfield
   {journal} {\bibinfo  {journal} {SciPost Phys.}\ }\textbf {\bibinfo {volume}
  {18}},\ \bibinfo {pages} {044} (\bibinfo {year} {2025})},\ \Eprint
  {https://arxiv.org/abs/2407.17913} {arXiv:2407.17913 [cond-mat.quant-gas]}
  \BibitemShut {NoStop}%
\bibitem [{\citenamefont {Siovitz}\ \emph {et~al.}(2023)\citenamefont
  {Siovitz}, \citenamefont {Lannig}, \citenamefont {Deller}, \citenamefont
  {Strobel}, \citenamefont {Oberthaler},\ and\ \citenamefont
  {Gasenzer}}]{Siovitz:2023ius}%
  \BibitemOpen
  \bibfield  {author} {\bibinfo {author} {\bibfnamefont {I.}~\bibnamefont
  {Siovitz}}, \bibinfo {author} {\bibfnamefont {S.}~\bibnamefont {Lannig}},
  \bibinfo {author} {\bibfnamefont {Y.}~\bibnamefont {Deller}}, \bibinfo
  {author} {\bibfnamefont {H.}~\bibnamefont {Strobel}}, \bibinfo {author}
  {\bibfnamefont {M.~K.}\ \bibnamefont {Oberthaler}},\ and\ \bibinfo {author}
  {\bibfnamefont {T.}~\bibnamefont {Gasenzer}},\ }\bibfield  {title} {\bibinfo
  {title} {{Universal Dynamics of Rogue Waves in a Quenched Spinor Bose
  Condensate}},\ }\href {https://doi.org/10.1103/PhysRevLett.131.183402}
  {\bibfield  {journal} {\bibinfo  {journal} {Phys. Rev. Lett.}\ }\textbf
  {\bibinfo {volume} {131}},\ \bibinfo {pages} {183402} (\bibinfo {year}
  {2023})},\ \Eprint {https://arxiv.org/abs/2304.09293} {arXiv:2304.09293
  [cond-mat.quant-gas]} \BibitemShut {NoStop}%
\bibitem [{\citenamefont {Shen}\ and\ \citenamefont
  {Berges}(2020)}]{Shen:2019jhl}%
  \BibitemOpen
  \bibfield  {author} {\bibinfo {author} {\bibfnamefont {L.}~\bibnamefont
  {Shen}}\ and\ \bibinfo {author} {\bibfnamefont {J.}~\bibnamefont {Berges}},\
  }\bibfield  {title} {\bibinfo {title} {{Spectral, statistical and vertex
  functions in scalar quantum field theory far from equilibrium}},\ }\href
  {https://doi.org/10.1103/PhysRevD.101.056009} {\bibfield  {journal} {\bibinfo
   {journal} {Phys. Rev. D}\ }\textbf {\bibinfo {volume} {101}},\ \bibinfo
  {pages} {056009} (\bibinfo {year} {2020})},\ \Eprint
  {https://arxiv.org/abs/1912.07565} {arXiv:1912.07565 [hep-ph]} \BibitemShut
  {NoStop}%
\bibitem [{\citenamefont {Heinen}\ \emph {et~al.}(2023)\citenamefont {Heinen},
  \citenamefont {Mikheev},\ and\ \citenamefont {Gasenzer}}]{Heinen:2022ham}%
  \BibitemOpen
  \bibfield  {author} {\bibinfo {author} {\bibfnamefont {P.}~\bibnamefont
  {Heinen}}, \bibinfo {author} {\bibfnamefont {A.~N.}\ \bibnamefont
  {Mikheev}},\ and\ \bibinfo {author} {\bibfnamefont {T.}~\bibnamefont
  {Gasenzer}},\ }\bibfield  {title} {\bibinfo {title} {{Anomalous scaling at
  nonthermal fixed points of the sine-Gordon model}},\ }\href
  {https://doi.org/10.1103/PhysRevA.107.043303} {\bibfield  {journal} {\bibinfo
   {journal} {Phys. Rev. A}\ }\textbf {\bibinfo {volume} {107}},\ \bibinfo
  {pages} {043303} (\bibinfo {year} {2023})},\ \Eprint
  {https://arxiv.org/abs/2212.01163} {arXiv:2212.01163 [cond-mat.quant-gas]}
  \BibitemShut {NoStop}%
\bibitem [{\citenamefont {Gasenzer}\ \emph {et~al.}(2010)\citenamefont
  {Gasenzer}, \citenamefont {Kessler},\ and\ \citenamefont
  {Pawlowski}}]{Gasenzer:2010rq}%
  \BibitemOpen
  \bibfield  {author} {\bibinfo {author} {\bibfnamefont {T.}~\bibnamefont
  {Gasenzer}}, \bibinfo {author} {\bibfnamefont {S.}~\bibnamefont {Kessler}},\
  and\ \bibinfo {author} {\bibfnamefont {J.~M.}\ \bibnamefont {Pawlowski}},\
  }\bibfield  {title} {\bibinfo {title} {{Far-from-equilibrium quantum
  many-body dynamics}},\ }\href
  {https://doi.org/10.1140/epjc/s10052-010-1430-3} {\bibfield  {journal}
  {\bibinfo  {journal} {Eur. Phys. J. C}\ }\textbf {\bibinfo {volume} {70}},\
  \bibinfo {pages} {423} (\bibinfo {year} {2010})},\ \Eprint
  {https://arxiv.org/abs/1003.4163} {arXiv:1003.4163} \BibitemShut {NoStop}%
\bibitem [{\citenamefont {Deng}\ \emph {et~al.}(2018)\citenamefont {Deng},
  \citenamefont {Schlichting}, \citenamefont {Venugopalan},\ and\ \citenamefont
  {Wang}}]{Deng:2018xsk}%
  \BibitemOpen
  \bibfield  {author} {\bibinfo {author} {\bibfnamefont {J.}~\bibnamefont
  {Deng}}, \bibinfo {author} {\bibfnamefont {S.}~\bibnamefont {Schlichting}},
  \bibinfo {author} {\bibfnamefont {R.}~\bibnamefont {Venugopalan}},\ and\
  \bibinfo {author} {\bibfnamefont {Q.}~\bibnamefont {Wang}},\ }\bibfield
  {title} {\bibinfo {title} {{Off-equilibrium infrared structure of
  self-interacting scalar fields: Universal scaling, Vortex-antivortex
  superfluid dynamics and Bose-Einstein condensation}},\ }\href
  {https://doi.org/10.1103/PhysRevA.97.053606} {\bibfield  {journal} {\bibinfo
  {journal} {Phys. Rev. A}\ }\textbf {\bibinfo {volume} {97}},\ \bibinfo
  {pages} {053606} (\bibinfo {year} {2018})},\ \Eprint
  {https://arxiv.org/abs/1801.06260} {arXiv:1801.06260 [hep-th]} \BibitemShut
  {NoStop}%
\bibitem [{\citenamefont {Namjoo}\ \emph {et~al.}(2018)\citenamefont {Namjoo},
  \citenamefont {Guth},\ and\ \citenamefont {Kaiser}}]{namjoo2018relativistic}%
  \BibitemOpen
  \bibfield  {author} {\bibinfo {author} {\bibfnamefont {M.~H.}\ \bibnamefont
  {Namjoo}}, \bibinfo {author} {\bibfnamefont {A.~H.}\ \bibnamefont {Guth}},\
  and\ \bibinfo {author} {\bibfnamefont {D.~I.}\ \bibnamefont {Kaiser}},\
  }\bibfield  {title} {\bibinfo {title} {Relativistic corrections to
  nonrelativistic effective field theories},\ }\href
  {https://doi.org/10.1103/PhysRevD.98.016011} {\bibfield  {journal} {\bibinfo
  {journal} {Phys. Rev. D}\ }\textbf {\bibinfo {volume} {98}},\ \bibinfo
  {pages} {016011} (\bibinfo {year} {2018})},\ \Eprint
  {https://arxiv.org/abs/1712.00445} {arXiv:1712.00445} \BibitemShut {NoStop}%
\bibitem [{\citenamefont {Berges}(2002)}]{Berges:2001fi}%
  \BibitemOpen
  \bibfield  {author} {\bibinfo {author} {\bibfnamefont {J.}~\bibnamefont
  {Berges}},\ }\bibfield  {title} {\bibinfo {title} {{Controlled
  nonperturbative dynamics of quantum fields out-of-equilibrium}},\ }\href
  {https://doi.org/10.1016/S0375-9474(01)01295-7} {\bibfield  {journal}
  {\bibinfo  {journal} {Nucl. Phys. A}\ }\textbf {\bibinfo {volume} {699}},\
  \bibinfo {pages} {847} (\bibinfo {year} {2002})},\ \Eprint
  {https://arxiv.org/abs/hep-ph/0105311} {arXiv:hep-ph/0105311} \BibitemShut
  {NoStop}%
\bibitem [{\citenamefont {Preis}\ \emph {et~al.}(2023)\citenamefont {Preis},
  \citenamefont {Heller},\ and\ \citenamefont {Berges}}]{Preis:2022uqs}%
  \BibitemOpen
  \bibfield  {author} {\bibinfo {author} {\bibfnamefont {T.}~\bibnamefont
  {Preis}}, \bibinfo {author} {\bibfnamefont {M.~P.}\ \bibnamefont {Heller}},\
  and\ \bibinfo {author} {\bibfnamefont {J.}~\bibnamefont {Berges}},\
  }\bibfield  {title} {\bibinfo {title} {{Stable and Unstable Perturbations in
  Universal Scaling Phenomena Far from Equilibrium}},\ }\href
  {https://doi.org/10.1103/PhysRevLett.130.031602} {\bibfield  {journal}
  {\bibinfo  {journal} {Phys. Rev. Lett.}\ }\textbf {\bibinfo {volume} {130}},\
  \bibinfo {pages} {031602} (\bibinfo {year} {2023})},\ \Eprint
  {https://arxiv.org/abs/2209.14883} {arXiv:2209.14883 [hep-ph]} \BibitemShut
  {NoStop}%
\bibitem [{\citenamefont {Berges}\ \emph {et~al.}(2004)\citenamefont {Berges},
  \citenamefont {Borsanyi},\ and\ \citenamefont {Wetterich}}]{Berges:2004ce}%
  \BibitemOpen
  \bibfield  {author} {\bibinfo {author} {\bibfnamefont {J.}~\bibnamefont
  {Berges}}, \bibinfo {author} {\bibfnamefont {S.}~\bibnamefont {Borsanyi}},\
  and\ \bibinfo {author} {\bibfnamefont {C.}~\bibnamefont {Wetterich}},\
  }\bibfield  {title} {\bibinfo {title} {{Prethermalization}},\ }\href
  {https://doi.org/10.1103/PhysRevLett.93.142002} {\bibfield  {journal}
  {\bibinfo  {journal} {Phys. Rev. Lett.}\ }\textbf {\bibinfo {volume} {93}},\
  \bibinfo {pages} {142002} (\bibinfo {year} {2004})},\ \Eprint
  {https://arxiv.org/abs/hep-ph/0403234} {arXiv:hep-ph/0403234} \BibitemShut
  {NoStop}%
\bibitem [{\citenamefont {Pr\"ufer}\ \emph {et~al.}(2018)\citenamefont
  {Pr\"ufer}, \citenamefont {Kunkel}, \citenamefont {Strobel}, \citenamefont
  {Lannig}, \citenamefont {Linnemann}, \citenamefont {Schmied}, \citenamefont
  {Berges}, \citenamefont {Gasenzer},\ and\ \citenamefont
  {Oberthaler}}]{Prufer:2018hto}%
  \BibitemOpen
  \bibfield  {author} {\bibinfo {author} {\bibfnamefont {M.}~\bibnamefont
  {Pr\"ufer}}, \bibinfo {author} {\bibfnamefont {P.}~\bibnamefont {Kunkel}},
  \bibinfo {author} {\bibfnamefont {H.}~\bibnamefont {Strobel}}, \bibinfo
  {author} {\bibfnamefont {S.}~\bibnamefont {Lannig}}, \bibinfo {author}
  {\bibfnamefont {D.}~\bibnamefont {Linnemann}}, \bibinfo {author}
  {\bibfnamefont {C.-M.}\ \bibnamefont {Schmied}}, \bibinfo {author}
  {\bibfnamefont {J.}~\bibnamefont {Berges}}, \bibinfo {author} {\bibfnamefont
  {T.}~\bibnamefont {Gasenzer}},\ and\ \bibinfo {author} {\bibfnamefont
  {M.~K.}\ \bibnamefont {Oberthaler}},\ }\bibfield  {title} {\bibinfo {title}
  {{Observation of universal dynamics in a spinor Bose gas far from
  equilibrium}},\ }\href {https://doi.org/10.1038/s41586-018-0659-0} {\bibfield
   {journal} {\bibinfo  {journal} {Nature}\ }\textbf {\bibinfo {volume}
  {563}},\ \bibinfo {pages} {217} (\bibinfo {year} {2018})},\ \Eprint
  {https://arxiv.org/abs/1805.11881} {arXiv:1805.11881 [cond-mat.quant-gas]}
  \BibitemShut {NoStop}%
\bibitem [{\citenamefont {Erne}\ \emph {et~al.}(2018)\citenamefont {Erne},
  \citenamefont {B\"ucker}, \citenamefont {Gasenzer}, \citenamefont {Berges},\
  and\ \citenamefont {Schmiedmayer}}]{Erne:2018gmz}%
  \BibitemOpen
  \bibfield  {author} {\bibinfo {author} {\bibfnamefont {S.}~\bibnamefont
  {Erne}}, \bibinfo {author} {\bibfnamefont {R.}~\bibnamefont {B\"ucker}},
  \bibinfo {author} {\bibfnamefont {T.}~\bibnamefont {Gasenzer}}, \bibinfo
  {author} {\bibfnamefont {J.}~\bibnamefont {Berges}},\ and\ \bibinfo {author}
  {\bibfnamefont {J.}~\bibnamefont {Schmiedmayer}},\ }\bibfield  {title}
  {\bibinfo {title} {{Universal dynamics in an isolated one-dimensional Bose
  gas far from equilibrium}},\ }\href
  {https://doi.org/10.1038/s41586-018-0667-0} {\bibfield  {journal} {\bibinfo
  {journal} {Nature}\ }\textbf {\bibinfo {volume} {563}},\ \bibinfo {pages}
  {225} (\bibinfo {year} {2018})},\ \Eprint {https://arxiv.org/abs/1805.12310}
  {arXiv:1805.12310 [cond-mat.quant-gas]} \BibitemShut {NoStop}%
\bibitem [{\citenamefont {Garc\'\i{}a-Orozco}\ \emph
  {et~al.}(2022)\citenamefont {Garc\'\i{}a-Orozco}, \citenamefont {Madeira},
  \citenamefont {Moreno-Armijos}, \citenamefont {Fritsch}, \citenamefont
  {Tavares}, \citenamefont {Castilho}, \citenamefont {Cidrim}, \citenamefont
  {Roati},\ and\ \citenamefont {Bagnato}}]{Garcia-Orozco:2021hkx}%
  \BibitemOpen
  \bibfield  {author} {\bibinfo {author} {\bibfnamefont {A.~D.}\ \bibnamefont
  {Garc\'\i{}a-Orozco}}, \bibinfo {author} {\bibfnamefont {L.}~\bibnamefont
  {Madeira}}, \bibinfo {author} {\bibfnamefont {M.~A.}\ \bibnamefont
  {Moreno-Armijos}}, \bibinfo {author} {\bibfnamefont {A.~R.}\ \bibnamefont
  {Fritsch}}, \bibinfo {author} {\bibfnamefont {P.~E.~S.}\ \bibnamefont
  {Tavares}}, \bibinfo {author} {\bibfnamefont {P.~C.~M.}\ \bibnamefont
  {Castilho}}, \bibinfo {author} {\bibfnamefont {A.}~\bibnamefont {Cidrim}},
  \bibinfo {author} {\bibfnamefont {G.}~\bibnamefont {Roati}},\ and\ \bibinfo
  {author} {\bibfnamefont {V.~S.}\ \bibnamefont {Bagnato}},\ }\bibfield
  {title} {\bibinfo {title} {{Universal dynamics of a turbulent superfluid Bose
  gas}},\ }\href {https://doi.org/10.1103/PhysRevA.106.023314} {\bibfield
  {journal} {\bibinfo  {journal} {Phys. Rev. A}\ }\textbf {\bibinfo {volume}
  {106}},\ \bibinfo {pages} {023314} (\bibinfo {year} {2022})},\ \Eprint
  {https://arxiv.org/abs/2107.07421} {arXiv:2107.07421 [cond-mat.quant-gas]}
  \BibitemShut {NoStop}%
\bibitem [{\citenamefont {Lannig}\ \emph {et~al.}()\citenamefont {Lannig},
  \citenamefont {Pr\"ufer}, \citenamefont {Deller}, \citenamefont {Siovitz},
  \citenamefont {Dreher}, \citenamefont {Gasenzer}, \citenamefont {Strobel},\
  and\ \citenamefont {Oberthaler}}]{Lannig:2023fzf}%
  \BibitemOpen
  \bibfield  {author} {\bibinfo {author} {\bibfnamefont {S.}~\bibnamefont
  {Lannig}}, \bibinfo {author} {\bibfnamefont {M.}~\bibnamefont {Pr\"ufer}},
  \bibinfo {author} {\bibfnamefont {Y.}~\bibnamefont {Deller}}, \bibinfo
  {author} {\bibfnamefont {I.}~\bibnamefont {Siovitz}}, \bibinfo {author}
  {\bibfnamefont {J.}~\bibnamefont {Dreher}}, \bibinfo {author} {\bibfnamefont
  {T.}~\bibnamefont {Gasenzer}}, \bibinfo {author} {\bibfnamefont
  {H.}~\bibnamefont {Strobel}},\ and\ \bibinfo {author} {\bibfnamefont {M.~K.}\
  \bibnamefont {Oberthaler}},\ }\href@noop {} {\bibinfo {title} {{Observation
  of two non-thermal fixed points for the same microscopic symmetry}}},\
  \Eprint {https://arxiv.org/abs/2306.16497} {arXiv:2306.16497
  [cond-mat.quant-gas]} \BibitemShut {NoStop}%
\bibitem [{\citenamefont {Moreno-Armijos}\ \emph {et~al.}(2025)\citenamefont
  {Moreno-Armijos}, \citenamefont {Fritsch}, \citenamefont
  {Garc\'{\i}a-Orozco}, \citenamefont {Sab}, \citenamefont {Telles},
  \citenamefont {Zhu}, \citenamefont {Madeira}, \citenamefont {Nazarenko},
  \citenamefont {Yukalov},\ and\ \citenamefont
  {Bagnato}}]{MorenoArmijos2024a.PhysRevLett.134.023401}%
  \BibitemOpen
  \bibfield  {author} {\bibinfo {author} {\bibfnamefont {M.~A.}\ \bibnamefont
  {Moreno-Armijos}}, \bibinfo {author} {\bibfnamefont {A.~R.}\ \bibnamefont
  {Fritsch}}, \bibinfo {author} {\bibfnamefont {A.~D.}\ \bibnamefont
  {Garc\'{\i}a-Orozco}}, \bibinfo {author} {\bibfnamefont {S.}~\bibnamefont
  {Sab}}, \bibinfo {author} {\bibfnamefont {G.}~\bibnamefont {Telles}},
  \bibinfo {author} {\bibfnamefont {Y.}~\bibnamefont {Zhu}}, \bibinfo {author}
  {\bibfnamefont {L.}~\bibnamefont {Madeira}}, \bibinfo {author} {\bibfnamefont
  {S.}~\bibnamefont {Nazarenko}}, \bibinfo {author} {\bibfnamefont {V.~I.}\
  \bibnamefont {Yukalov}},\ and\ \bibinfo {author} {\bibfnamefont {V.~S.}\
  \bibnamefont {Bagnato}},\ }\bibfield  {title} {\bibinfo {title} {Observation
  of relaxation stages in a nonequilibrium closed quantum system: Decaying
  turbulence in a trapped superfluid},\ }\href
  {https://doi.org/10.1103/PhysRevLett.134.023401} {\bibfield  {journal}
  {\bibinfo  {journal} {Phys. Rev. Lett.}\ }\textbf {\bibinfo {volume} {134}},\
  \bibinfo {pages} {023401} (\bibinfo {year} {2025})},\ \Eprint
  {https://arxiv.org/abs/2407.11237} {arXiv:2407.11237 [cond-mat.quant-gas]}
  \BibitemShut {NoStop}%
\bibitem [{\citenamefont {Sab}\ \emph {et~al.}()\citenamefont {Sab},
  \citenamefont {Moreno-Armijos}, \citenamefont {García-Orozco}, \citenamefont
  {Fernandes}, \citenamefont {Zhu}, \citenamefont {Fritsch}, \citenamefont
  {Perrin}, \citenamefont {Nazarenko},\ and\ \citenamefont
  {Bagnato}}]{sab2026universalbehaviorrelaxationdynamics}%
  \BibitemOpen
  \bibfield  {author} {\bibinfo {author} {\bibfnamefont {S.}~\bibnamefont
  {Sab}}, \bibinfo {author} {\bibfnamefont {M.~A.}\ \bibnamefont
  {Moreno-Armijos}}, \bibinfo {author} {\bibfnamefont {A.~D.}\ \bibnamefont
  {García-Orozco}}, \bibinfo {author} {\bibfnamefont {G.~V.}\ \bibnamefont
  {Fernandes}}, \bibinfo {author} {\bibfnamefont {Y.}~\bibnamefont {Zhu}},
  \bibinfo {author} {\bibfnamefont {A.~R.}\ \bibnamefont {Fritsch}}, \bibinfo
  {author} {\bibfnamefont {H.}~\bibnamefont {Perrin}}, \bibinfo {author}
  {\bibfnamefont {S.}~\bibnamefont {Nazarenko}},\ and\ \bibinfo {author}
  {\bibfnamefont {V.~S.}\ \bibnamefont {Bagnato}},\ }\href@noop {} {\bibinfo
  {title} {Universal behavior on the relaxation dynamics of
  far-from-equilibrium quantum fluids}},\ \Eprint
  {https://arxiv.org/abs/2603.02182} {arXiv:2603.02182 [cond-mat.quant-gas]}
  \BibitemShut {NoStop}%
\bibitem [{\citenamefont {Liang}\ \emph {et~al.}()\citenamefont {Liang},
  \citenamefont {Wu}, \citenamefont {Paranjape}, \citenamefont {Schittenkopf},
  \citenamefont {Li}, \citenamefont {Schmiedmayer},\ and\ \citenamefont
  {Erne}}]{liang2025universalnonthermalfixedpoint}%
  \BibitemOpen
  \bibfield  {author} {\bibinfo {author} {\bibfnamefont {Q.}~\bibnamefont
  {Liang}}, \bibinfo {author} {\bibfnamefont {R.}~\bibnamefont {Wu}}, \bibinfo
  {author} {\bibfnamefont {P.}~\bibnamefont {Paranjape}}, \bibinfo {author}
  {\bibfnamefont {B.}~\bibnamefont {Schittenkopf}}, \bibinfo {author}
  {\bibfnamefont {C.}~\bibnamefont {Li}}, \bibinfo {author} {\bibfnamefont
  {J.}~\bibnamefont {Schmiedmayer}},\ and\ \bibinfo {author} {\bibfnamefont
  {S.}~\bibnamefont {Erne}},\ }\href@noop {} {\bibinfo {title} {{Universal
  non-thermal fixed point for quasi-1D Bose gases}}},\ \Eprint
  {https://arxiv.org/abs/2505.20213} {arXiv:2505.20213 [cond-mat.quant-gas]}
  \BibitemShut {NoStop}%
\bibitem [{\citenamefont {Kofman}\ \emph {et~al.}(1994)\citenamefont {Kofman},
  \citenamefont {Linde},\ and\ \citenamefont {Starobinsky}}]{Kofman:1994rk}%
  \BibitemOpen
  \bibfield  {author} {\bibinfo {author} {\bibfnamefont {L.}~\bibnamefont
  {Kofman}}, \bibinfo {author} {\bibfnamefont {A.~D.}\ \bibnamefont {Linde}},\
  and\ \bibinfo {author} {\bibfnamefont {A.~A.}\ \bibnamefont {Starobinsky}},\
  }\bibfield  {title} {\bibinfo {title} {{Reheating after inflation}},\ }\href
  {https://doi.org/10.1103/PhysRevLett.73.3195} {\bibfield  {journal} {\bibinfo
   {journal} {Phys. Rev. Lett.}\ }\textbf {\bibinfo {volume} {73}},\ \bibinfo
  {pages} {3195} (\bibinfo {year} {1994})},\ \Eprint
  {https://arxiv.org/abs/hep-th/9405187} {arXiv:hep-th/9405187} \BibitemShut
  {NoStop}%
\bibitem [{\citenamefont {Kofman}\ \emph {et~al.}(1997)\citenamefont {Kofman},
  \citenamefont {Linde},\ and\ \citenamefont {Starobinsky}}]{Kofman:1997yn}%
  \BibitemOpen
  \bibfield  {author} {\bibinfo {author} {\bibfnamefont {L.}~\bibnamefont
  {Kofman}}, \bibinfo {author} {\bibfnamefont {A.~D.}\ \bibnamefont {Linde}},\
  and\ \bibinfo {author} {\bibfnamefont {A.~A.}\ \bibnamefont {Starobinsky}},\
  }\bibfield  {title} {\bibinfo {title} {{Towards the theory of reheating after
  inflation}},\ }\href {https://doi.org/10.1103/PhysRevD.56.3258} {\bibfield
  {journal} {\bibinfo  {journal} {Phys. Rev. D}\ }\textbf {\bibinfo {volume}
  {56}},\ \bibinfo {pages} {3258} (\bibinfo {year} {1997})},\ \Eprint
  {https://arxiv.org/abs/hep-ph/9704452} {arXiv:hep-ph/9704452} \BibitemShut
  {NoStop}%
\bibitem [{\citenamefont {Sikivie}\ and\ \citenamefont
  {Yang}(2009)}]{Sikivie:2009qn}%
  \BibitemOpen
  \bibfield  {author} {\bibinfo {author} {\bibfnamefont {P.}~\bibnamefont
  {Sikivie}}\ and\ \bibinfo {author} {\bibfnamefont {Q.}~\bibnamefont {Yang}},\
  }\bibfield  {title} {\bibinfo {title} {{Bose-Einstein Condensation of Dark
  Matter Axions}},\ }\href {https://doi.org/10.1103/PhysRevLett.103.111301}
  {\bibfield  {journal} {\bibinfo  {journal} {Phys. Rev. Lett.}\ }\textbf
  {\bibinfo {volume} {103}},\ \bibinfo {pages} {111301} (\bibinfo {year}
  {2009})},\ \Eprint {https://arxiv.org/abs/0901.1106} {arXiv:0901.1106
  [hep-ph]} \BibitemShut {NoStop}%
\bibitem [{\citenamefont {Arias}\ \emph {et~al.}(2012)\citenamefont {Arias},
  \citenamefont {Cadamuro}, \citenamefont {Goodsell}, \citenamefont {Jaeckel},
  \citenamefont {Redondo},\ and\ \citenamefont {Ringwald}}]{Arias:2012az}%
  \BibitemOpen
  \bibfield  {author} {\bibinfo {author} {\bibfnamefont {P.}~\bibnamefont
  {Arias}}, \bibinfo {author} {\bibfnamefont {D.}~\bibnamefont {Cadamuro}},
  \bibinfo {author} {\bibfnamefont {M.}~\bibnamefont {Goodsell}}, \bibinfo
  {author} {\bibfnamefont {J.}~\bibnamefont {Jaeckel}}, \bibinfo {author}
  {\bibfnamefont {J.}~\bibnamefont {Redondo}},\ and\ \bibinfo {author}
  {\bibfnamefont {A.}~\bibnamefont {Ringwald}},\ }\bibfield  {title} {\bibinfo
  {title} {{WISPy Cold Dark Matter}},\ }\href
  {https://doi.org/10.1088/1475-7516/2012/06/013} {\bibfield  {journal}
  {\bibinfo  {journal} {JCAP}\ }\textbf {\bibinfo {volume} {06}},\ \bibinfo
  {pages} {013}},\ \Eprint {https://arxiv.org/abs/1201.5902} {arXiv:1201.5902
  [hep-ph]} \BibitemShut {NoStop}%
\bibitem [{\citenamefont {Berges}\ and\ \citenamefont
  {Jaeckel}(2015)}]{Berges:2014xea}%
  \BibitemOpen
  \bibfield  {author} {\bibinfo {author} {\bibfnamefont {J.}~\bibnamefont
  {Berges}}\ and\ \bibinfo {author} {\bibfnamefont {J.}~\bibnamefont
  {Jaeckel}},\ }\bibfield  {title} {\bibinfo {title} {{Far from equilibrium
  dynamics of Bose-Einstein condensation for Axion Dark Matter}},\ }\href
  {https://doi.org/10.1103/PhysRevD.91.025020} {\bibfield  {journal} {\bibinfo
  {journal} {Phys. Rev. D}\ }\textbf {\bibinfo {volume} {91}},\ \bibinfo
  {pages} {025020} (\bibinfo {year} {2015})},\ \Eprint
  {https://arxiv.org/abs/1402.4776} {arXiv:1402.4776 [hep-ph]} \BibitemShut
  {NoStop}%
\bibitem [{\citenamefont {Bloch}\ \emph {et~al.}(2008)\citenamefont {Bloch},
  \citenamefont {Dalibard},\ and\ \citenamefont {Zwerger}}]{Bloch_2008}%
  \BibitemOpen
  \bibfield  {author} {\bibinfo {author} {\bibfnamefont {I.}~\bibnamefont
  {Bloch}}, \bibinfo {author} {\bibfnamefont {J.}~\bibnamefont {Dalibard}},\
  and\ \bibinfo {author} {\bibfnamefont {W.}~\bibnamefont {Zwerger}},\
  }\bibfield  {title} {\bibinfo {title} {Many-body physics with ultracold
  gases},\ }\href {https://doi.org/10.1103/revmodphys.80.885} {\bibfield
  {journal} {\bibinfo  {journal} {Rev. Mod. Phys.}\ }\textbf {\bibinfo {volume}
  {80}},\ \bibinfo {pages} {885–964} (\bibinfo {year} {2008})}\BibitemShut
  {NoStop}%
\end{thebibliography}%
\end{document}